\documentclass[structabstract]{aa}
\usepackage{graphicx}
\usepackage{txfonts}
\usepackage{natbib}
\usepackage{hyperref}
\usepackage{color}

\begin{document}

   \title{Modeling optical and UV polarization of AGNs}

   \subtitle{II. Polarization imaging and complex reprocessing}

   \author{F.~Marin\inst{1}\thanks{\email{frederic.marin@astro.unistra.fr}}
     \and R.~W. Goosmann\inst{1}
     \and C.~M. Gaskell\inst{2}
     \and D.~Porquet\inst{1}
     \and M.~Dov{\v c}iak\inst{3}
     }

   \institute{Observatoire Astronomique de Strasbourg, Universit\'e de Strasbourg,
     CNRS, UMR 7550, 11 rue de l'Universit\'e, 67000 Strasbourg, France
     \and Centro de Astrof\'isica de Valpara\'iso y Departamento de
     F\'isica y Astronom\'ia, Universidad de Valpara\'iso, Av. Gran
     Breta\~na 1111, Valpara\'iso, Chile
     \and Astronomical Institute of the Academy of Sciences, Bo{\v c}ni
     II 1401, 14131 Prague, Czech Republic }

   \date{Received June 4th, 2012; Accepted ?? ??th, 2012}

  \abstract
  {The innermost parts of active galactic nuclei (AGNs) are
    believed to be comprised of several emission and scattering media
    coupled by radiative processes. These regions generally cannot be spatially resolved.
    Spectropolarimetric observations give important information about
    the reprocessing geometry.}
  {We aim to obtain a coherent model of the polarization signature resulting
    from the radiative coupling between the components, to compare our
    results with polarimetry of thermal AGNs and thereby to put
    constraints on the geometry.}
  {We use a new public version of {\sc stokes}, a Monte Carlo radiative
    transfer code presented in the first paper of this series. The code
    has been significantly improved for computational speed and polarization
    imaging has been implemented. The imaging capability helps to improve understanding of 
    the contributions of different components to the spatially-integrated flux.  
    We couple continuum sources with a variety
    of reprocessing regions such as equatorial scattering
    regions, polar outflows, and toroidal obscuring dust and we
    study the resulting polarization. We
    explore combinations and compute a grid of thermal AGN models
    for different half-opening angles of the torus and polar
    winds. We also consider a range of optical depths for
    equatorial and polar electron scattering and investigate how
    the model geometry influences the type-1/type-2
    polarization dichotomy for thermal AGNs (type-1 AGNs tending to have 
    polarization parallel to the axis of the torus and type-2 AGNs tending to 
    have polarization perpendicular to it).}
  {We put new constrains on the inflowing medium within
    the inner walls of the torus. To
    reproduce the observed polarization in type-1 objects,
    the inflow should be confined to the common equatorial plane of
    the torus and the accretion disc and have a radial optical
    depth of $1 < \tau < 3$. Our modeling of type-1 AGNs
    indicates that the torus is more likely to have a large ($\sim
    60^\circ$) half-opening angle. Polarization perpendicular to the axis of the torus may
    arise at a type-1 viewing angle for a torus
    half-opening angle of $30^\circ$ -- $45^\circ$ or polar outflows
    with an optical depth near unity. Our modeling suggests that
    most Seyfert-2 AGN must have a half-opening angle $> 60^\circ$
    to match the level of perpendicular polarization expected. If outflows are
    collimated by the torus inner walls, they must not be
    optically thick ($\tau < 1$) in order to preserve the
    polarization dichotomy. The wind's optical depth is found not
    to play a critical role for the degree of polarization of type-2
    thermal AGNs but it has a significant impact on the
    type-1/type-2 polarization dichotomy when the optical depth
    exceeds $\tau = 0.3$.}
  {}

\keywords{Galaxies: active -- Galaxies: Seyfert -- Polarization --
	 Radiative transfer -- Scattering}

\maketitle


\section{Introduction}

Active galactic nuclei (AGN) are divided observationally into a number of
(sub-)classes based on the optical spectrum and radio properties.  If the broad-line region
(BLR) is directly visible in the optical an AGN is called a ``type-1'', while if it is not it 
is called a ``type-2''. AGNs can also be usefully divided into ``radio-loud'' and ``radio-quiet'' 
depending on the relative strength of the kpc-scale radio jets and lobes.

\citet{Blandford1978} suggested that BL Lac objects were radio-loud AGNs 
with their jets aimed close to our line of sight. \citet{Keel1980} discovered that nearby 
type-1 AGNs are preferentially face-on while nearby type-2 AGNs are randomly oriented.  
These discoveries showed that the viewing angle of the observer is a key element in AGN classification
\citep{Antonucci1993}. The apparent absence of a BLR in type-2 AGNs is explained by a dusty medium that hides 
the BLR at certain viewing angles from the observer. The ionizing continuum source and its surrounding 
BLR sit inside the funnel of the torus. In such a scenario the half-opening angle 
of the torus can be estimated from the ratio of type-1 to type-2 AGNs in an isotropically-selected 
sample and from the ``infra-red calorimeter'' -- the relative strength of the mid-IR continuum which 
arises from dust reprocessing of the higher-energy continuum (see \citealt{Antonucci2011}).
There is evidence that the opening angle is a function of luminosity \citep[]{Lawrence1991} 
or accretion history \citep{Wang2007}.

The distinction between radio-loud and radio-quiet objects is not orientation-independent
and thus it cannot be explained as a geometry effect.
Furthermore, it has been argued by \citet{Antonucci2011} that there is a fundamental
difference in the dominant energy generation between high-accretion-rate sources, which are believed to 
release the bulk of their radiated energy via thermal disk emission in the far ultraviolet 
(see \citealt{Gaskell2008b} and low-accretion-rate AGNs, which are dominated by non-thermal emission. 
The latter class of AGNs would include the so-called ``naked'' AGNs \citep{Georgantopoulos2003, Gliozzi2007, Bianchi2008}
\footnote{``Naked'' AGNs are a subclass of radio-quiet Seyfert 2 galaxies (so-called ``non-hidden-BLR'' Seyfert 2s = NHBLR Seyfert 2s)
but show a smaller accretion regime than their parents. See \citet{Zhang2006} for a parallel between NHBLR Seyfert 2s and 
Narrow Line Seyfert 1 AGN (NLS1) which both show high accretion rates.}. 
The transition between the two modes of dominant energy generation would take place at an Eddington ratio 
of $\sim 10^{-3}$ \citep{Antonucci2011}. Following \citet{Antonucci2011} we will refer to these two 
types of AGNs as ``thermal AGNs'' and ``non-thermal AGNs'' respectively\footnote{In 
the literature on possible role of AGN feedback on galaxy growth the two modes are sometimes referred 
to as ``quasar mode'' and ``radio mode'' but we prefer the labels ``thermal'' and ``non-thermal''
since historically the word ``quasar'' refers to a high-luminosity, radio-loud AGN. While all 
``quasars'' are thermal AGNs, the converse is not true as the majority of thermal AGNs are of low luminosity.}. 
The thermal/non-thermal dichotomy is the causal consequence of the accretion mode (related to the accretion rate)
that has an impact on the presence of the accretion disc and thus on the eventual observation of its thermal emission.
By definition, thermal AGNs always have a big blue bump. As far as is known (see \citealt{Antonucci2011}) 
they also always have a BLR, although the BLR might be hidden, as in type-2 AGNs
\footnote{There is on-going debate over where there are NHBLR Seyfert 2s. \citet{Antonucci2011} argues that we cannot 
say from existing evidence that NHBLR Seyfert 2s exist.}. 
Non-thermal AGNs are FR I radio galaxies and LINERs. 
It is important to note that the thermal/non-thermal distinction is {\em not} the same as the 
radio-quiet/radio-loud distinction. A radio-loud AGN can be either a thermal AGN 
(for example, 3C~273) or a non-thermal AGN (such as M87). For completeness we also point out 
that optically violently variable (OVV) behavior in blazars can arise from both thermal and non-thermal AGNs. 
All that matters is that we view the AGN with a jet aimed close to our line of sight. 
Because of the strong intrinsic polarization of the synchrotron emission in the jets we do 
not consider blazars here.

In this work, we focus on polarization modeling of radio-quiet AGNs that
release the bulk of their radiated energy in the far UV. Our goal
is to infer clues about the geometry and composition of the different AGN components from
the observed polarization properties. We assume that in thermal AGNs the emission comes
from an optically-thick accretion disc \citep{Lynden-Bell1969}. Closely associated with
the accretion disc, and occupying approximately the same range of radius, is the BLR -- a
supposedly flattened region of high-density ($n_e \sim 10^{10}$ cm$^{-3}$), rapidly-moving
($v \sim 0.05 - 0.001 c$) gas which is optically thin in the continuum for $h\nu < 13.6$ eV
\citep[see][for a review]{Gaskell2009}. 

Not all matter spiraling into the gravitational potential of the black hole becomes accreted.
In many radio-quiet AGN strong winds are seen in the X-ray and UV spectrum
\citep{Mathur1994,Mathur1995,Costantini2010}. Supposedly, these winds are expelled very
close to the black hole and find a spatial continuation in the observed polar ionization
cones. In some objects the outflowing gas can be seen as broad absorption lines \citep{Weymann1991,Knigge2008}. 
Immediately outside the BLR and accretion disc is the geometrically-thick, dusty torus \citep[but see][and references
therein for the interpretation of the equatorial obscuration as a wind]{Elvis2000,Elvis2004,Kazanas2012}.
Finally, starting on a scale larger than the BLR and often extending to distances much greater than the size of 
the torus there is the lower-density, lower-velocity gas of the so-called ``narrow-line region'' (NLR).

The accretion flow at the outer accretion disc and the inner boundary
of the torus funnel is difficult to assess observationally. At a distance of
about a thousand gravitational radii from the black hole the accretion flow should
become gravitationally unstable and become fragmented \citep[see e.g.][]{Lodato2007}.
There are indications that the medium is continuous while being inhomogeneous in density
\citep{Elitzur2007}. In this context the
interaction between the primary radiation from the accretion disc with the
surrounding media causes scattering-induced polarization signatures. In a single
electron or dust scattering event, the resulting polarization degree and
position angle depend on the scattering angle and therefore the resulting
net polarization of the collective reprocessing in an AGN allows us, when
accurately modeled, to draw conclusions about the geometry of the scattering
regions. Some evidence for a flattened geometry of the matter just inside the torus comes from optical polarimetry.
For many type-1 objects the $\vec E$-vector of the continuum
radiation aligns with the projected axis of the (small scale) radio structures
\citep{Antonucci1982,Antonucci1983}. Identifying the radio structures with collimated
outflows progressing along the symmetry axis of the dusty torus, such a polarization
state is most easily explained by scattering in a flattened, equatorial scattering
region \citep{Antonucci1984}. The polarization state of the radiation thus brings
informations about the reprocessing geometry. Based on an analysis of the polarization
structure across broad emission lines, \citet{Smith2004} suggested that the equatorial
scattering region may be partly intermixed with or lying slightly further out than the BLR. 

This is the second paper of a series, in which we systematically examine the polarization
response due to the combination of different scattering regions. We take several
steps in building up models representative of, double-component AGNs such like
''naked'', ''bare'', FR I or LINERS-like galaxies to then present a multi-component
model for thermal AGNs and examine how the polarization spectra and images are
influenced by the different AGN constituents. We apply the latest, publicly available
version 1.2 of the radiative transfer code {\sc stokes}. Source codes, executables
and a manual of {\sc stokes}~1.2 can be found on the web\footnote{http://www.stokes-program.info/}.

The remainder of the paper is organized as follows: In Sect.~\ref{sec:stokes} we
briefly resume our previous modeling work on AGN and we describe the new elements
of {\sc stokes}~1.2. In Sect.~\ref{sec:indiscat}, polarization images for individual
AGN scattering regions are studied. Consistent models for the combination of two
scattering regions are presented in Sect.~\ref{sec:combined} and a three-component
model approximating the unified AGN model is explored in Sect.~\ref{sec:AGN}.
In Sect.~\ref{sec:discuss} we discuss our results and relate them to the work
done by other groups before we draw some conclusions in Sect.~\ref{sec:conclu}.

\section{Applying the radiative transfer code STOKES}
\label{sec:stokes}

In  \citep{Goosmann2007}, hereafter refereed as paper~I, we published polarization modeling for 
reprocessing by different individual scattering components. In paper 1 we studied centrally 
illuminated dusty tori, polar outflows, and equatorial scattering regions with different
geometries. The spectral flux and polarization were computed and discussed
in the context of observed spectropolarimetric data and of the modeling work
done by other groups. In subsequent work, we have explored the polarization
induced by different dust compositions in the torus or the polar outflow
\citep{Goosmann2007c,Goosmann2007b}, ompared two competitive scenarios 
explaining the emergence of the broad Fe K$\alpha$ line in the Seyfert 1 galaxy
MCG-6-30-15 \citep{marin2012b}, investigated time-dependent
polarization due to reverberation \citep{Goosmann2008,Gaskell2012}, and investigated the
effect of an inflow velocity of the scattering medium on the shape of AGN
emission lines \citep{Gaskell2008}.

In the above mentioned studies we focussed on the effects of single scattering regions. 
It is important to consider the effects of combinations of scattering regions. For example, by putting
together a dusty torus, ionized polar outflows
and equatorial scattering it is possible to reproduce the observed dichotomy
of the polarization position angle between type-1 and type-2 AGN
\citep{Goosmann2007d} and even to derive predictions for this dichotomy
in the X-ray range (\citealt{Marin2011a, Marin2012}).
In the present paper, we therefore systematically investigate the polarization due to
scattering in various combinations of reprocessing regions. We use the
latest, publicly available version, {\sc stokes 1.2}, of our radiative
transfer code. In this section, we describe the most significant
improvements and changes made with respect to the previous version
{\sc stokes}~1.0.

\subsection{Improved random number generator}

The standard deviation of a Monte Carlo calculation is usually proportional to
the square root of the number of samples. In radiative transfer
modeling, enough photons need to be sampled to suppress the Poisson
noise with respect to significant spectral or polarization features.
In many of the models shown below we therefore sampled $10^9$ photons
or more. The generation of such long series of random numbers may run
into numerical problems. In {\sc stokes}~1.0 a linear congruential
generator (LCG) was used, which is fast and efficient only for short
series. For the large sampling numbers that are typically applied in
the complex modeling presented here, the LCG tends to loop back on
series of values it has sampled before. In {\sc stokes}~1.2, we
therefore implemented a version of the Mersenne Twister Generator
(MTG) algorithm \citep{Matsumoto1998}. The MTG generates
pseudo-random numbers using a so-called Twisted Generalized Feedback
Shift Register. The most common version, MT~19937, has a very high
period of $2^{19937}~-~1$ and provides a 623-dimensional
equidistribution up to an accuracy of 32 bits. The MTG is more
efficient than most other random number generators and passes the
Diehard tests described in \citet{Marsaglia1985}. A detailed analysis
of the MTG is given in \citet{Matsumoto1998} and references
therein. Using the MTG in {\sc stokes} substantially improves the
statistics for a given sampling number of photons. Although each call
of the MTG requires 20\% more computation time compared to the LCG,
far fewer photons are needed to obtain the same limit on the
Poissonian fluctuations. Therefore, there is a significant net gain in
computation time and the results converge faster.

  \begin{figure}
   \centering 
      \includegraphics[width=10cm]{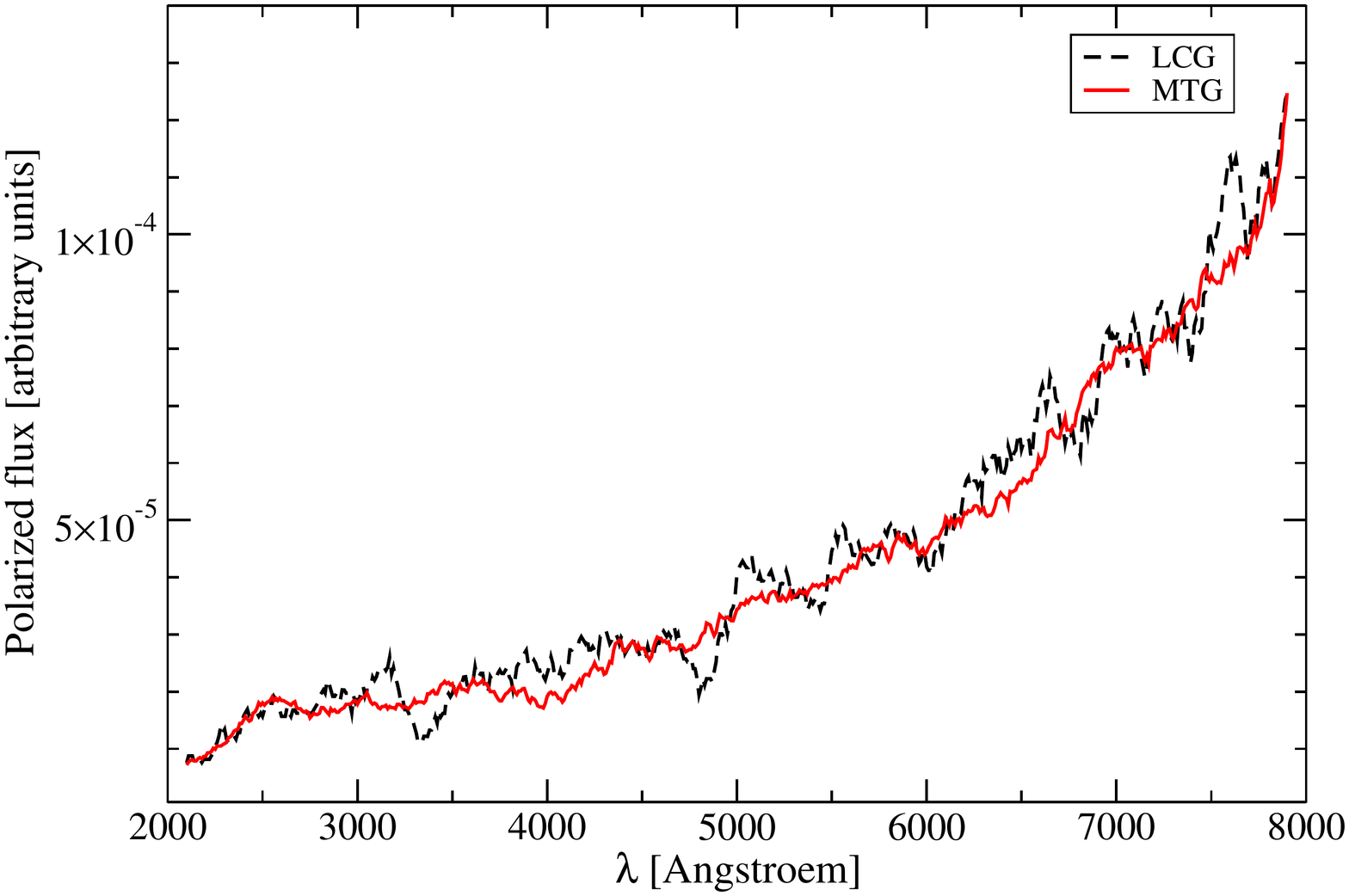}
      \caption{Modeling a central, isotropic source irradiating a
        dusty, equatorial scattering torus with a half-opening angle
        of $30^\circ$ (see text). The spectrum represents the total
        flux, \textit{F/F$_{\rm *}$}, normalized to the source flux
        \textit{F$_{\rm *}$}, for a viewing angle of $i$~=~$72^\circ$.
        The spectra were calculated sampling the same number of
        photons but using two different random number generators.
        \textit{Black dashed line}: The LCG algorithm as implemented
        in {\sc stokes}~1.0; \textit{Red solid line}: The MTG
        algorithm as implemented in {\sc stokes}~1.2}.	
       \label{Fig1}
  \end{figure}

We show a comparison of both methods in Fig.\ref{Fig1}. We model an
AGN obscuring torus sampling a total of $3.10^8$ photons, alternately
using the LCG random number generating algorithm of {\sc stokes}~1.0
and the MTG algorithm implemented in {\sc stokes}~1.2. We assume a
large, dusty torus with an elliptical cross-section, constant density
and a $V$-band optical depth of $\tau_{\rm V}$ $\sim 600$. The
half-opening angle of the scattering region with respect to the
symmetry axis of the torus equals $\theta_0$~=~$30^\circ$. The dust
has a ``Milky Way'' composition (see section 4.2 of Paper~I for more
details). We chose this model to illustrate the effects of the random
number generator because the spectra suffer from heavy absorption and
only multiply scattered photons escape from the funnel of the torus. As a
result, the statistics at the particular viewing angle of
$i$~=~$72^\circ$ tends to be low. Fig.\ref{Fig1} shows that, for the
same number of simulated photons, the simulation using the LCG
algorithm suffers from much worse Poissonian noise. To obtain a
curve as smooth as the one obtained with the MTG algorithm LCG would
have generate a dozen times as many photons.

\subsection{Polarization imaging}

In order to understand scattering by various regions in AGNs (see next subsection) it
is very useful to be able to see polarization maps. These can also potentially be
compared with future observations. With {\sc stokes}~1.2 it is now possible to 
generate polarization images.
For this, each photon is spatially ``located'' before its escape from
the modeling region and projected onto the observer's plane of the sky.
The resulting polarization maps can be compared to polarization imaging
observations of spatially resolved objects. In unresolved objects, the model
images can help to study the interplay of several scattering components.

\begin{figure}
   \centering
   \includegraphics[trim = 0mm 40mm 0mm 40mm, clip, width=10.5cm]{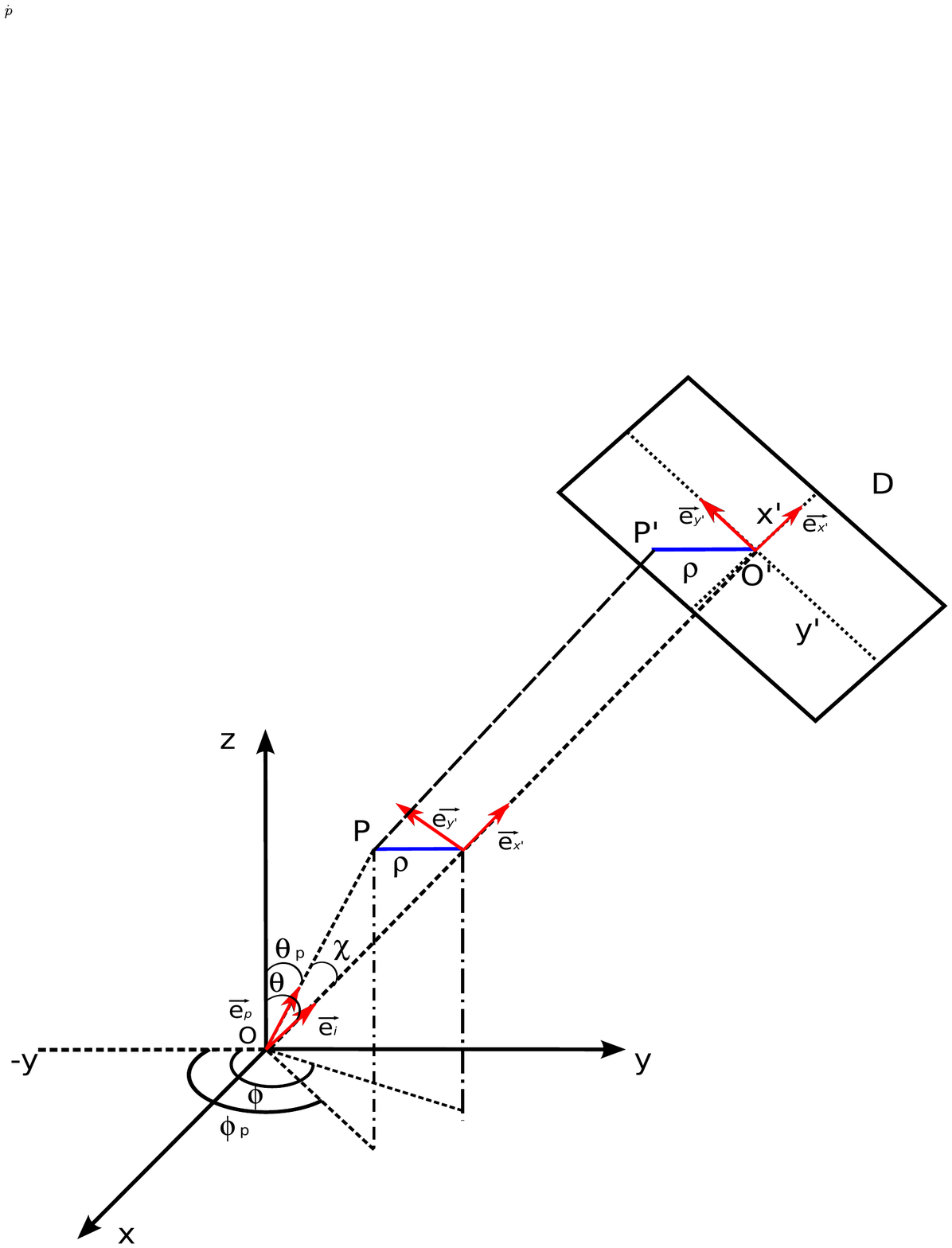}
      \caption{Spatial coordinates that determine the position of a photon $P$
	       and its projection $P'$ onto the plane of the sky, $D$.}
      \label{Fig2}%
   \end{figure}

We project the position of an escaping photon onto a distant plane, $D$,
being orthogonal to the line of sight (Fig.~\ref{Fig2}). The photon position
$P$ is determined by its distance, $r_p$, from the origin $O$ of the model
space and by the two angles $\theta_p$ and $\phi_p$. The center of the projection
plane $O'$ is connected to $O$ by the segment $OO'$, which determines the
polar and azimuthal viewing angles $\phi$ and $\theta$. The distance between
$O$' and the projected photon position $P'$ is denoted by the vector
$\overrightarrow \varrho$. By expressing $\overrightarrow \varrho$ in the
local frame of $D$, we obtain the projected photon coordinates $x'$ and
$y'$ in the plane of the sky.

The unit vectors $\overrightarrow{e_p}$ and $\overrightarrow{e_i}$ are expressed
in spherical coordinates:

   \begin{equation}
	\overrightarrow{e_p} = \left(
	\begin{array}{ccc}
	  \sin\phi_p\sin\theta_p \\ -\cos\phi_p\sin\theta_p \\ \cos\theta_p
        \end{array} \right),
   \end{equation}

   \begin{equation}
	\overrightarrow{e_i} = \left(
	\begin{array}{ccc}
	  \sin\phi \sin i \\ -\cos\phi \sin i \\ \cos i
        \end{array}
        \right),
   \end{equation}

while the unit vectors of the observer's plane of the sky can be written as

   \begin{equation}
  	\overrightarrow{e_{x'}} = \left(
	\begin{array}{ccc}
	  \cos\phi \\ \sin\phi \\ 0
        \end{array} \right), \label{x}
   \end{equation}

   \begin{equation}
	\overrightarrow{e_{y'}} = \left(
	\begin{array}{ccc}
	  -\sin\phi \cos i \\ \cos\phi \cos i \\ \sin i
        \end{array}
        \right). \label{y}
   \end{equation}

We then construct the vector
$\overrightarrow \varrho = r_p[\overrightarrow{e_p} - \cos\chi \overrightarrow{e_i}]$
and obtain its components in spherical coordinates:

   \begin{equation}
	\varrho_x = r_p[\sin\phi_p \sin\theta_p - \cos\chi \sin\phi \sin i ],
   \end{equation}

   \begin{equation}
	\varrho_y = r_p[\cos\chi \cos\phi \sin i - \cos\phi_p \sin\theta_p],
   \end{equation}

   \begin{equation}
	\varrho_z = r_p[\cos\theta_p - \cos\chi \cos i],
   \end{equation}

with $\chi$ being the angle between $\overrightarrow{e_p}$ and $\overrightarrow{e_i}$.
Using equations (\ref{x}) and (\ref{y}), we finally obtain the projected coordinates
$x' = \overrightarrow{\varrho} \overrightarrow{e_{x'}}$ and
$y' = \overrightarrow{\varrho} \overrightarrow{e_{y'}}$ of the photon in the plane of the sky:

   \begin{equation}
	x' = - r_p \sin\theta_p \sin(\phi-\phi_p),
   \end{equation}

   \begin{equation}
	y' = r_p[\sin i \cos\theta_p - \cos i \sin\theta_p \cos(\phi-\phi_p)].
   \end{equation}


\section{Polarization maps of individual scattering regions}
\label{sec:indiscat}

In a model with multiple reprocessing components, theoretical polarization maps are
very useful for understanding the impact of individual scattering regions. We first test
the new imaging routines of {\sc stokes} by reanalyzing some of the individual
reprocessing regions presented in Paper~I. For the remainder of this paper,
we define a {\it type-1} view of a thermal AGN by a line of sight towards the central
source that does not intercept the torus. In this case the viewing angle (measured from the pole)
is smaller than the half-opening angle of the torus. If this is not the case we call the
AGN a type-2.
Polarization is described as ``parallel'' when the $\vec E$-vector
is aligned with the projected torus axis (polarization position angle
$\gamma = 90^\circ$). 
Sometimes we denote the difference between parallel and perpendicular polarization by 
the sign of the polarization percentage, $P$: a negative value of $P$ stands for parallel polarization, 
a positive $P$ for perpendicular one. Finally, both the spectra and the maps will show the total
(linear plus circular) polarization, $P$. Due to dust scattering the $V$ Stokes parameter 
can be non-zero, but in all our models the circular polarization was found to be
a hundred times lower than the linear polarization and thus it does not have an impact on the polarization results.

\subsection{Modeling a large, dusty torus}
\label{sec:inditorus}

The spectral properties of such large, dusty scattering torus were presented in
Section 4.2 of Paper~I. For all the models presented below we define an isotropic,
point-like source emitting an unpolarized spectrum with a power-law spectral energy distribution
$F_{\rm *}~\propto~\nu^{-\alpha}$ and $\alpha = 1$ at the centre of the torus. The inner and outer radii of the torus are set to
0.25~pc and 100~pc, respectively. The radial optical depth of $\tau_{\rm V}$ measured
inside the equatorial plane is taken to be $\sim 750$ in the $V$-band. The half-opening
angle of the torus is set to $\theta_0 = 30^\circ$ with respect to the
vertical axis.

\subsubsection{Wavelength-integrated polarization images}

We sample a total of 10$^9$ photons and model spectra and images at 20 polar viewing
angles \textit{i} and 40 azimuthal angles $\phi$; \textit{i} and $\phi$ are the
viewing angles defined as in Fig.\ref{Fig2}. The angle \textit{i} is measured between
the line of sight and the \textit{z} axis; $\phi$ is measured between the projection of
the line of sight onto the $xy$-plane and the \textit{x} axis. A rather fine stratification
in viewing angle is necessary to limit the image distortion that occurs preferentially at
polar angles. The meshes of the coordinate grid have a different shape at the poles,
where they are more ``trapezium-like'', than at the equator where they are almost
``square-like''. The spectra are presented in terms of $\cos \textit{i}$ providing
an equal flux per angular bin for an isolated, isotropic source located at the
center of the model space. Since the model is symmetric with respect to the torus
axis, we average all Stokes parameters over $\phi$ and thereby improve the statistics.
As in Fig.\ref{Fig3.2} we will present imaging results at three different polar angles :
\textit{i} $\sim 18^\circ$ (near to pole-on view), \textit{i} $\sim 45^\circ$
(intermediate viewing angle), and \textit{i} $\sim 87^\circ$ (edge-on view). These lines
of sight roughly represent AGN of type-1, of an intermediate type between type-1/type-2,
and of type-2, respectively. We define a spatial resolution of 30 bins for the $x'$ and
$y'$ axes so that the photon flux is divided into 900 pixels. Each of these pixels is
labeled by the coordinates $x'$ and $y'$ (in parsecs) and stores the spectra of the
four Stokes parameters across a wavelength range of 1800~\AA~to 8000~\AA. Ultimately,
each pixel contains the same type of spectral information that is provided by the
previous, non-imaging version of {\sc stokes}.

   \begin{figure}
   \centering 
      \includegraphics[width=10cm]{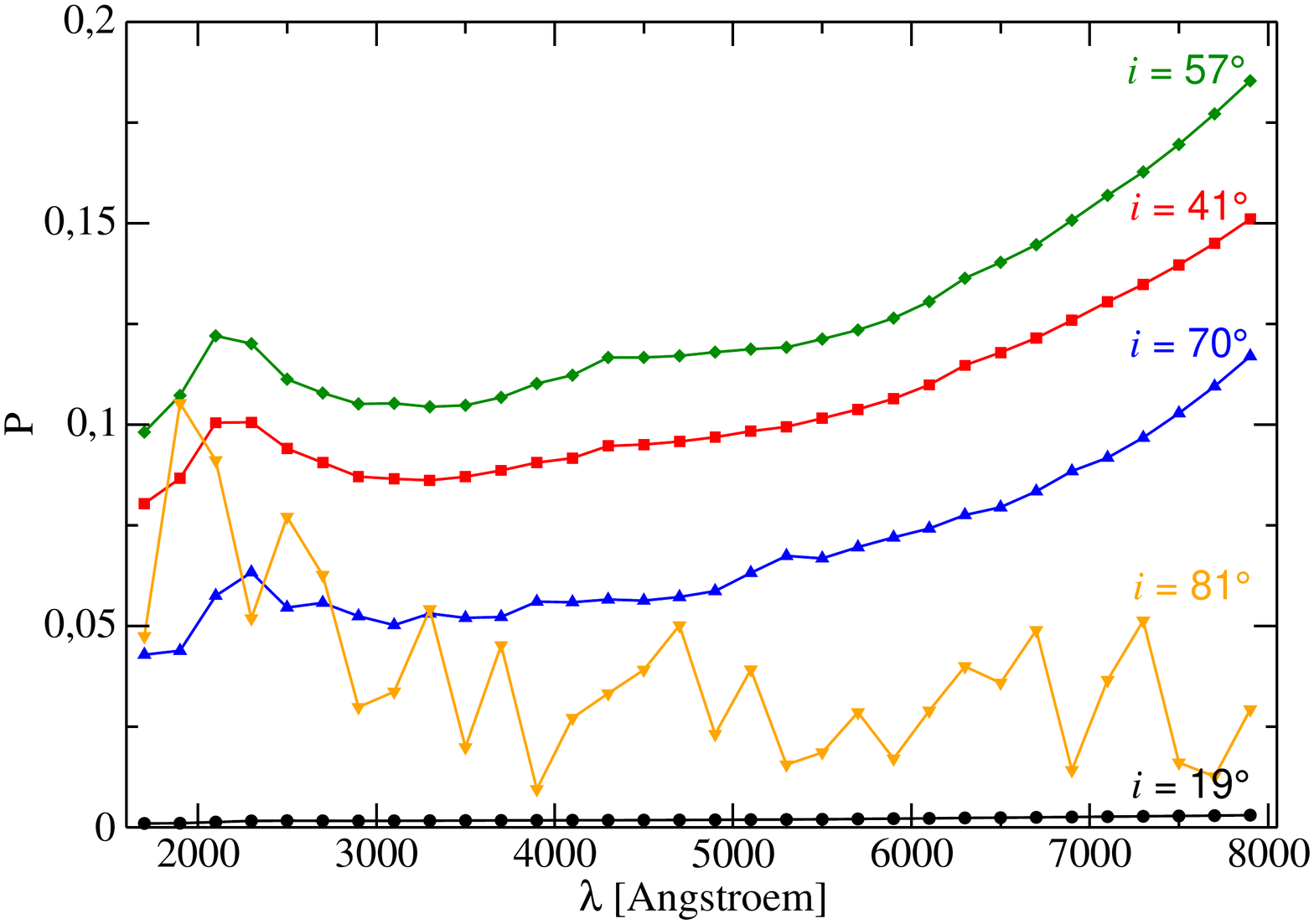}
      \includegraphics[width=10cm]{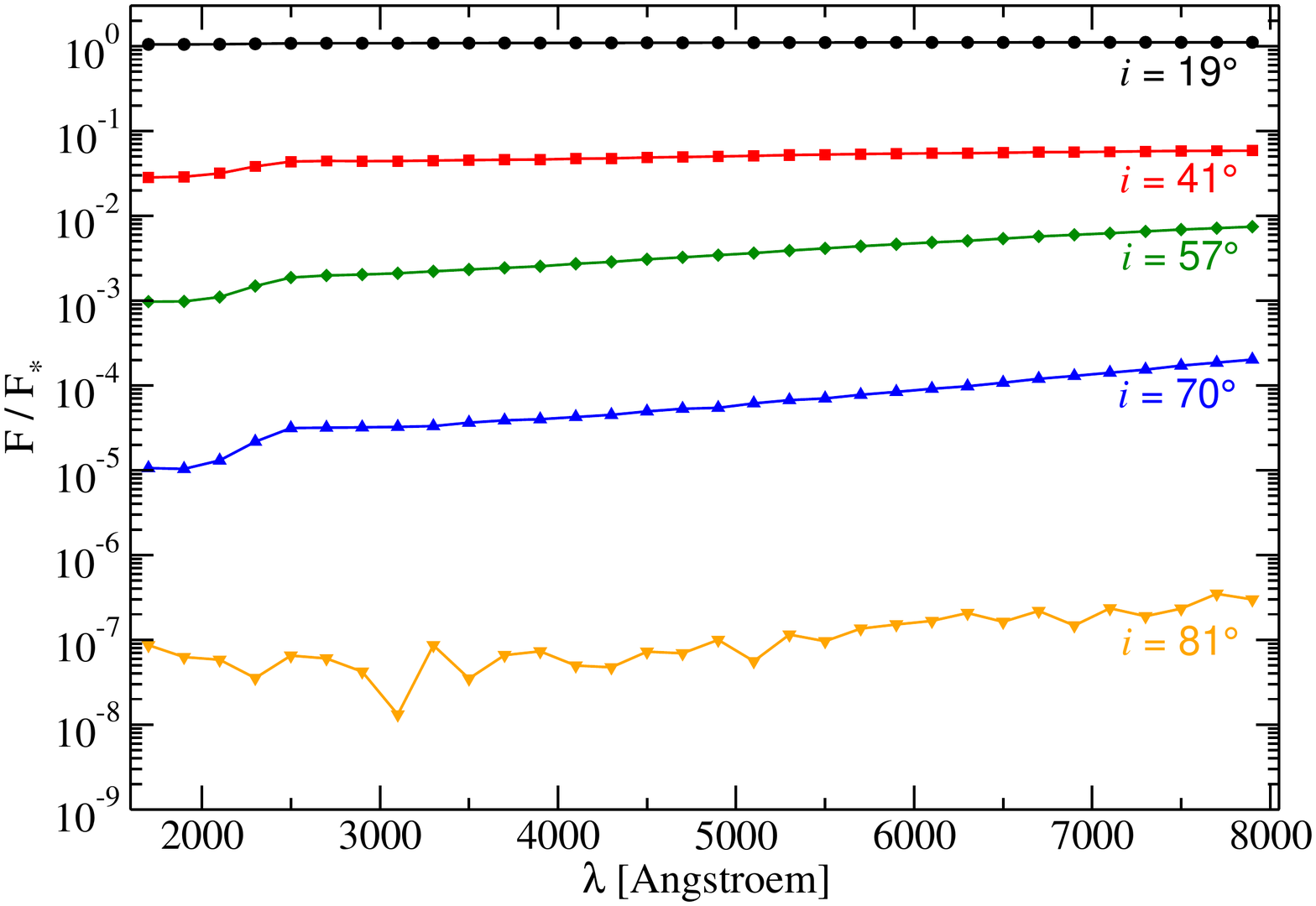}
      \caption{Modeling an optically-thick elliptically-shaped torus
        with $\theta_0 = 30^\circ$ measured relative to the symmetry axis.
        The spatially integrated polarization \textit{P} (upper panel) and
	the fraction $F/F_{\rm *}$ of the central flux (lower panel) are seen
	at different viewing inclinations, \textit{i}.}
     \label{Fig3.1}%
   \end{figure}
%

The polarization and flux spectra integrated over $x'$ and $y'$ are shown in
Fig.~\ref{Fig3.1}. As expected, they are in agreement with previous results
(see section 4.2 and Fig.~6 of Paper~I for more a details). The polarization
images are shown in Fig.~\ref{Fig3.2}. Due to the large size of the torus we
apply a zoom on the imaging and analyze in particular the torus funnel from
where most scattered radiation emerges. The maps simultaneously show the
polarized flux, $PF/F_{*}$, the polarization position angle, $\gamma$, and
the percentage of polarization, $P$; $\gamma$ and $P$ are represented by
black vectors drawn in the center of each spatial bin. A vertical vector indicates
a polarization of $\gamma = 90^\circ$, a slash to the right denotes
$90^\circ > \gamma > 0^\circ$ and a horizontal vector stands for an angle
of $\gamma = 0^\circ$. The length of the vector is proportional to $P$.
The Stokes parameters are integrated over all wavelengths
and all azimuthal viewing angles $\phi$. Note that in all images the polarized
flux $PF/F_{*}$ is normalized to the central flux $F_{\rm *}$ that is emitted
into the same viewing direction.

When the torus is seen pole-on (Fig.~\ref{Fig3.2} top, $i \backsim 18^\circ$), most of the
polarized flux comes from its inner edge, which is best exposed to the source.
The medium is optically thick, and so the scattering occurs mainly close to the
surface of the scattering region. The unpolarized light from the source has a
depolarizing effect and so does the spatial distribution of the inner edge of
the torus that is almost symmetric with respect to the line of sight. As our
``pole-on'' viewing angle $i = 18^\circ$ is effectively off-axis (different from
$i = 0^\circ$), the shape of the scattering region is slightly deformed in the
projection process and does not appear axis-symmetric. Also, due to the slight
inclination, the image does not record much of the flux being scattered off of
the nearest surface of the inner torus walls.

The image in Fig.~\ref{Fig3.2} (top) illustrates the discussion on
the net polarization induced by scattering off of the inner surfaces of
a dusty torus (see \citet{Kartje1995} and Paper~I): the surfaces parallel
to the observer's line of sight produce a polarization vector tending towards
an orientation of $90^\circ$, while the surfaces oriented perpendicularly
with respect to the line of sight produce a polarization vector at $0^\circ$.
At an intermediate viewing angle (Fig.~\ref{Fig3.2} bottom), the effects of extinction
by dust become very strong. The inner wall opposite the observer is still
visible but the walls on the side and the nearest inner surface disappear
below the torus horizon. Often, the photons must undergo multiple scatterings
inside the torus funnel before they escape and are observed.
As the absorption probability increases with the number of scattering events,
we thus observe a lower polarized flux than at pole-on view. For a line of
sight near the equator, no flux is observed due to complete absorption by
the optically-thick dust. Therefore, we do not present the polarization map
at an angle \textit{i} of $\sim 87^\circ$.

   \begin{figure}
   \centering
      \includegraphics[trim = 5mm 5mm 0mm 10mm, clip, width=8cm]{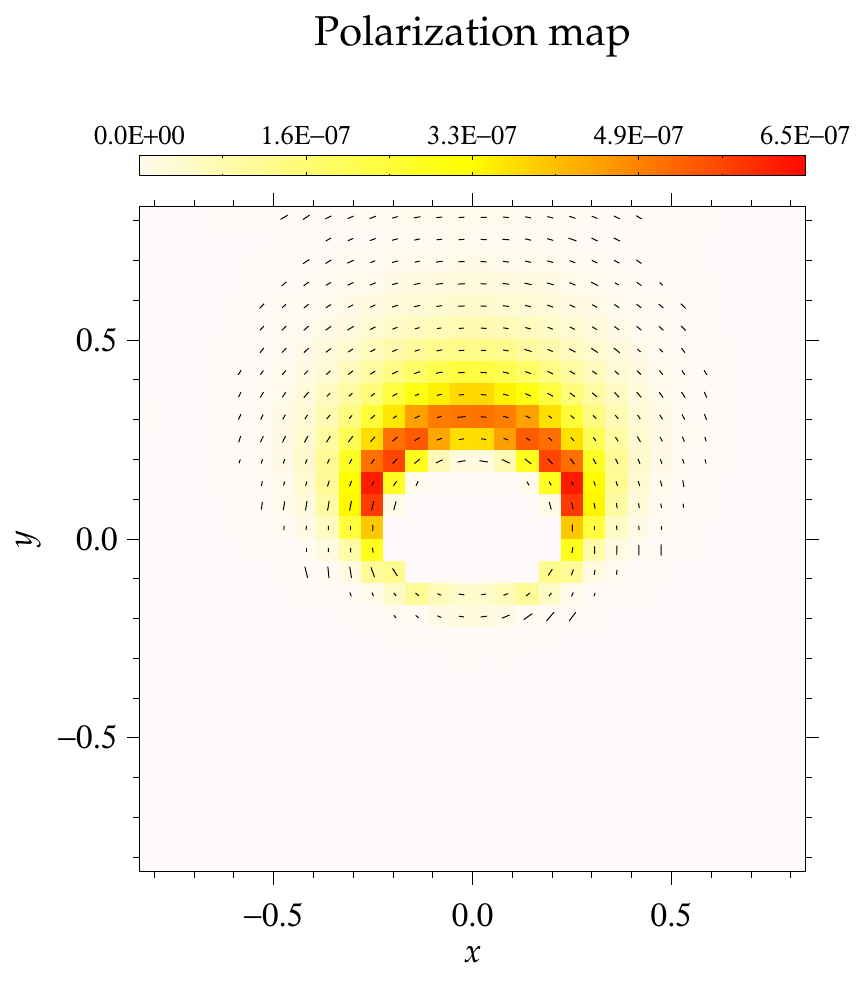}
      \includegraphics[trim = 5mm 5mm 0mm 10mm, clip, width=8cm]{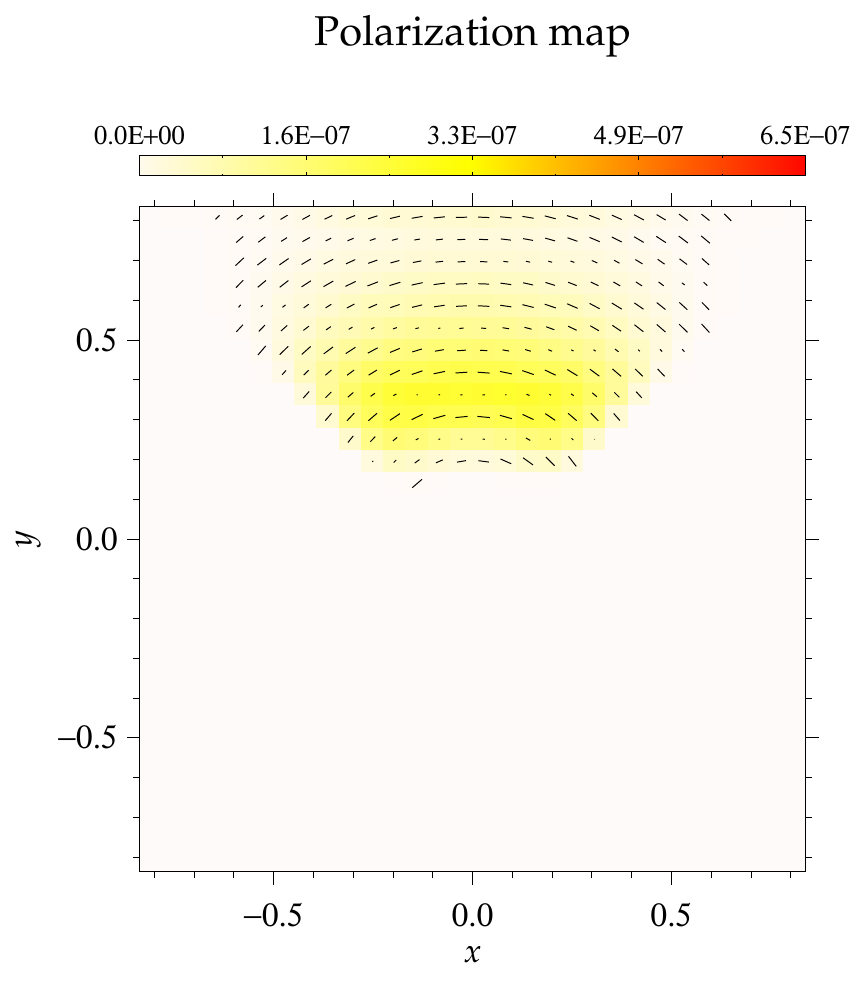}
      \caption{Model image of the polarized flux, $PF/F_{*}$, for a
        centrally-irradiated optically-thick, elliptically-shaped
        torus with $\theta_0$~=~$30^\circ$ measured relative to the
        symmetry axis; $PF/F_{*}$ is color-coded and integrated
        over the wavelength band covered.
	\textit{Top}: face-on view at $i \backsim 18^\circ$;
        \textit{Bottom}: view at $i \backsim 45^\circ$.}
     \label{Fig3.2}
   \end{figure}

\subsubsection{UV to optical polarization images}
\label{first:torus_wav}

\begin{figure}
   \centering 
      \includegraphics[trim = 5mm 5mm 0mm 10mm, clip, width=8cm]{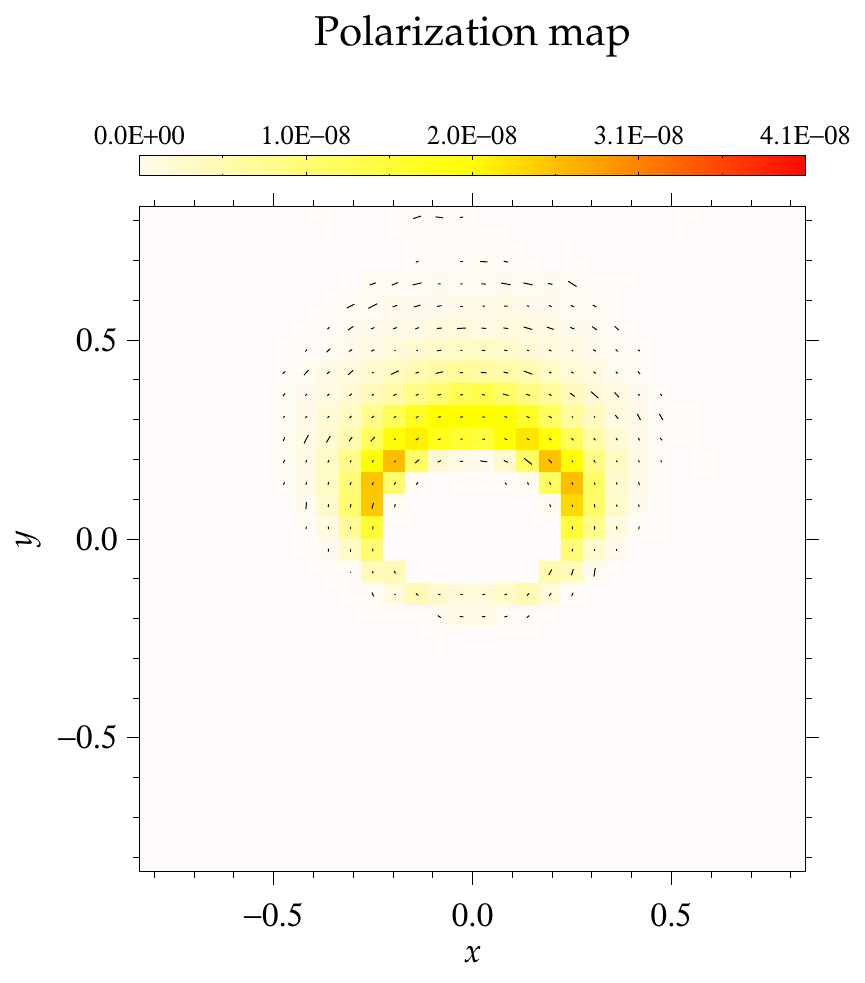}
      \includegraphics[trim = 5mm 5mm 0mm 10mm, clip, width=8cm]{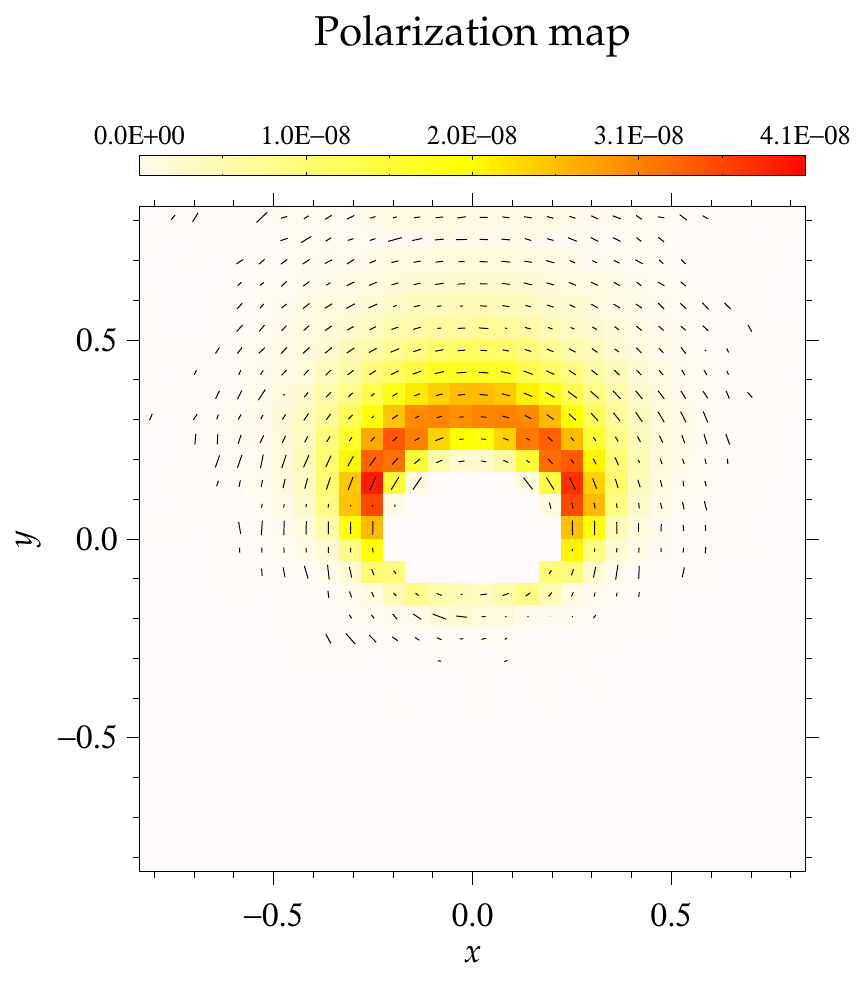}
      \caption{Modeled image of the polarized flux, $PF/F_{*}$, for a
        centrally irradiated, optically-thick torus with
        $\theta_0$~=~$30^\circ$ measured relative to the axis of
        symmetry; $PF/F_{*}$ is color-coded and integrated over the
        wavelength band.
	\textit{Top}: face-on image at $\backsim$ 2175~\AA;
	\textit{Bottom}: face-on image at $\backsim$ 7500~\AA.}
     \label{Fig3.3}
   \end{figure}

The net polarized flux coming from a given position on the torus inner walls
is determined by a complex interplay between the dust albedo and the scattering
phase function. Both of these properties are a function of wavelength.
We therefore illustrate the wavelength-dependence of the polarization map in
Fig.~\ref{Fig3.3}. The two maps are taken at the specific wavelengths of
2175~\AA~and 7500~\AA, which correspond to a characteristic extinction feature
in the UV and to a plateau region of the extinction cross-section in the optical.
The spatial distribution in polarized flux differs significantly between the two
wavebands, with the 7500~\AA~map showing a stronger $PF/F_{*}$. At 2175~\AA,
the scattering phase function greatly favors forward scattering over scattering
to other directions. Since the torus is optically thick and its funnel is narrow,
the photons are more likely to be absorbed. At 7500~\AA, the scattering phase
function is less anisotropic and it therefore allows scattered (i.e., polarized)
photons to escape more easily from the funnel. The fact that optical photons also
encounter a slightly higher albedo in the dust grains than UV photons does work in the same direction.
The combination of these effects explains, why the spatial distribution in
polarized flux in the optical reaches out to larger distances from the central
source than in the UV waveband.

\subsection{Modeling polar outflows}

Polar outflows are a major constituent when explaining the AGN polarization
behavior. They allowed the initial discovery of Seyfert-1 nuclei in Seyfert-2
objects by the means of spectropolarimetry \citep{Antonucci1985}.
An approximative geometry of the polar scattering regions corresponds to an
hourglass shape with a central break where the photon source is located.
Below the dust sublimation radius, where it is believed that the polar wind is mainly
composed of ionized gas whereas beyond this radius, the gas can coexist with dust.
We thus model both, an electron-filled double-cone for the ionized material closer
to the source and then a pair of more distant, dusty outflows.

\subsubsection{Modeling polar electron scattering}
\label{sec:e-cone}

   \begin{figure}
   \centering
      \includegraphics[width=10.5cm]{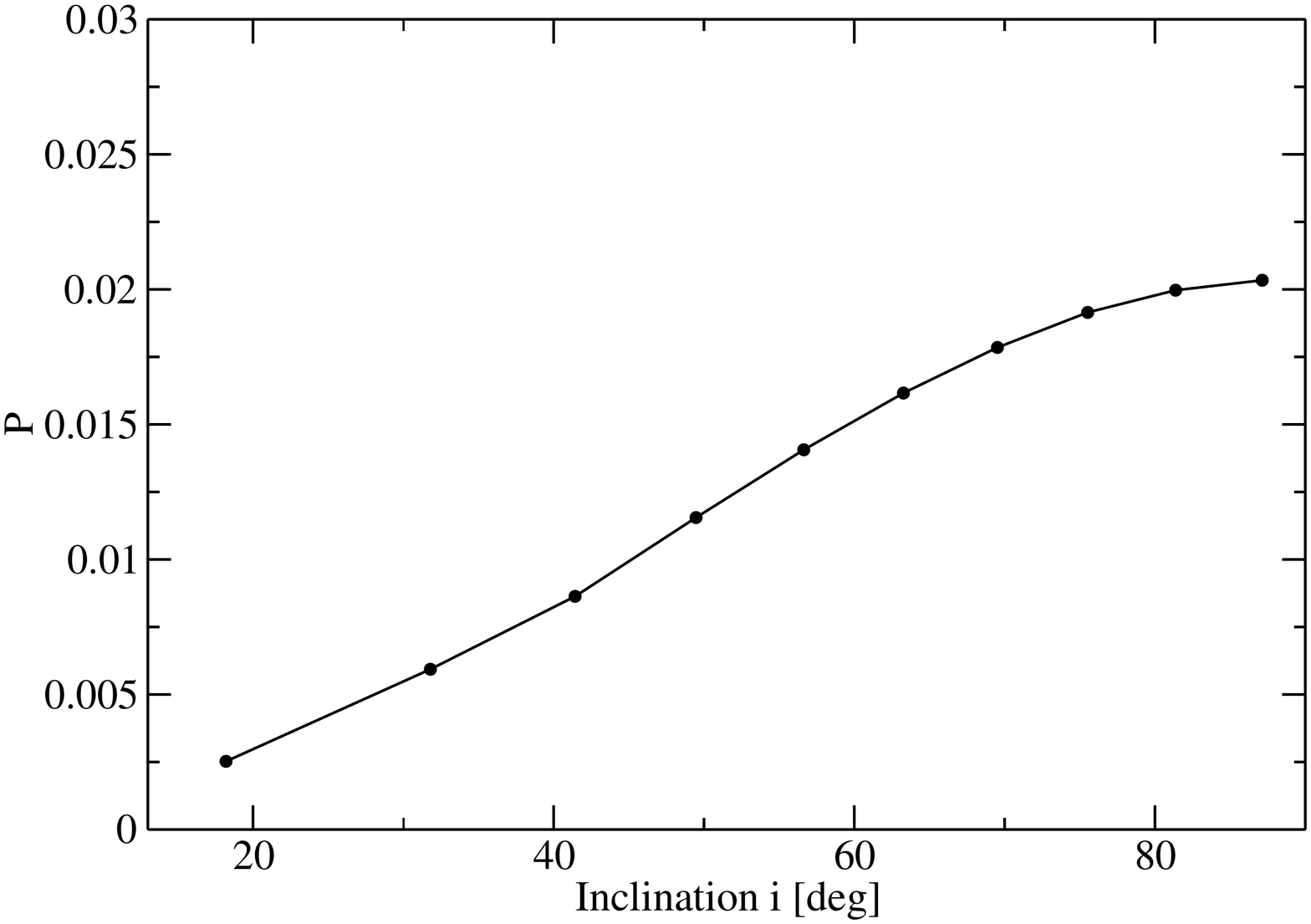}
      \caption{Modeling an electron-filled, scattering double-cone with the
        half-opening angle $\theta_{c}~=~30^\circ$ measured relative
        to the symmetry axis. The net polarization, $P$, is plotted versus the
        observer's inclination \textit{i}.}
     \label{Fig4.1}%
   \end{figure}

Using the formalism introduced by \citet{Brown1977}, \citet{Miller1990}
and \citet{Miller1991} were the first to compute the polarization from
an AGN double-cone composed of electrons. Based on these results,
\citet{Wolf1999} and \citet{Watanabe2003} conducted Monte Carlo simulations
that also include the effects of multiple scattering not taken into account
in the analytical formula of \citet{Brown1977}. Paper~I repeated and confirmed
these studies. It turned out that a radial gradient of the electron density
inside the double cone does not have a major impact on the resulting polarization.
In most cases, a uniform density leads to very similar results. Here,
we resume this investigation and add our new results for the polarization
imaging.

We implement the same polar wind characteristics as in Paper~I and model a
double cone filled with electrons. The electron density is adjusted to
achieve a radial Thomson optical depth of $\tau \sim 1$ being measured
in the vertical direction between the inner and outer surfaces of a single
cone. The half-opening angle of the double cone is $30^\circ$ measured
from the vertical axis. Unlike the modeling in Paper~I, we here do not
restrict the emission angle of the central source but consider isotropic
emission.

   \begin{figure}
   \centering
      \includegraphics[trim = 5mm 5mm 0mm 10mm, clip, width=8cm]{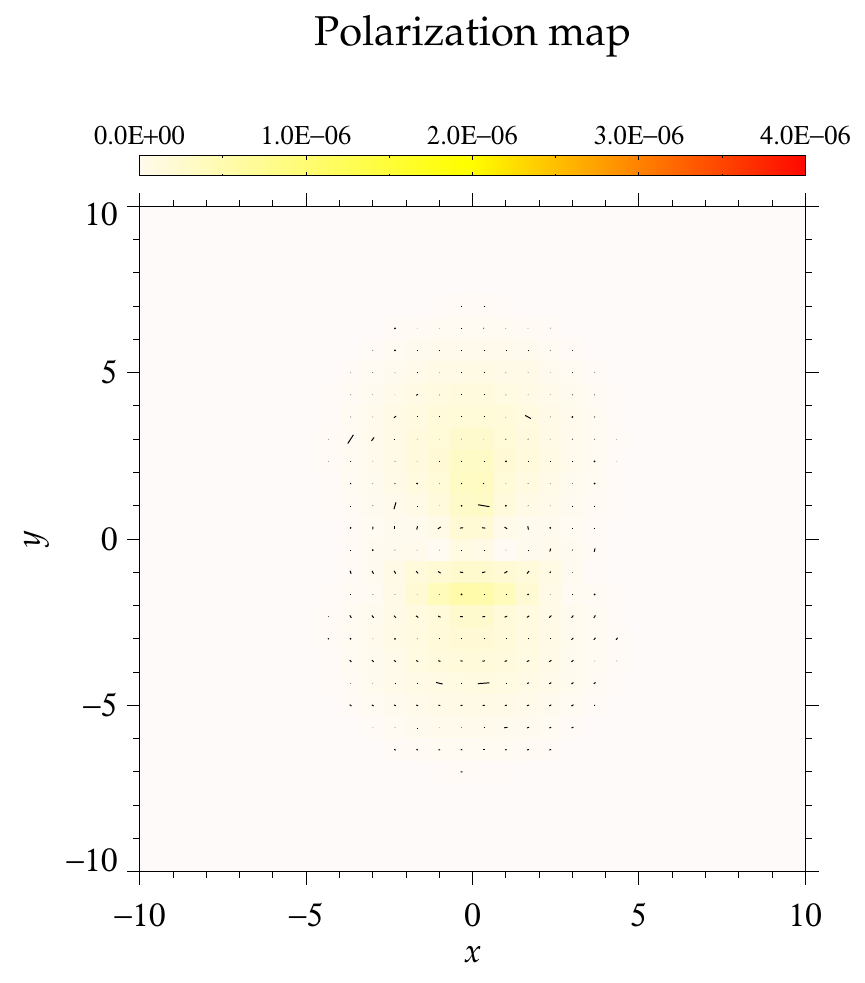}
      \includegraphics[trim = 5mm 5mm 0mm 10mm, clip, width=8cm]{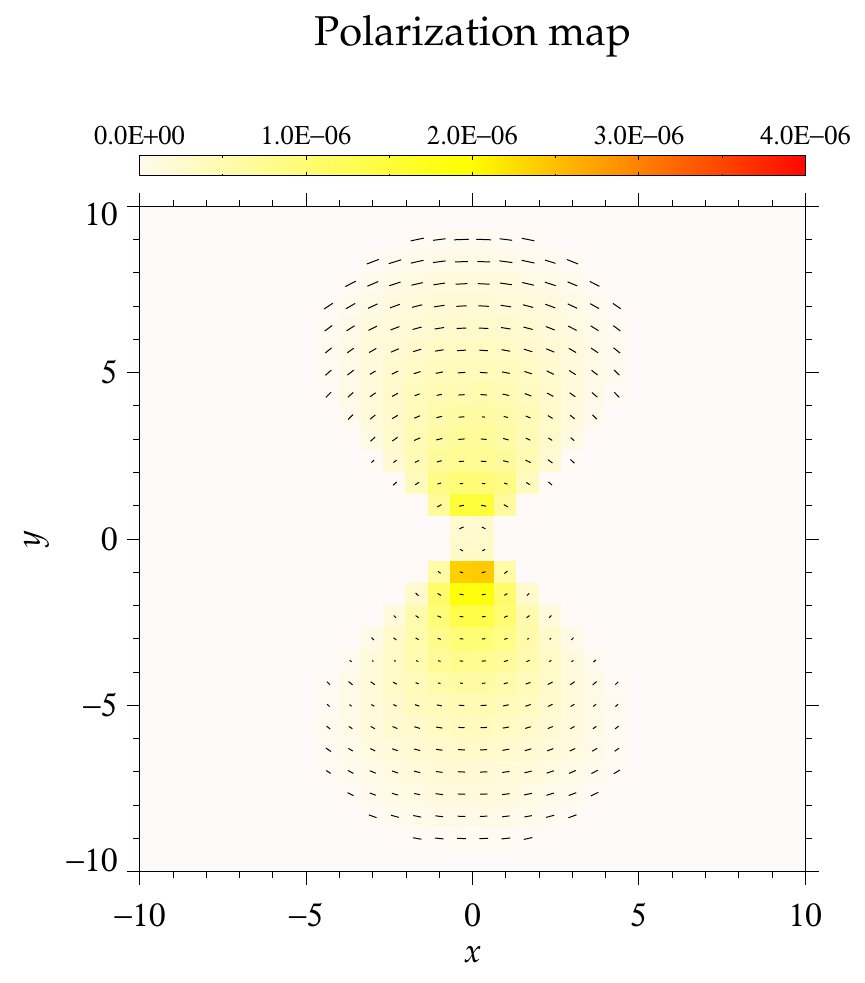}
      \includegraphics[trim = 5mm 5mm 0mm 10mm, clip, width=8cm]{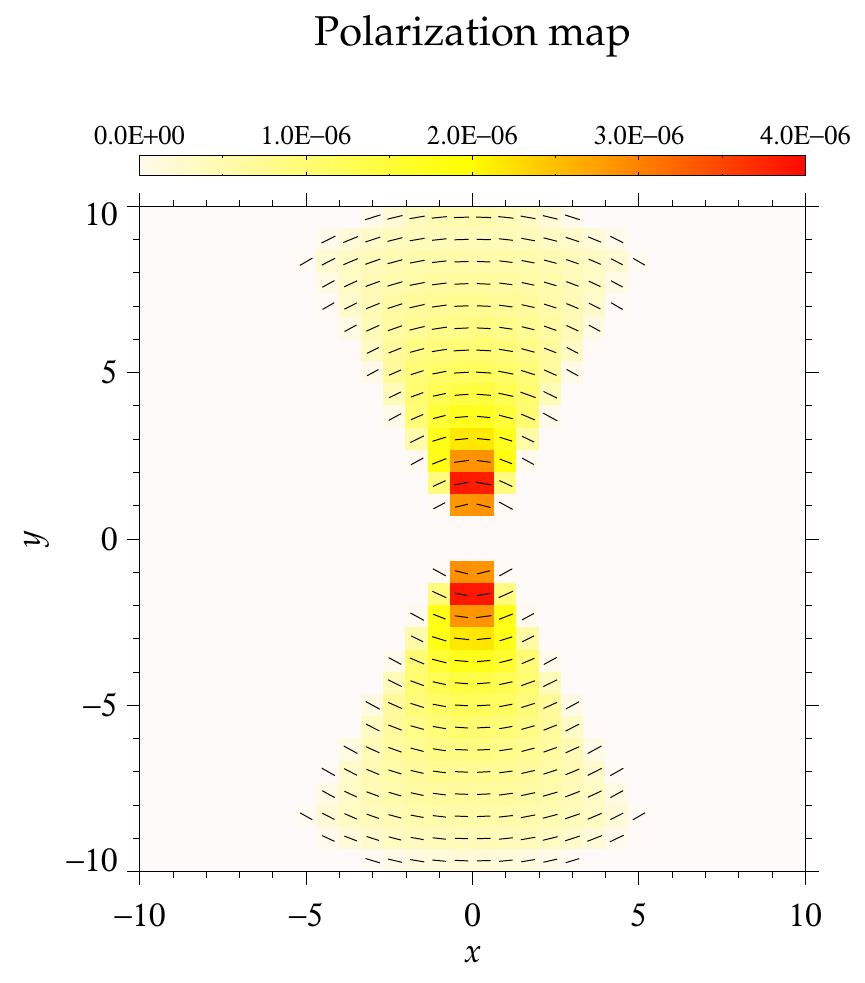}
      \caption{Modeled images of the $PF/F_{*}$ for a electron-filled,
	       scattering double-cone composed with the half-opening angle
	       $\theta_{c}~=~30^\circ$ measured relative to the symmetry axis.
               The polarized flux, $PF/F_{*}$, is color-coded and
	       integrated over the wavelength band.
               \textit{Top}: face-on image;
               \textit{Middle}: image at $i \backsim 45^\circ$;
               \textit{Bottom}: edge-on image.}
     \label{Fig4.2}%
   \end{figure}

The resulting polarization as function of inclination is shown in Fig.\ref{Fig4.1}.
The polarization curve is very similar to the one found in Paper~I (Section 5.1),
but its normalization has drastically diminished. This is due to the strong dilution
by the unpolarized radiation coming directly from the source that we
excluded in our previous modeling. Note that a second source of polarization dilution,
not taken into account as we focus on the continuum emission, might also come from
the recombination lines and continua. Such a radiative mechanism should then decrease
the observed emission-line polarization and e.g., the Balmer continuum and will be subject of our future work.

Polarization maps are given in Fig.\ref{Fig4.2}. From a pole-on view, the highest $PF/F_{*}$ is concentrated
below the (projected) center of the model space. It is dominated by photons
back-scattered at the base of the far cone. Above the center, a secondary,
dimmer maximum appears that originates from forward-scattering in the near cone.
The slightly oval shape of the figure is a projection effect as the line of sight
is inclined by $18^\circ$ relative to the symmetry axis. The approximate symmetry
at pole-on viewing position explains why the net polarization at this viewing angle is rather
low. The distribution of polarization position angles across the image shows that
the polarization produced on the left side of the line of sight partly cancels
with that produced on the right side. The net $\gamma$ in the pole-on view,
is oriented perpendicularly to the symmetry axis. Increasing \textit{i}
diminishes the spatial symmetry and thereby leads to a growth of the polarized flux.
Note the presence of a flux gradient along the vertical axis. Due to the conical
geometry and uniform electron density, photons that penetrate farther
into the cone see a larger optical depth before escaping into an
intermediate or edge-on viewing direction. Multiple scattering therefore
diminishes the resulting polarization produced farther away from the center.

The strong radial gradient in polarized flux we obtain in our modeling
relates to the question of how strong the effect of a density gradient inside the double
cone is on the net polarization (see \citealt{Watanabe2003} and Paper~I).
Indeed, if the density gradient is steep near the source, the polarized flux
produced at the base of the cones should diminish, but, at the same time, the
polarization from the optically thinner, outer parts of the cones rises because
multiple-scattering becomes less important. It is thus not trivial to use
polarimetry to constrain the radial density profile in conical scattering regions.

\subsubsection{Polar dust scattering}

Beyond the dust-sublimation radius of an AGN 
dust particles can survive or form. We therefore have to study the effect
of dust absorption and scattering on the polarization induced by polar
outflows. We use again the dust composition based on the Milky Way model
(see Paper I and reference within) and define a dusty double cone with a
half-opening angle equal to $30^\circ$ from the vertical axis. The inner
boundary of the scattering region was fixed at 10~pc above the source and
the double cone extends to a distance of 100~pc. The optical depth in
the $V$-band was fixed at $\tau_{V}$ $\sim 0.3$.

   \begin{figure}
   \centering 
      \includegraphics[width=10.5cm]{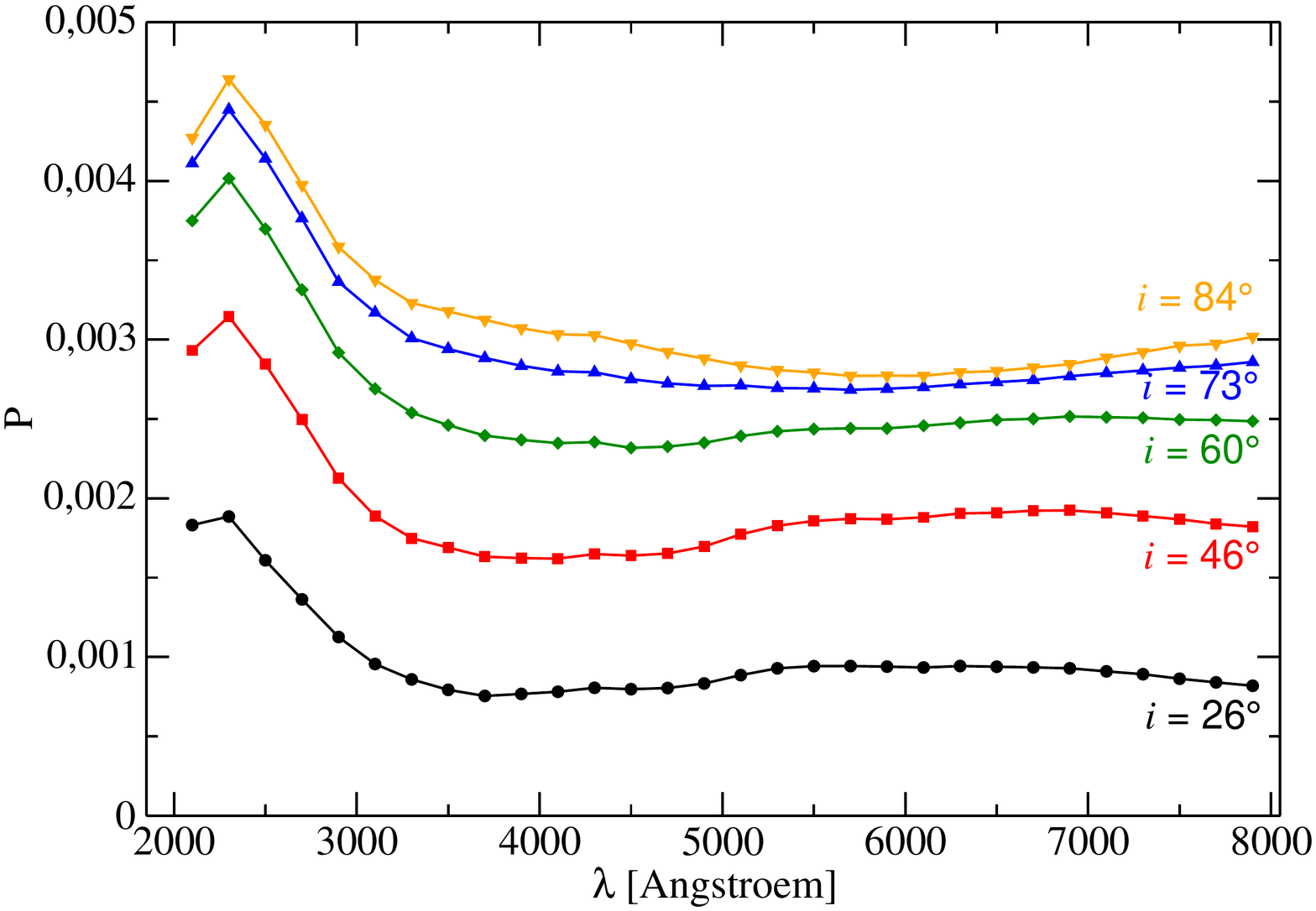}
      \caption{Modeling a dusty double-cone of half-opening angle
        $\theta_{c}~=~30^\circ$ measured relative to the axis of
        symmetry. The spectrum of the polarization percentage
        \textit{P} is plotted for different viewing inclinations,
        \textit{i}.}
     \label{Fig4.3}%
   \end{figure}


   \begin{figure}
   \centering
      \includegraphics[trim = 5mm 5mm 0mm 10mm, clip, width=8cm]{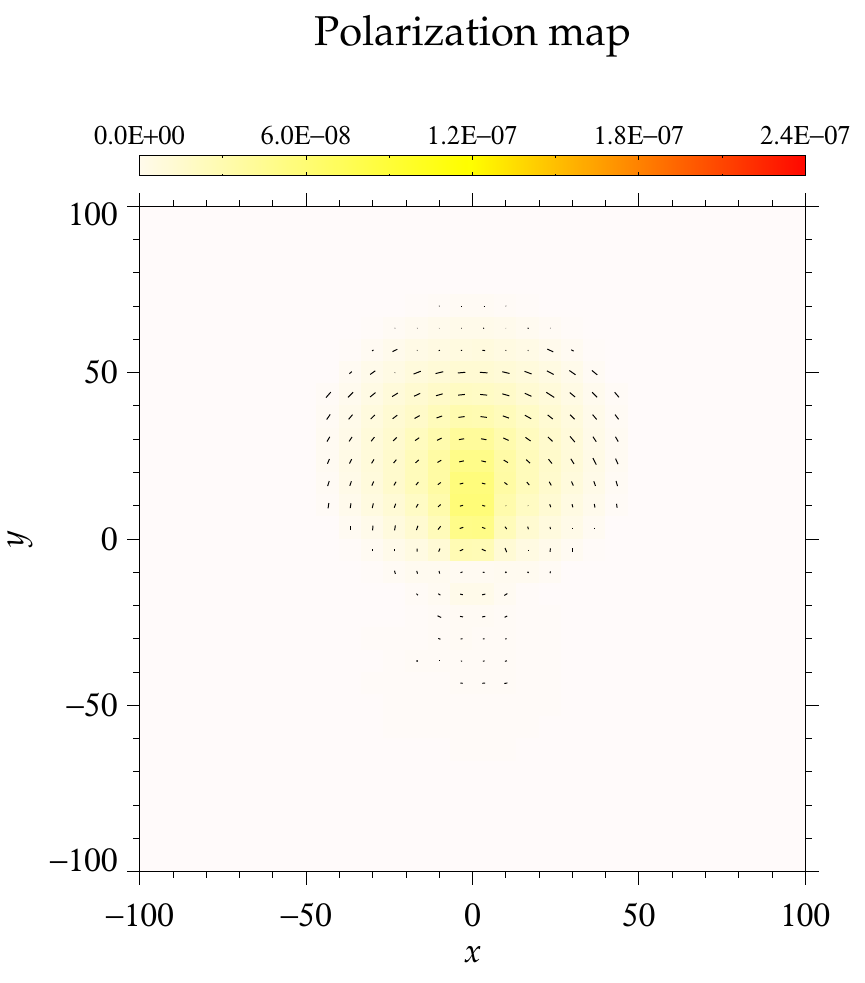}
      \includegraphics[trim = 5mm 5mm 0mm 10mm, clip, width=8cm]{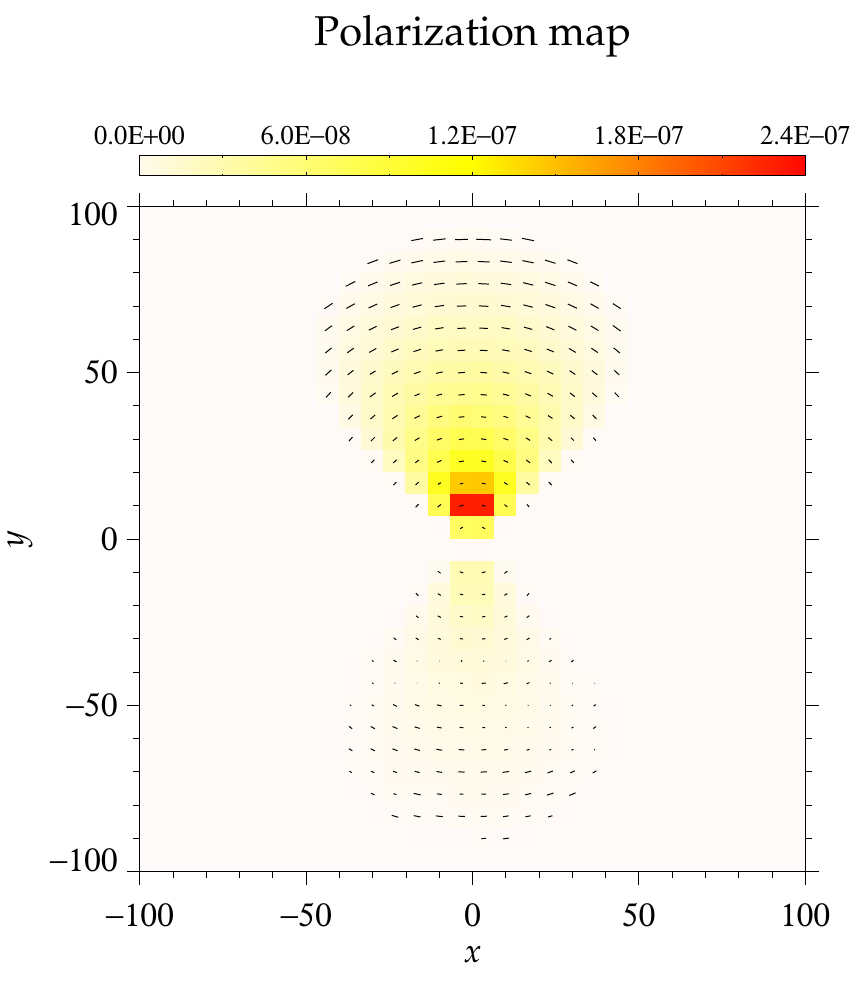}
      \includegraphics[trim = 5mm 5mm 0mm 10mm, clip, width=8cm]{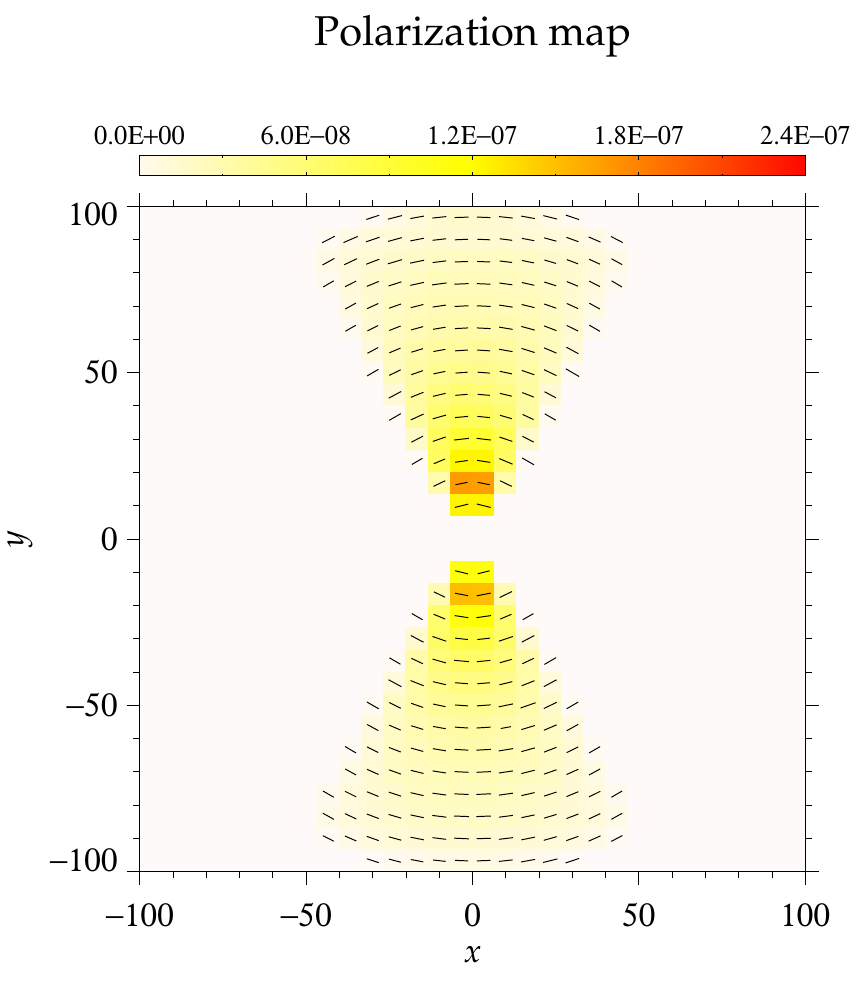}
      \caption{Modeled images of the polarized flux, $PF/F_{*}$, for a
        dusty double-cone of half-opening angle
        $\theta_{c}~=~30^\circ$ measured relative to the axis of
        symmetry; $PF/F_{*}$ is color-coded and integrated over the
        wavelength band. \textit{Top}: face-on image;
        \textit{Middle}: image at $i \backsim 45^\circ$;
        \textit{Bottom}: edge-on image.}
     \label{Fig4.4}%
   \end{figure}

The polarization spectra are shown in Fig.\ref{Fig4.3}. They are similar
to the results of Paper~I (except that the Poissonian fluctuations are
smoother due to better statistics with the newly-implemented MTG).
We show the images of the polarized flux in Fig.\ref{Fig4.4}. Unlike
the case of the double cones of scattering electrons, the face-on view shows only one
spatial maximum that corresponds to forward scattering in the near cone.
The dust phase function favors forward scattering over backscattering,
and therefore the far cone is dominated by absorption and cannot be seen
in polarized flux. However, the same characteristics of the polarization
position angle, $\gamma$, can be observed for both types of scattering media.
At an intermediate viewing angle, we observe a larger polarized flux but
still mainly dominated by the lower part of the upper cone. The behavior of the
net polarization is again related to the spatial symmetry that changes
with viewing angle, just as for the case of the electron-scattering double cone
described in Sect.~\ref{sec:e-cone}. At an edge-on viewing position, mostly perpendicular
polarization with a radial gradient is seen. Significant polarized flux still
emerges close to the outer limits of the reprocessing region.
For wavelengths above 2200~\AA, the scattering cross section decreases
monotonically so that photons can travel farther into the medium before
being scattered or absorbed.

\subsubsection{The effects of the optical depth and wavelength}

\begin{figure}
   \centering
      \includegraphics[trim = 5mm 5mm 0mm 10mm, clip, width=8cm]{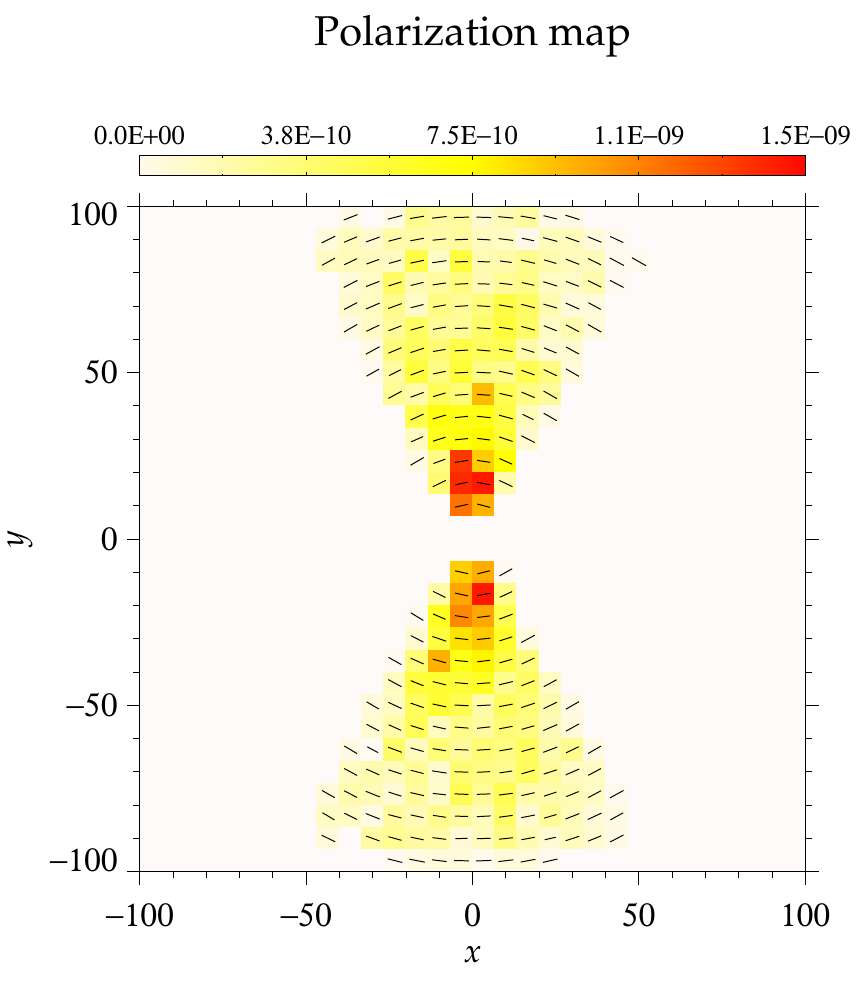}
      \includegraphics[trim = 5mm 5mm 0mm 10mm, clip, width=8cm]{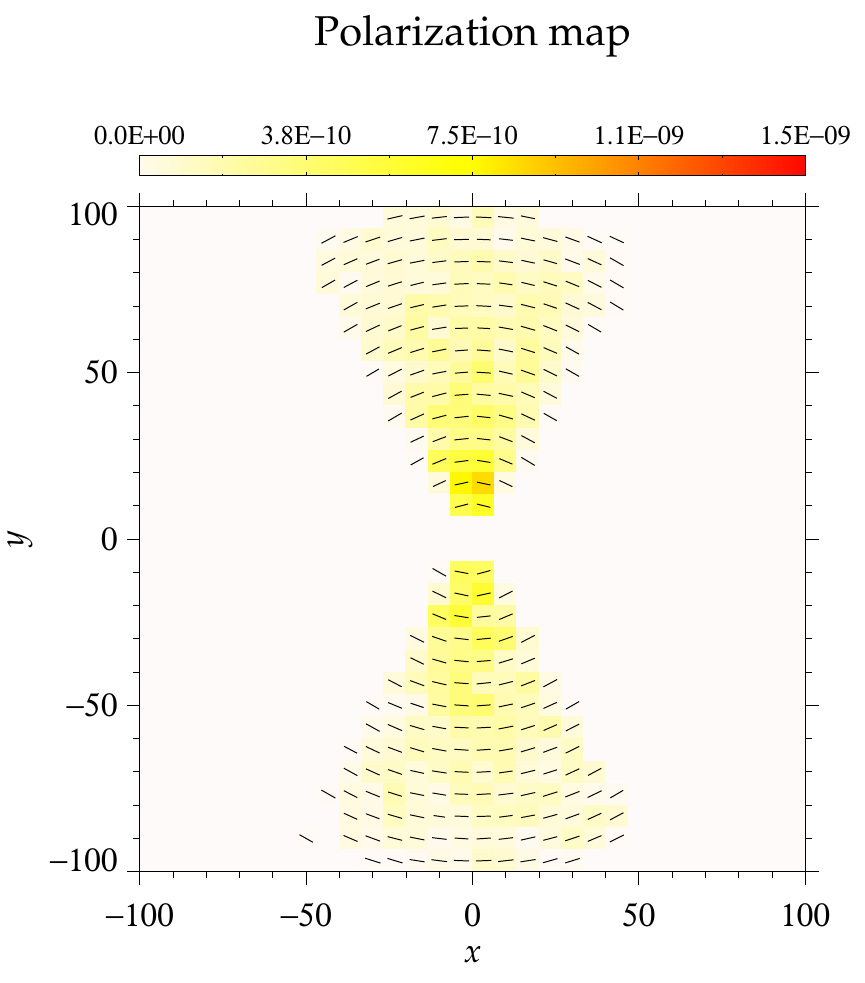}
      \caption{Modeled images of the polarized flux, $PF/F_{*}$, for a dusty
	       double-cone with the half-opening angle $\theta_{c} = 30^\circ$
	       measured relative to the symmetry axis and a radial optical
	       depth of $\tau_{V} \sim 0.03$; $PF/F_{*}$ is colour-coded and
	       integrated over the wavelength band.
	       \textit{Top}: edge-on image at $\backsim$ 2175~\AA;
	       \textit{Bottom}: edge-on image at $\backsim$ 7500~\AA.}
     \label{Fig4.5}
   \end{figure}

\begin{figure}
   \centering
      \includegraphics[trim = 5mm 5mm 0mm 10mm, clip, width=8cm]{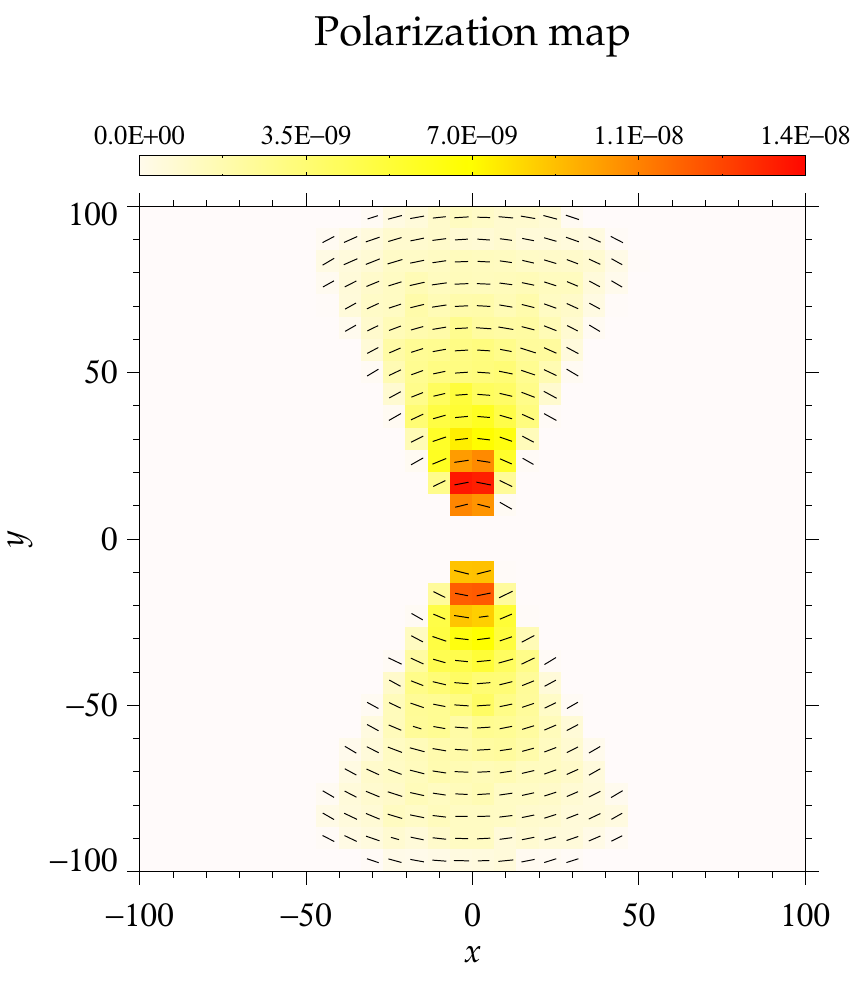}
      \includegraphics[trim = 5mm 5mm 0mm 10mm, clip, width=8cm]{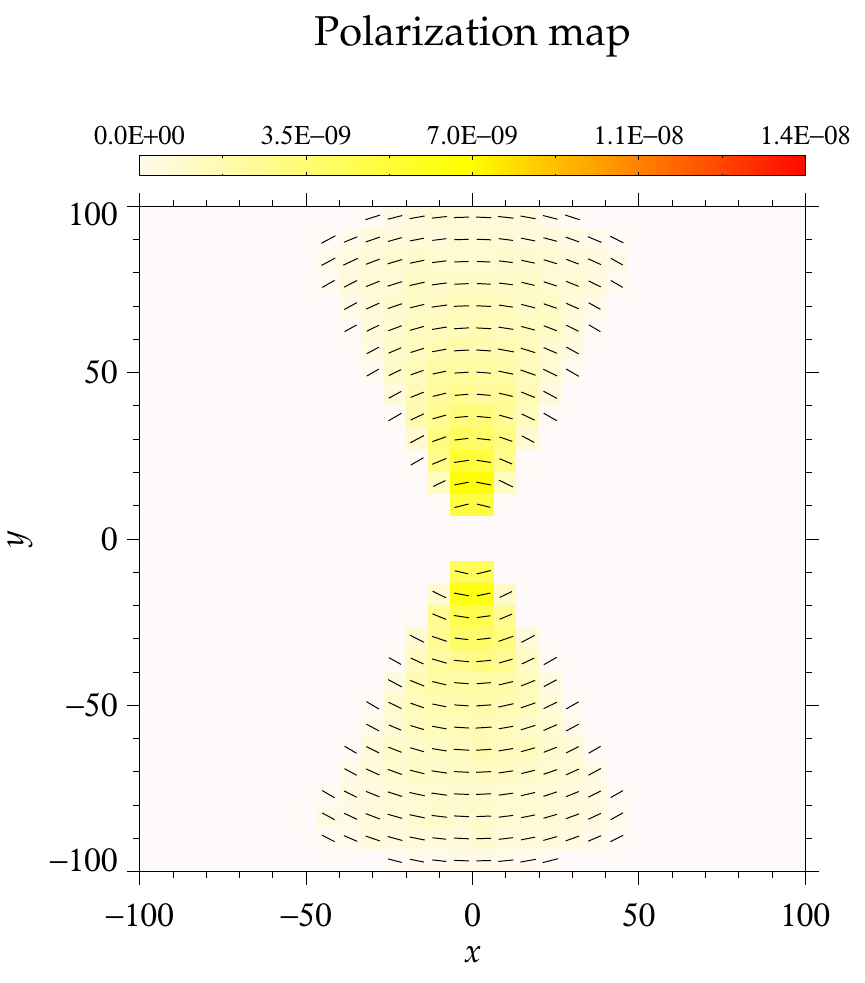}
      \caption{Modeled images of the polarized flux, $PF/F_{*}$, for a
	       dusty double-cone with the half-opening angle $\theta_{c} = 30^\circ$
	       measured relative to the symmetry axis and a radial optical depth
	       of $\tau_{V} \sim 0.3$; $PF/F_{*}$ is color-coded and integrated
	       over the wavelength band.
	       \textit{Top}: edge-on image at $\backsim$ 2175~\AA;
	       \textit{Bottom}: edge-on image at $\backsim$ 7500~\AA.}
     \label{Fig4.6}
   \end{figure}

\begin{figure}
   \centering
      \includegraphics[trim = 5mm 5mm 0mm 10mm, clip, width=8cm]{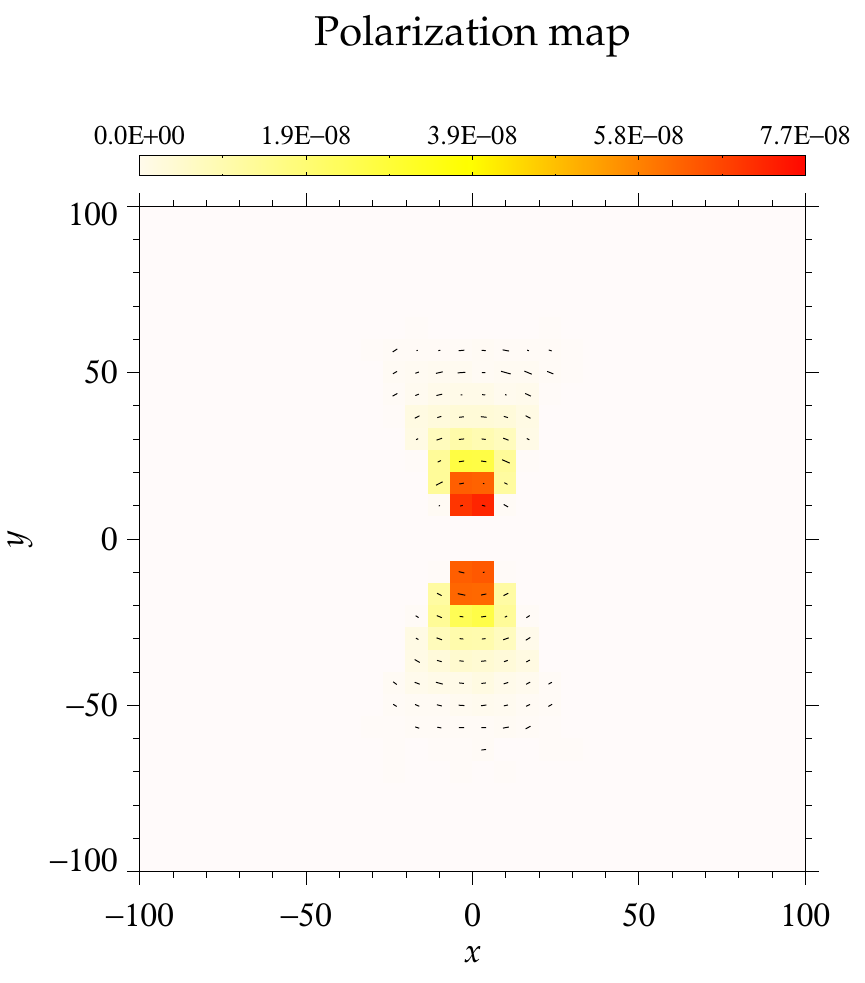}
      \includegraphics[trim = 5mm 5mm 0mm 10mm, clip, width=8cm]{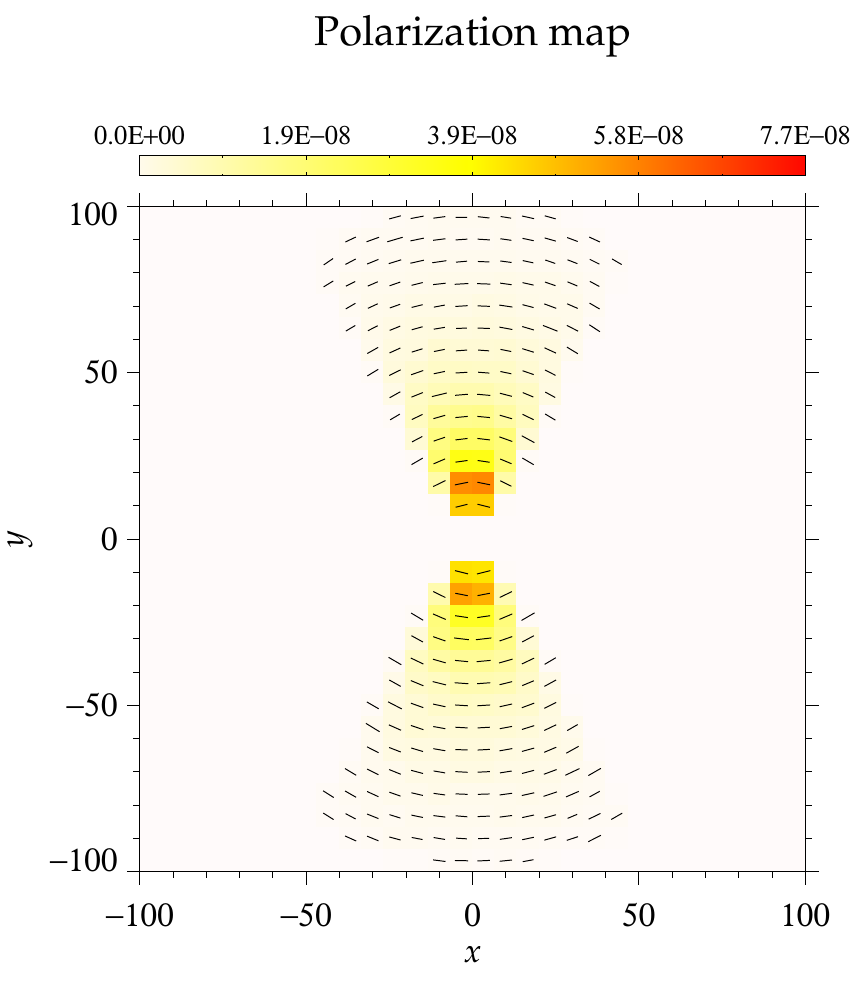}
      \caption{Modeled images of the polarized flux, $PF/F_{*}$, for a dusty
	       double-cone with the half-opening angle $\theta_{c} = 30^\circ$
	       measured relative to the symmetry axis and a radial optical depth
	       of $\tau_{V} \sim 3$; $PF/F_{*}$ is color-coded and integrated
	       over the wavelength band.
	       \textit{Top}: edge-on image at $\backsim$ 2175~\AA;
	       \textit{Bottom}: edge-on image at $\backsim$ 7500~\AA.}
     \label{Fig4.7}
   \end{figure}

We studied the dependence of the polarization on wavelength and on the
optical depth of the dusty outflows. We varied the optical depth of the medium
in the $V$-band, $\tau_{V}$, between 0.03, 0.3, and 3 and we show the
imaging results in Figs.~\ref{Fig4.5}, \ref{Fig4.6}, and \ref{Fig4.7},
respectively. Maps are shown for an edge-on viewing angle and at
wavelengths of 2175~\AA~and 7500~\AA.

It appears from the colour scale of the figures that up to a range of
$\tau_{V} < 3$ the $PF/F_{*}$ increases with $\tau_{V}$. As the optical
depth determines the average number of scattering events before escape
or absorption, increasing it leads to more scattering and therefore to
more $PF/F_{*}$, at least before the optical depth becomes sufficiently
high to have depolarizing multiple-scattering effects becoming predominant.
The systematic rise in $PF/F_{*}$ seems to be similar at both wavelengths
considered. Photons with larger wavelengths (i.e., $\backsim$7500 \AA),
pass through a larger portion of material before being scattered and
escaping. This effect is related to the evolution of the extinction
cross section with wavelength as mentioned earlier.

For our modeling of the dusty outflows the optical depth is always much
lower (between 0.03 and 3) than the one assumed for the optically opaque
torus ($\backsim$750). The photons scattering off the outflows thus do
not remain close to the surface of the scattering region but can penetrate
deep into the medium before being scattered or absorbed. At 2175 \AA,
the photon phase function favors forward over backscattering. Photons
that have already scattered and progressed in the direction of the observer
are thus less likely to be scattered out of the line of sight. In general,
the $PF/F_{*}$ is therefore greater at UV wavelengths than in the optical band.

\subsection{Equatorial scattering in a radiation-supported disc}
\label{sec:equatdisc}

To explain the presence of parallel polarization (with respect to
the projected symmetry axis), a
third type of scattering region has been proposed(\citealt{Antonucci1984}): a radiation
supported, geometrically-thin scattering disc lying in the equatorial
plane (\citealt{Chandrasekhar1960}, \citealt{Angel1969}), \citealt{Sunyaev1985}).
Most often the geometry of a flared wedge is assumed for this scattering region
(see Fig. 17 in Paper I for a schematic review of the possible geometries).
Following the suggestions of \citet{Goodrich1994} and the simulations of
 \citet{Young2000} and \citet{Smith2004,Smith2005}, in Paper~I we showed that
the flared disc geometry can be replaced by a torus geometry without
noticeable changes in polarization. This substitution is consistent for
a flared disc half-opening angle lower than $30^\circ$. Here, we simulate
an equatorial scattering region using a geometrically thin torus composed
of electrons with a Thomson optical depth of $\tau \sim 1$ and a half-opening
angle of $10^\circ$ with respect to the equatorial plane. The inner radius
of the thin torus is $3 \times 10^{-4}$~pc and its outer radius is
$5 \times 10^{-4}$~pc.

   \begin{figure}
   \centering 
      \includegraphics[width=10.5cm]{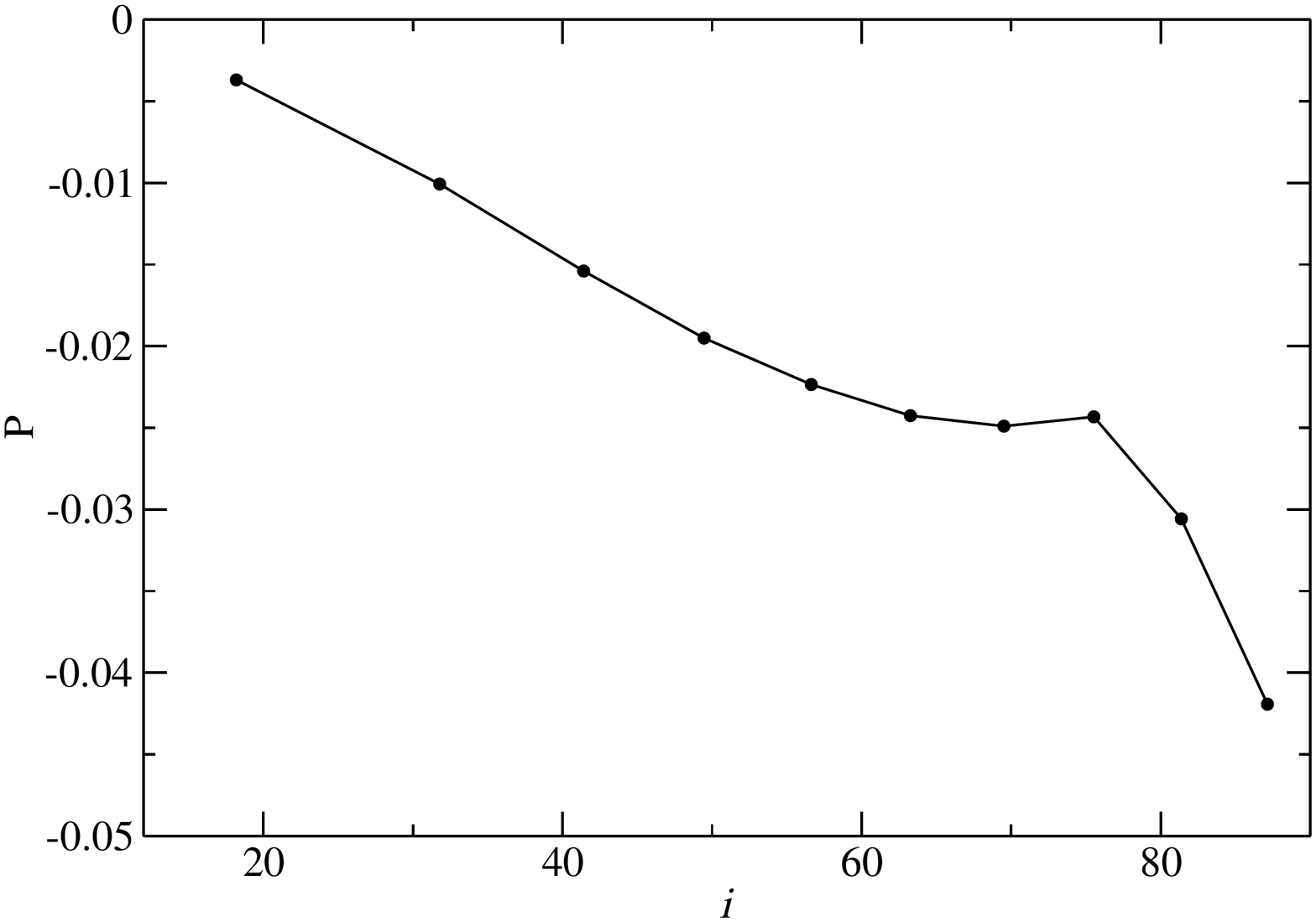}
      \caption{Modeling of an equatorial, electron-filled scattering disc with a
        half-opening angle $\theta_0$~=~$10^\circ$ measured with respect
        to the equatorial plane. Polarization \textit{P} is plotted
        against the inclination \textit{i} along the axis of the
        observer. Negative values for \textit{P} indicate a parallel
        polarization.}
     \label{Fig5.1}%
   \end{figure}

\begin{figure}
   \centering
      \includegraphics[trim = 5mm 5mm 0mm 10mm, clip,width=8cm]{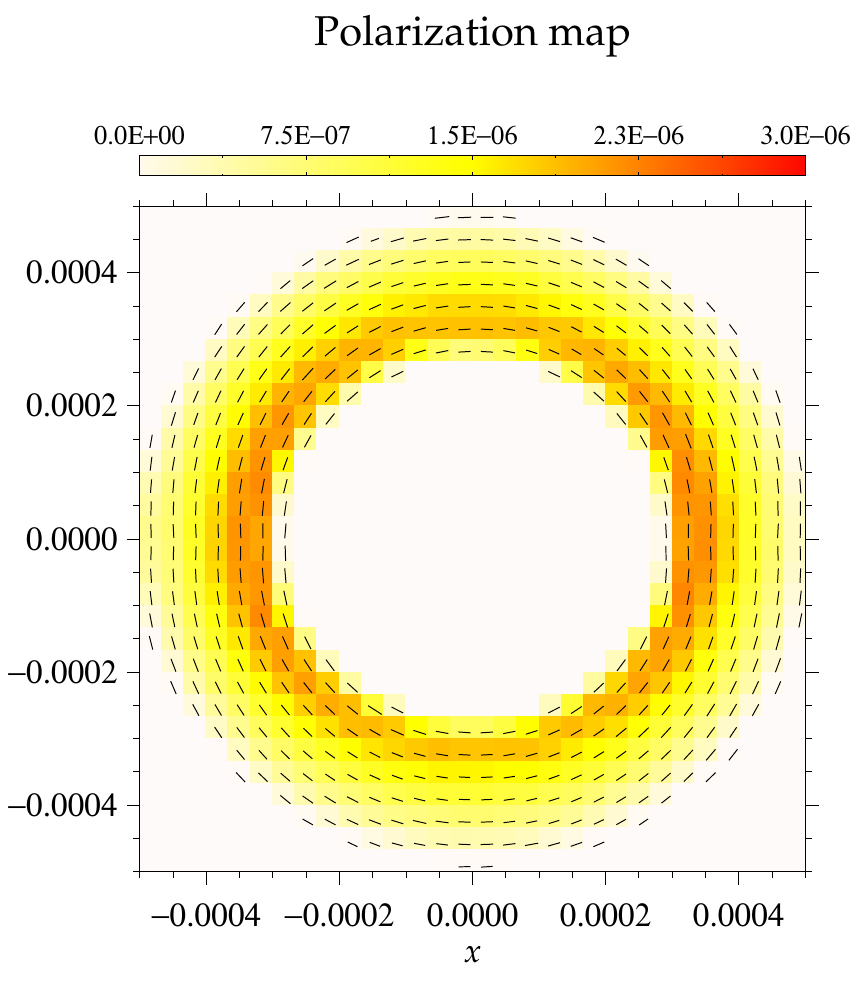}
      \includegraphics[trim = 5mm 5mm 0mm 10mm, clip, width=8cm]{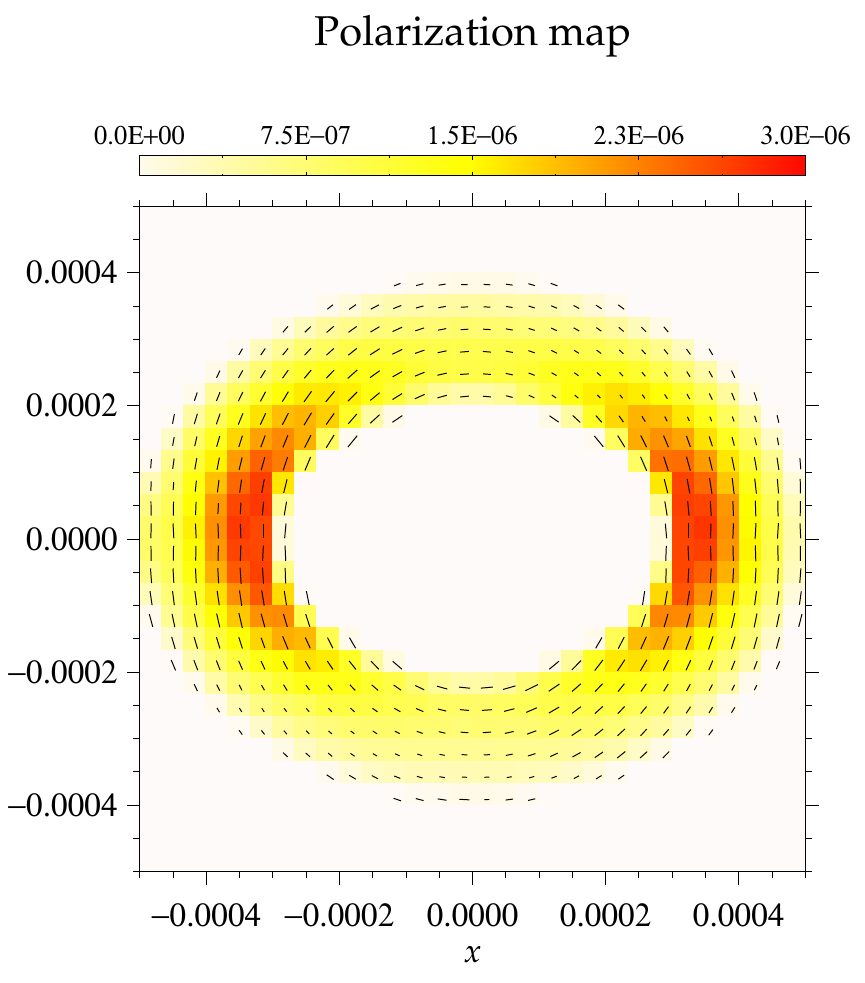}
      \includegraphics[trim = 5mm 5mm 0mm 10mm, clip, width=8cm]{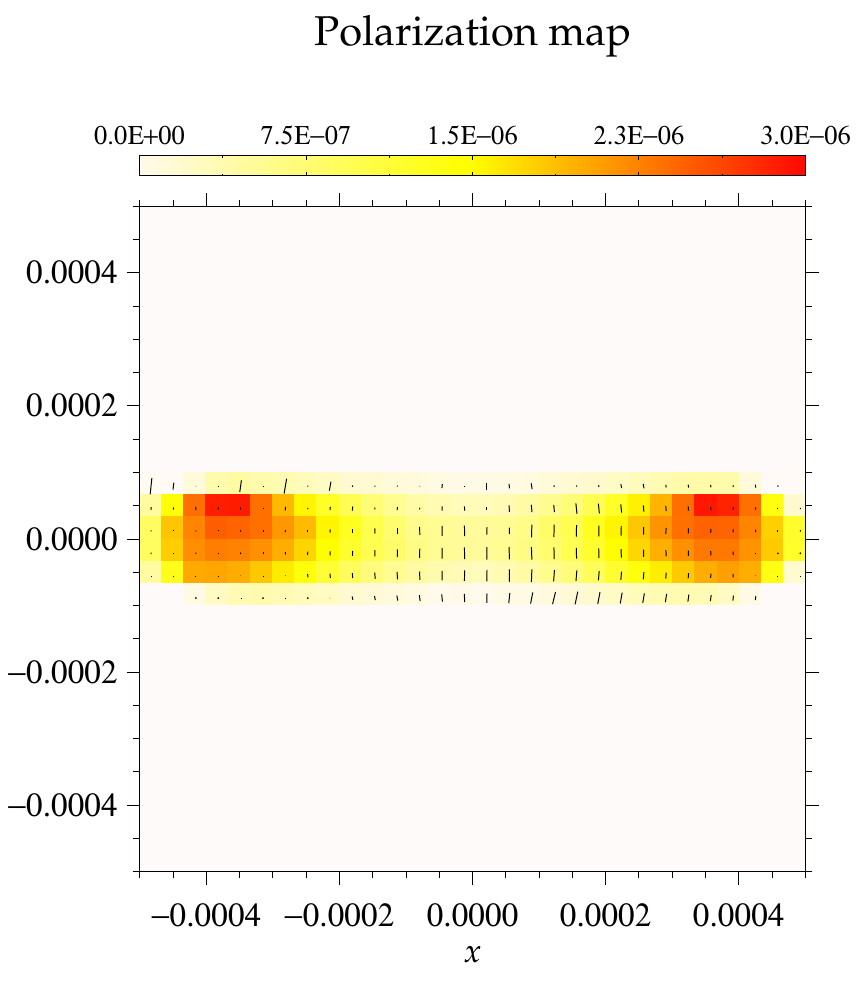}
      \caption{Modeled image of the polarized flux, $PF/F_{*}$, for an
        electron-filled, equatorial scattering disc with a half
        opening angle $\theta_0$~=~$10^\circ$ measured with respect to
        the equatorial plane; $PF/F_{*}$ is color-coded and integrated
        over the wavelength band. \textit{Top}: face-on image;
        \textit{Middle}: image at $i \backsim 45^\circ$;
        \textit{Bottom}: edge-on image.}
     \label{Fig5.2}%
   \end{figure}

The polarization spectra as a function of the inclination \textit{i}
are shown in Fig.\ref{Fig5.1}. A negative value of \textit{P} indicates
parallel polarization at all viewing angles. The spectra are very
similar to the ones presented in Paper~I for the geometry of a flared
disc. Figure~\ref{Fig5.2} presents the polarization maps for the
simulated equatorial scattering disc. For a pole-on view, the disc is
divided into concentric rings of polarized flux with polarization
vectors oriented tangentially to the rings. By integrating the entire
region we obtain a polarization position angle of $90^\circ$ for a
nearly face-on viewing angle. The inner surface of the equatorial disc
causes a lower polarized flux than the rest of the reprocessing region.
The torus geometry implies that the inner region has a height
(relative to the equatorial plane) that is less important than the
central region. However, if the density is uniform throughout the medium;
we then deduce that the vertical column density is lower on the edges
of the torus. Hence, the $PF/F_{*}$ is greater in the central part
where the vertical column density and therefore the scattering
probability is maximum. When we tilt the line of sight to an
intermediate position, we observe a polarized flux greater on
the torus surfaces that are parallel to the line of sight. These
surfaces retain a strong polarization as the scattering angle with
respect to the observer's line of sight stays close to $90^\circ$.
Surfaces orthogonal to the line of sight cause a lower $PF/F_{*}$
compared to the pole-on view. Therefore, the net polarization
remains parallel and is even stronger at an intermediate view than
at the near face-on view. For angles \textit{i} at nearly edge-on
view, the scattering geometry only allows for parallel polarization
for all parts of the equatorial disc.

On the $PF/F_{*}$-plots the photon source seems hidden by the medium as
it is seen only in transmission which induces very low polarization.
However, this strong, low-polarization flux dilutes more significant
polarized flux coming from other areas of the scattering region.
A second reason why the net polarized flux remains moderate is that
at the optical depth $\tau_{\rm V} = 1$ considered here multiple
scattering events occur inside the medium.

\section{Exploring the radiative coupling between two reprocessing regions}
\label{sec:combined}

Having modeled the polarization induced by individual scattering regions,
we now include the effects of radiative coupling between them. Especially
for significant optical depths, the coupling turns out to be important and
should be properly addressed by Monte-Carlo methods \citep[see the discussions
in][and in Paper I]{Goodrich1994}. We apply a step-by-step method combining
first only two scattering regions at a time. Then, we approach a more complete
AGN model that is composed of three individual scattering regions
(see Sect.~\ref{sec:AGN}). In the following we will not include the
previously studied dusty outflows as the $NLR$ regions responsible for dust
signatures in polar scattering objects are situated farther away from the central
engine than the three other reprocessing regions.

All three models presented here feature the same unpolarized, isotropic central
source that was described previously. The first model in this section
consists of an equatorial, electron-filled disc and polar, electron-filled
outflows (Sect.~\ref{sec:combequatwind}). The second model is composed
of the equatorial disc and an optically-thick, dusty torus (Sect.~\ref{sec:combequattorus}).
The last model comprises an optically-thick torus and polar outflows
(Sect.~\ref{sec:combtoruswind}).

\subsection{Equatorial scattering disc and electron-filled outflows}
\label{sec:combequatwind}

The equatorial scattering disc is again simulated by a electron-filled,
geometrically-thin torus as described in Sect.~\ref{sec:equatdisc}.
The polar electron-filled outflows are modeled according to
Sect.~\ref{sec:e-cone}. Such a model setup may be applicable to
non-thermal AGN with very low toroidal absorption. Such claims have been
made for FR~I radio galaxies \citep{Chiaberge1999,Whysong2004} and
LINERs \citep{Maoz2005}.

   \begin{figure}
   \centering
      \includegraphics[width=10.5cm]{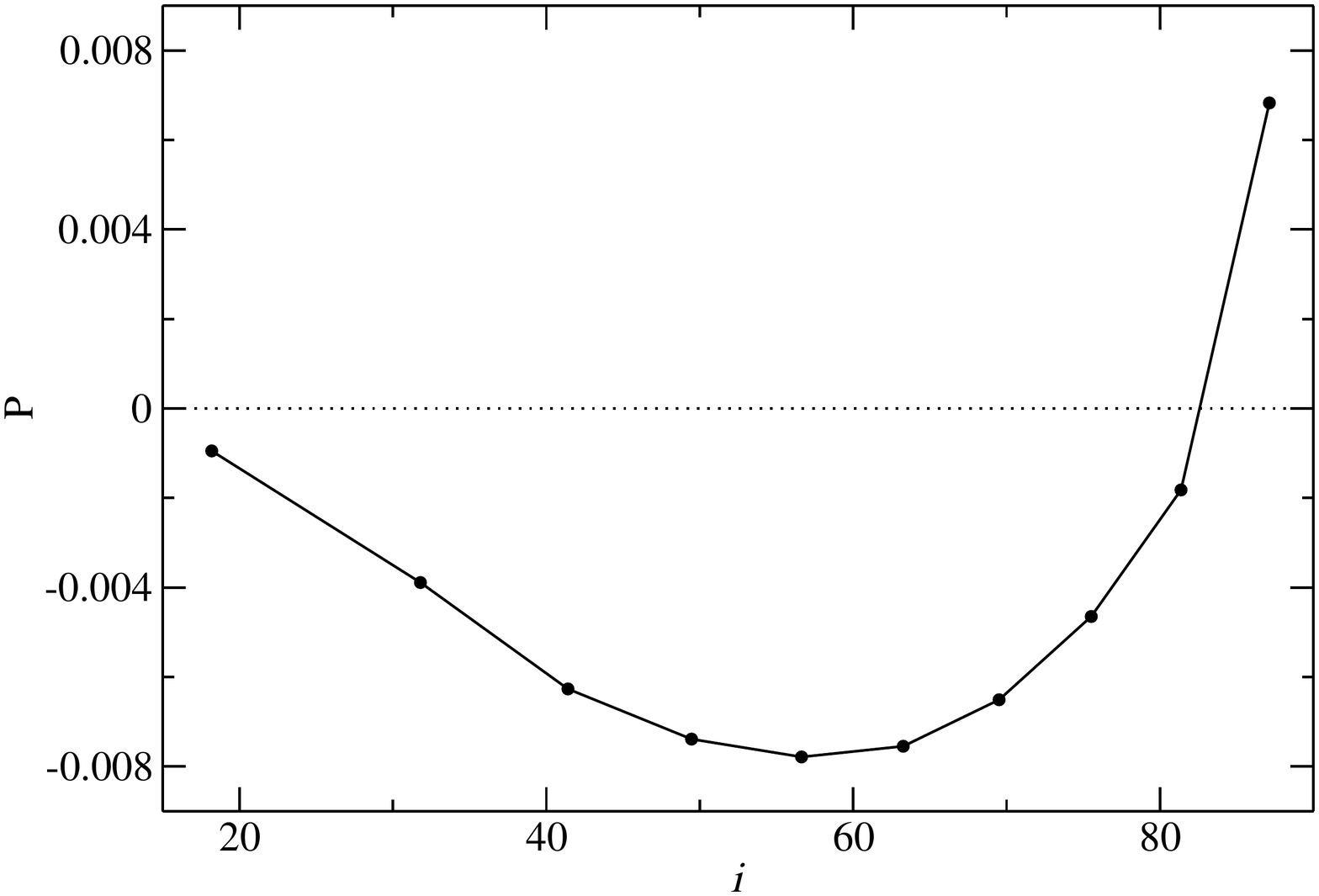}
      \caption{Modeling electron-filled, polar outflows with a
        half-opening angle $\theta_{c} = 30^\circ$ relative to the
        symmetry axis combined with an electron-filled, equatorial
        disc with a half-opening angle $\theta_0 = 10^\circ$
        measured relative to the equatorial plane. The net
        polarization is plotted versus the inclination \textit{i} of
        the observer.}
     \label{Fig6.1}%
   \end{figure}

The resulting polarization percentage as a function of the viewing
angle is shown in Fig.~\ref{Fig6.1}. At pole-on and intermediate
viewing angles, \textit{P} is negative (parallel net polarization).
This is due to the predominance of the equatorial disc producing a
polarization angle of $\gamma = 90^\circ$ regardless of the line of
sight. At pole-on view, the parallel polarization that the photons
acquire in the equatorial disc is preserved during their passage through
the optically thin polar winds. Only towards the largest viewing angles
the scattering polarization induced by the polar winds dominates and
gives a net polarization angle of $\gamma = 0^\circ$. Note that
\textit{P} is always constant in wavelength as in this particular
model we only consider Thomson scattering.

   \begin{figure}
   \centering
      \includegraphics[trim = 5mm 5mm 0mm 10mm, clip, width=8cm]{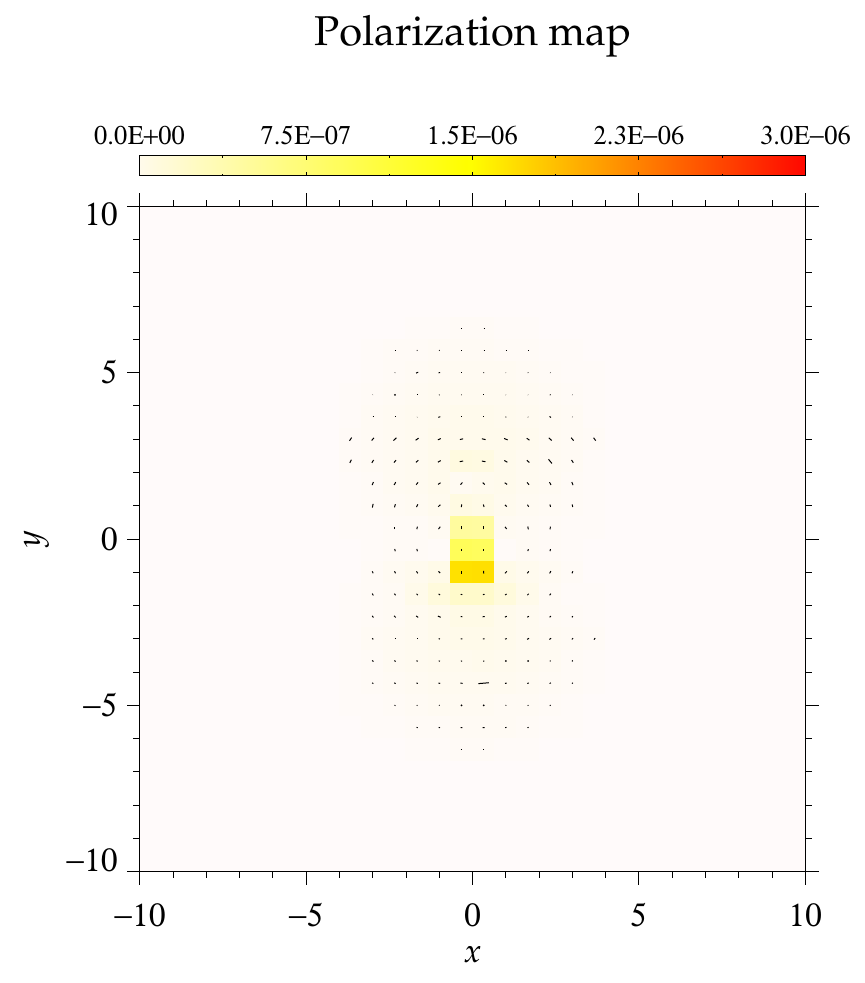}
      \includegraphics[trim = 5mm 5mm 0mm 10mm, clip, width=8cm]{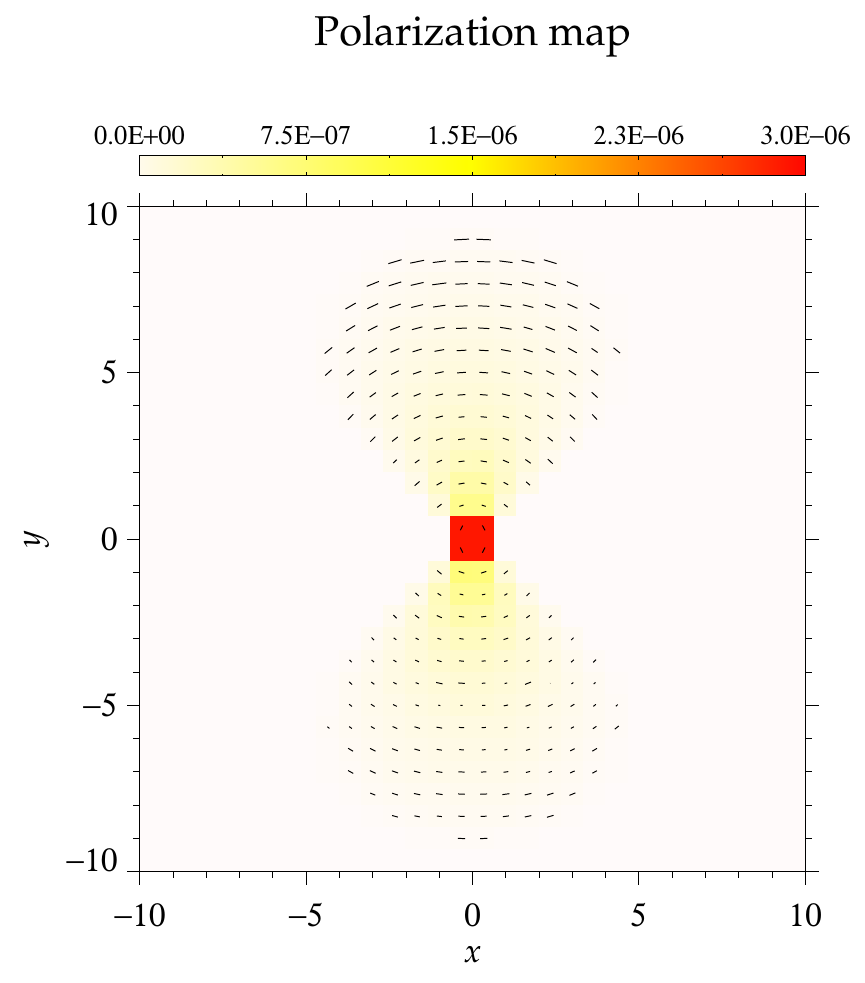}
      \includegraphics[trim = 5mm 5mm 0mm 10mm, clip, width=8cm]{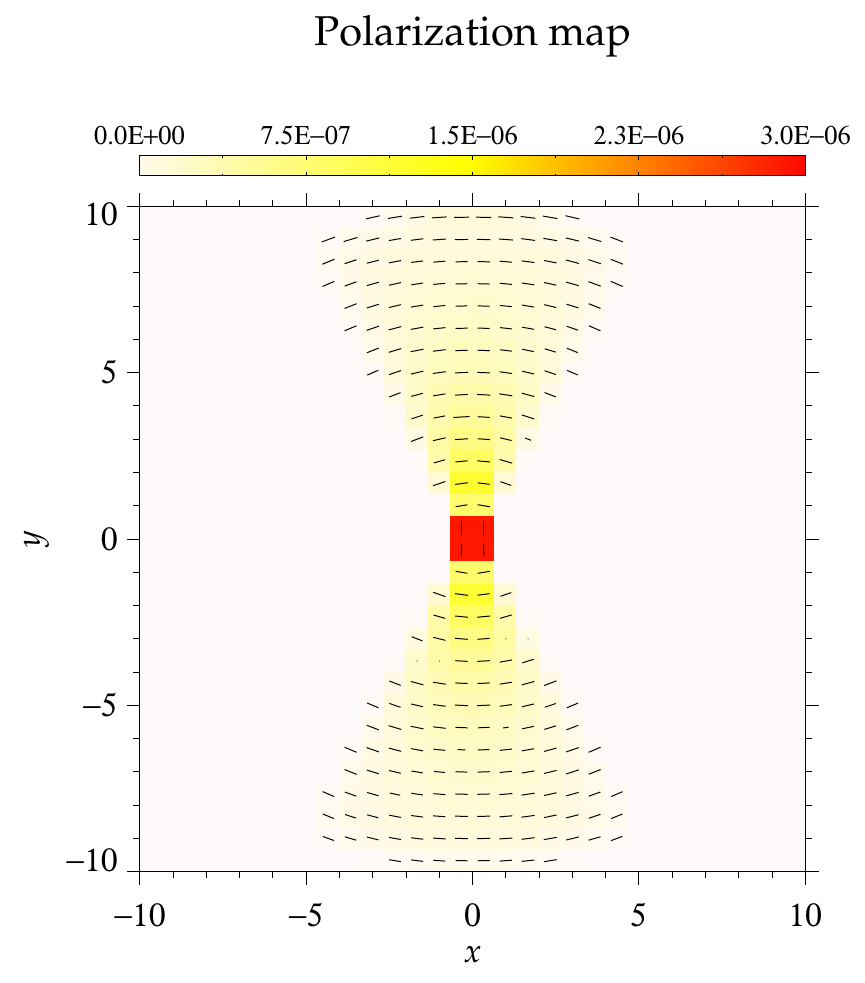}
      \caption{Modeled image of the polarized flux, $PF/F_{*}$, for
        the combination of an electron-filled, equatorial disc with
        electron-filled, polar outflows; The wavelength-independent
        $PF/F_{*}$ is color-coded and integrated over the complete
        wavelength range.
        \textit{Top}: face-on image;
        \textit{Middle}: image at $i \backsim 45^\circ$;
        \textit{Bottom}: edge-on image.}
     \label{Fig6.2}%
   \end{figure}

The polarization maps in Fig.~\ref{Fig6.2} further illustrate our
discussion. At pole-on view (Fig.~\ref{Fig6.2} top), the double cone
is visible in transmission and reflection. In many areas of the image,
the polarization angle is at $\gamma = 0^\circ$, however, the associated
polarized flux remains weak compared to that coming from the equatorial
disc and having $\gamma = 90^\circ$. The integrated $PF/F_{*}$ therefore
has $\gamma = 90^\circ$. The polarization images at intermediate viewing
angles (Fig.~\ref{Fig6.2} middle) are partly in agreement with the results obtained for scattering in
polar outflows alone (see Sect.~\ref{sec:e-cone}). Again, the two maxima in
polarized flux induced by scattering in the far and the near cone are seen
and they show perpendicular polarization. At the center, the impact of the equatorial
disc is visible, and $\gamma$ rotates toward $90^\circ$. Finally, when seen
edge-on (Fig.~\ref{Fig6.2} bottom), the net polarization has switched to
$\gamma = 0^\circ$. The scattered photons from the polar winds are now strongly
polarized and dominate the net polarized flux.

\subsection{Radiation-supported disc and obscuring torus}
\label{sec:combequattorus}

Next, we consider a combination of the equatorial scattering disc with
an optically-thick, dusty torus. The parameterization of the reprocessing
regions is as described before (see Sect.~\ref{sec:equatdisc} and
\ref{sec:inditorus}, respectively). The absence of polar outflows
highlights a particular sub-class of thermal Seyfert-1 AGNs that are characterized
by a very weak or absent amount of intrinsic warm absorption \citep{Patrick2011},
a subclass also known as ``bare'' AGN.

   \begin{figure}
   \centering
      \includegraphics[width=10.5cm]{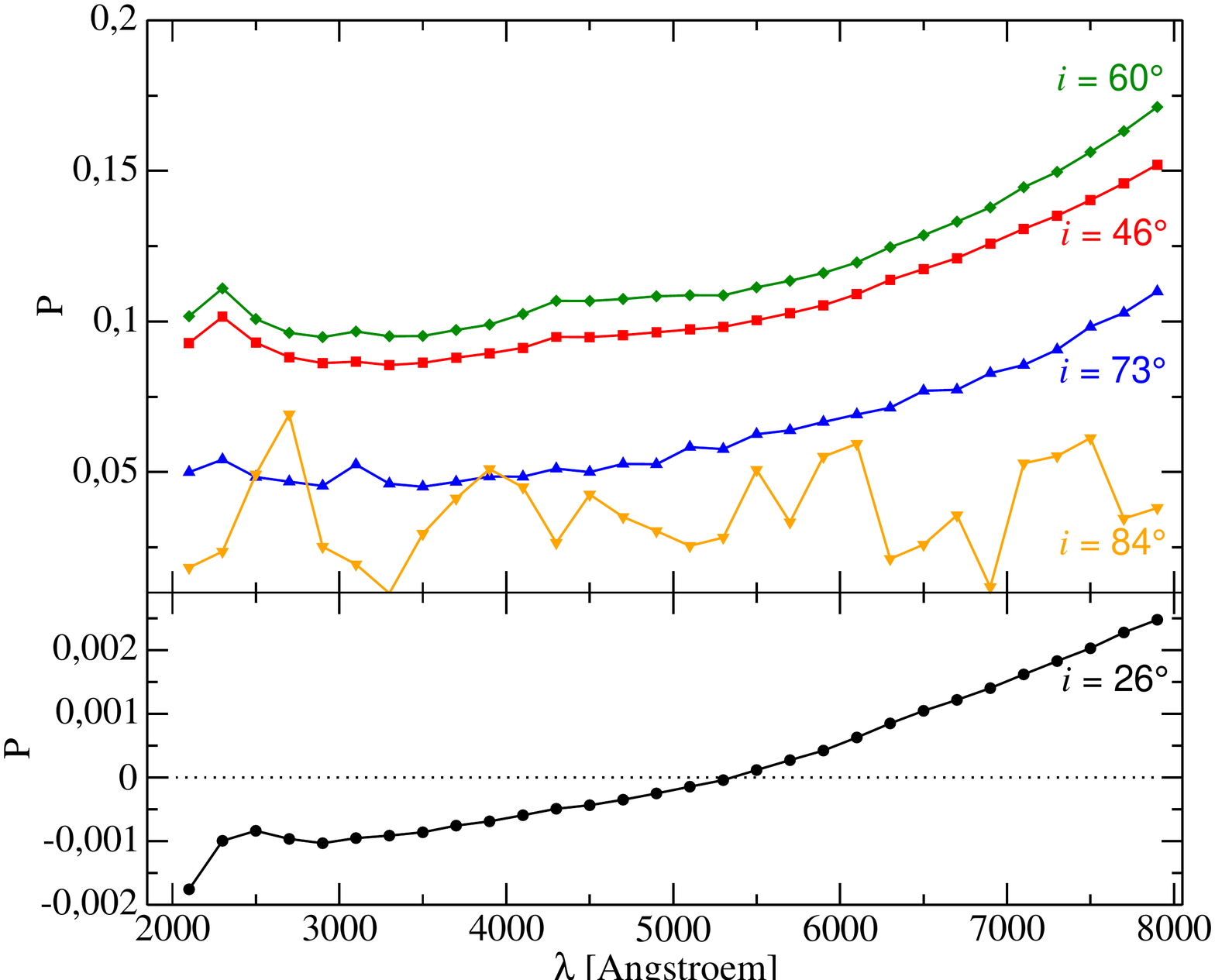}
      \includegraphics[width=10.5cm]{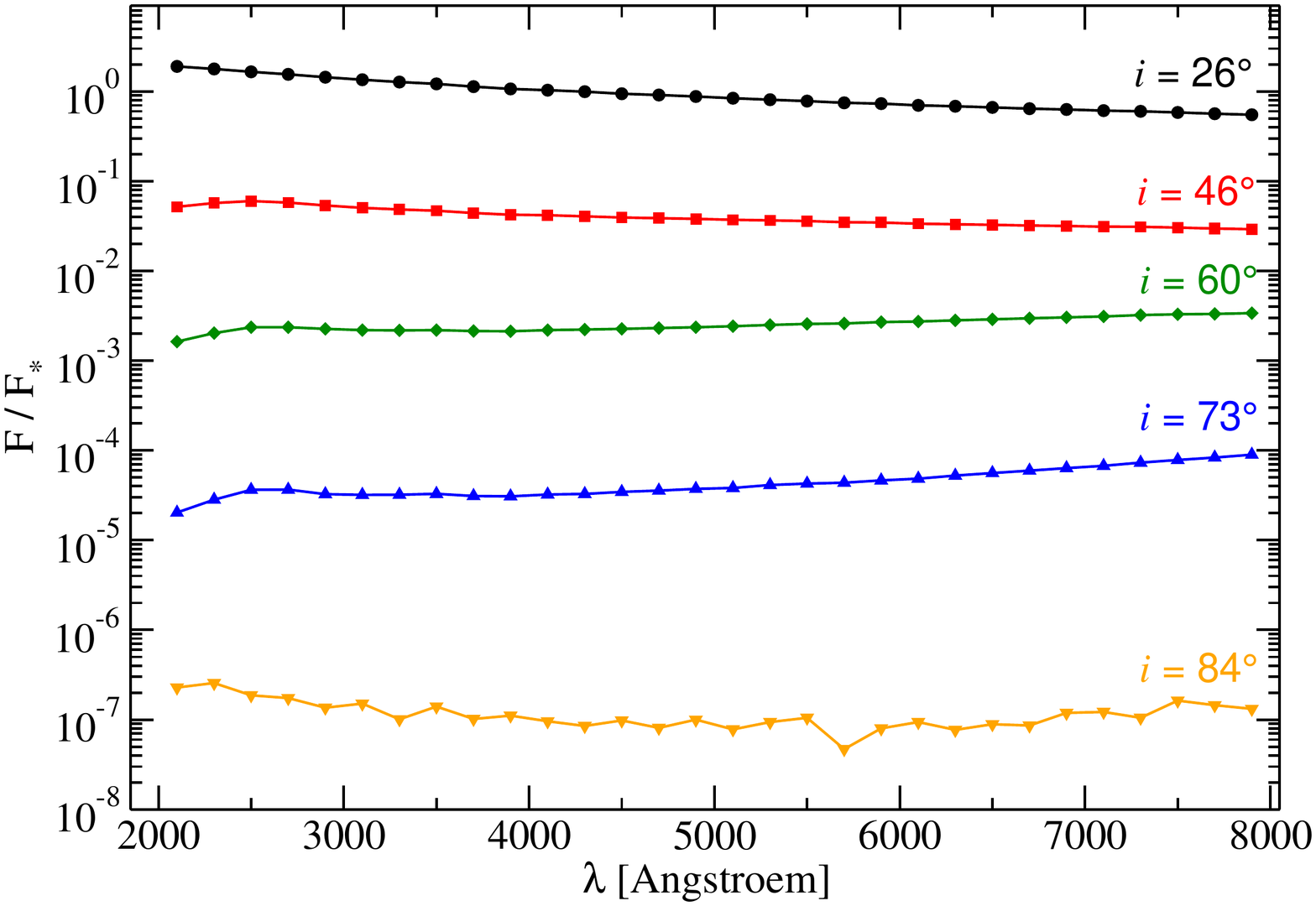}
      \includegraphics[width=10.5cm]{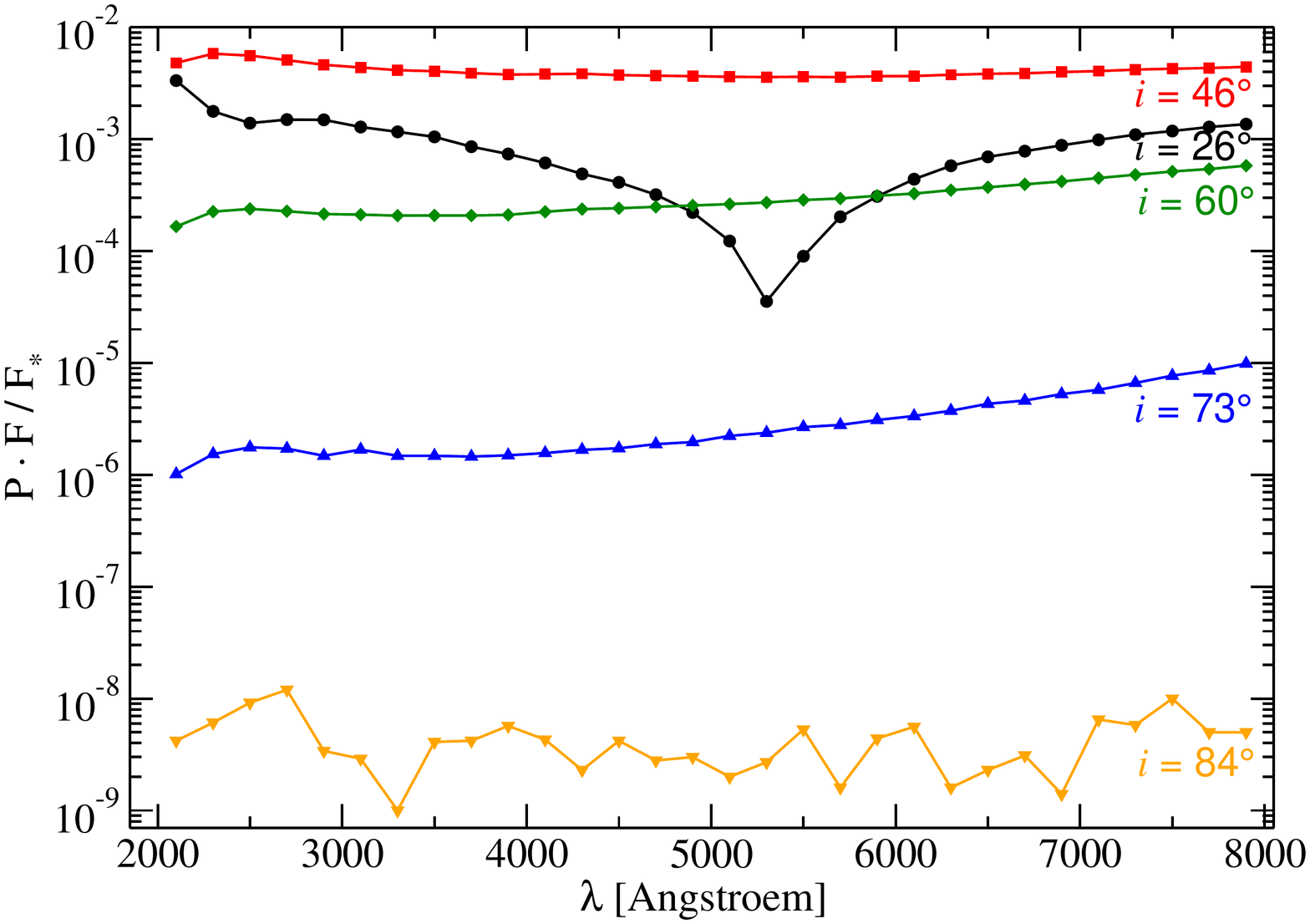}
      \caption{Modeling an optically-thick dusty torus with the
        half-opening angle $\theta_{c}~=~30^\circ$ relative to the
        symmetry axis combined with an electron-filled, equatorial
        disc with half-opening angle $\theta_0$~=~$10^\circ$ measured
        from equatorial plane. \textit{Top}: Polarization, \textit{P},
        as seen at different viewing inclinations, \textit{i};
        \textit{Middle}: the fraction, F/$F_{\rm *}$ of the central
        flux.}
     \label{Fig6.3}%
   \end{figure}

Figure~\ref{Fig6.3} shows, from top to bottom, the polarization
percentage, the spectral flux, and the normalized polarized flux
as a function of wavelength. At pole-on view, the behavior of
\textit{P} is particularly interesting: The polarization degree is
negative in the UV-band indicating parallel polarization. Then,
with increasing wavelength, the polarization weakens and undergoes
a sign inversion around 5200~\AA. For larger wavelengths, the
polarization becomes again significant but perpendicular. This behavior
is due to the competition in polarized flux between the equatorial
disc and the torus. The two scattering regions produce polarization
at opposite signs, but the albedo and scattering phase function of
the dust change systematically with wavelength. Below 5200~\AA,
the phase function of Mie scattering strongly promotes forward
scattering over scattering towards other directions. Note also,
that the dust albedo in the ultraviolet is slightly lower than at
longer wavelengths so that, in total, bluer photons hitting the inner
walls of the torus have a higher probability to be absorbed than redder
photons. In the UV, the polarized flux emerging from the equatorial disc
thus predominates. Above 5200~\AA, the Mie scattering phase function
is less anisotropic and the polarized flux scattered off of the torus
inner walls and towards a pole-on observer becomes more important.
The net polarization is then dominated by the torus and perpendicular.

At higher inclination, the equatorial scattering disc is hidden behind
the torus and therefore its polarized flux with $\gamma = 90^\circ$
is not directly visible. The net polarization is now perpendicular across
the whole waveband and rises towards longer wavelengths as it does for the
reprocessing of a dusty torus alone (see Fig.~\ref{Fig3.1}). The interplay
between electron and dust scattering is also visible in the F/$F_{\rm *}$
spectrum (Fig.~\ref{Fig6.3}, middle). With increasing $i$, the spectral
slope in the optical changes gradually. This behavior is again determined
by the combined effect of the wavelength-dependent albedo and scattering
phase function as well as the specific scattering geometry chosen in this model.
 Comparison with the spectral flux obtained for a dusty torus alone
(see Fig.~\ref{Fig3.1}) reveals that the negative slope for low viewing
angles is an effect of the additional electron scattering happening inside
the torus funnel. The scattering feature of carbonaceous dust in the UV
is seen only at intermediate and edge-on viewing angles. At pole-on view,
the feature is blended by the direct flux from the source and by scattered
radiation coming from the equatorial disc.

It is instructive to also discuss the polarized flux spectrum
(Fig.~\ref{Fig6.3}, bottom) for this modeling case. At pole-on view,
the $PF/F_{*}$ shows a minimum across the switch of the polarization
position angle around 5200~\AA~ and reduces the polarized flux by a large
factor (note that when constructing the polarized flux we always take the
absolute value of $P$). We should point out that the presence of the flip
in polarization strongly depends on the exact model geometry and the scattering
efficiency of the equatorial scattering disc. The strongest polarized flux
occurs at a line of sight slightly below the torus horizon. Towards higher
inclinations, the flux drops rapidly and so does $PF/F_{*}$.

   \begin{figure}
   \centering
      \includegraphics[trim = 5mm 5mm 0mm 10mm, clip, width=8cm]{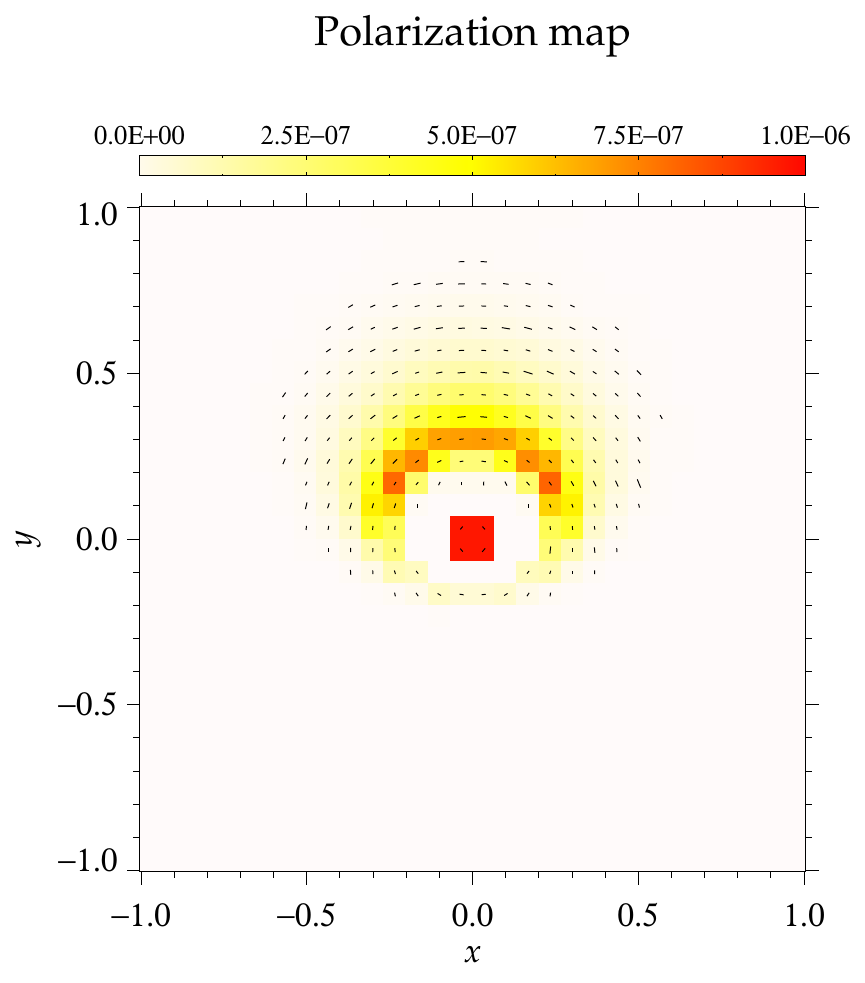}
      \includegraphics[trim = 5mm 5mm 0mm 10mm, clip, width=8cm]{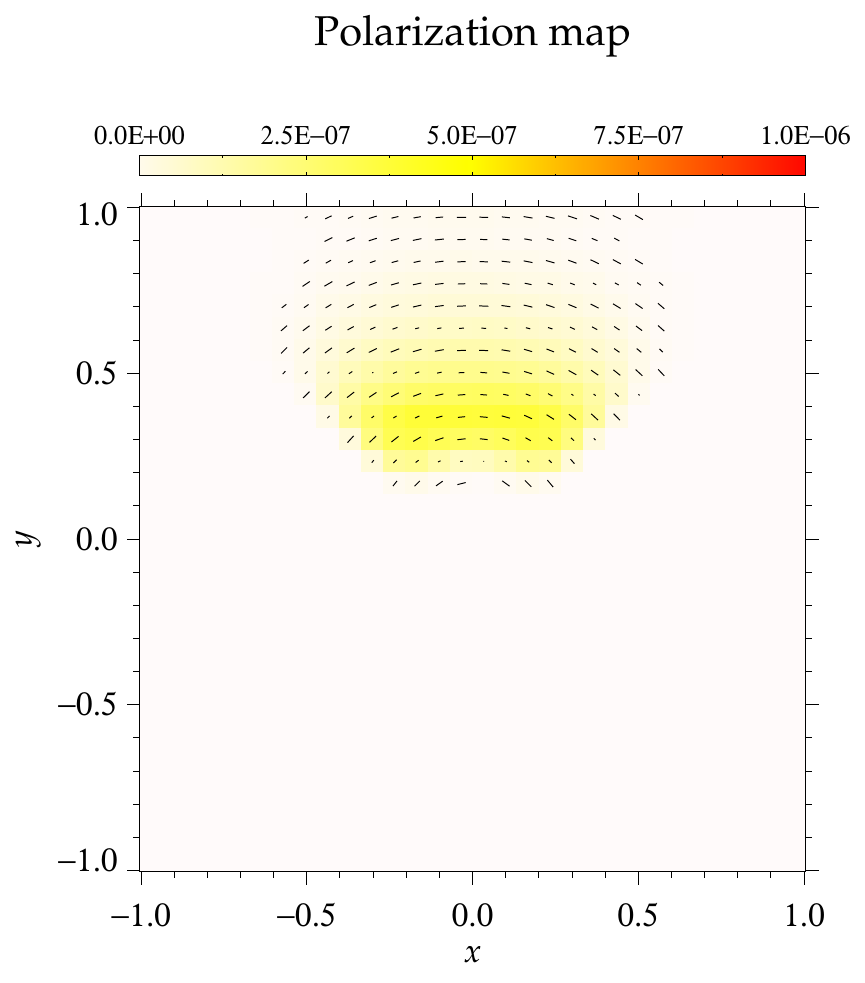}
      \caption{Modeled image of the polarized flux, $PF/F_{*}$, for an
        electron-filled, equatorial disc combined with an optically
        thick, dusty torus; $PF/F_{*}$ is color-coded and integrated
        over the wavelength band. \textit{Top}: face-on image;
        \textit{Bottom}: image at $i \backsim 45^\circ$.}
     \label{Fig6.4}
   \end{figure}

The polarization maps are presented in Fig.~\ref{Fig6.4}. Due to
the large difference in spatial scale between the two scattering regions,
we restrict the mapping to the funnel of the dusty torus.
At pole-on view (Fig.~\ref{Fig6.4}, top), the (wavelength-integrated)
map shows that the polarized flux with $\gamma = 0^\circ$ coming from
the dusty torus is almost of the same order of magnitude than the one
emerging in the equatorial disc and carrying $\gamma = 90^\circ$. One has
to take into account that the polarized flux spectrum reported in
Fig.~\ref{Fig6.3} is integrated over all scattering surfaces; with
respect to the equatorial scattering disc, the polarized flux coming
from a given position on the torus inner wall is smaller, but this is
compensated by a larger integration surface. As discussed above,
the exact outcome of the competition between the two components and
the resulting polarization position angle depends on the wavelength.
The situation is much clearer at an intermediate viewing angle
(Fig.~\ref{Fig6.4} bottom). Here, the polarized flux mostly emerges
from the far-sided inner wall of the torus that induces
$\gamma = 0^\circ$. The near-sided wall and the walls on the side of
the line of sight are barely visible as they are covered by the torus
body. The equatorial scattering disc is hidden in the torus funnel so
that the net polarization can only be equal to $\gamma = 0^\circ$.
We do not show the edge-on image for this model because almost all
radiation is blocked by the optically-thick torus.

\subsection{Electron polar outflows and obscuring torus}
\label{sec:combtoruswind}

Finally, we construct a reprocessing model that combines the
electron-filled polar outflows and the optically-thick, dusty torus as
presented previously in Sects.~\ref{sec:e-cone} and \ref{sec:inditorus},
respectively. Such a model setup would feature an AGN that (temporarily)
lacks a material connection between the inner boundaries of the
dusty torus and the outer parts of the accretion disc. The so-called
``naked'' AGN \citep{Panessa2002,Hawkins2004,Panessa2009,Tran2011} that seem
to lack a BLR could fall into this category.

   \begin{figure}
   \centering
      \includegraphics[width=10.5cm]{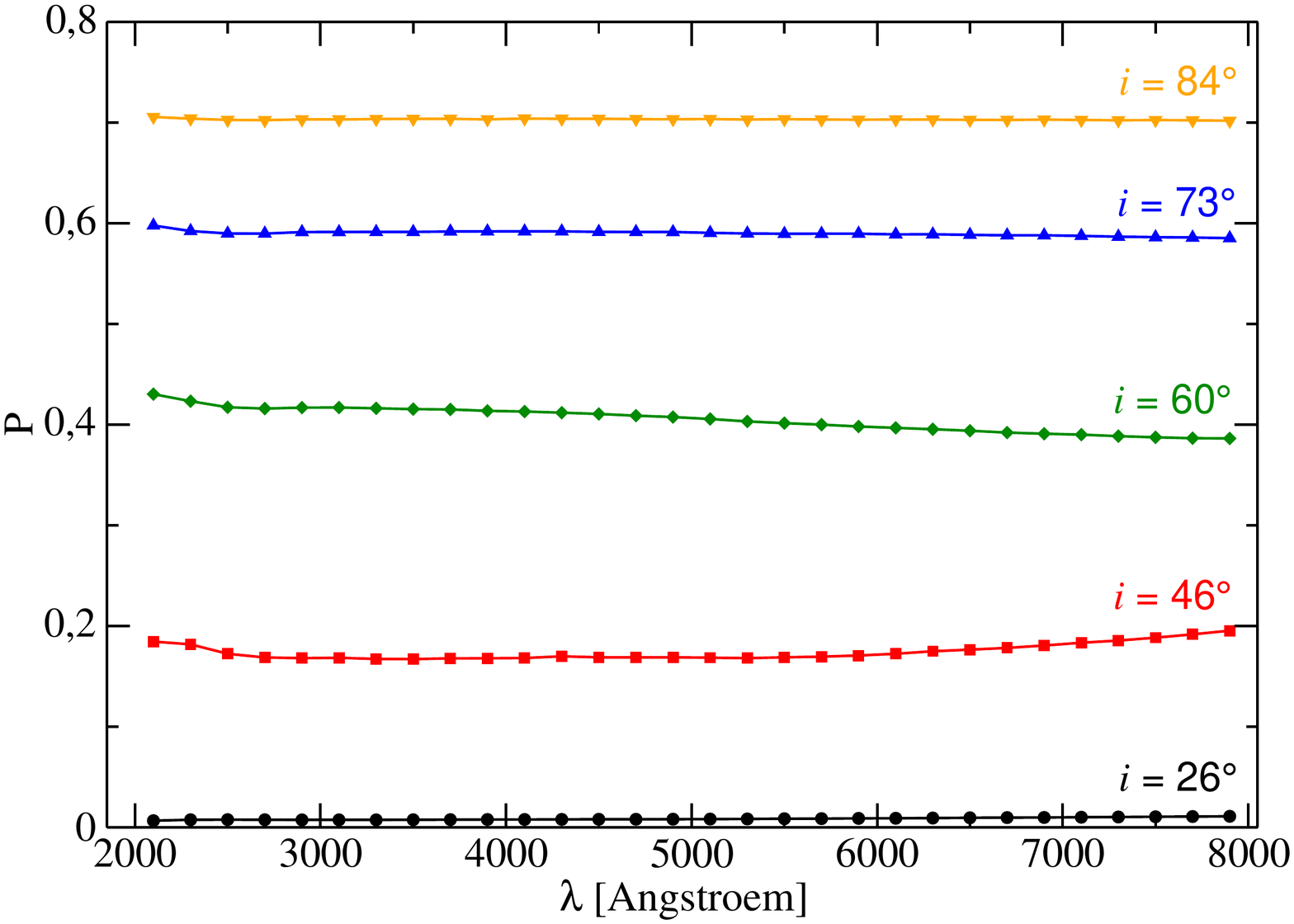}
      \includegraphics[width=10.5cm]{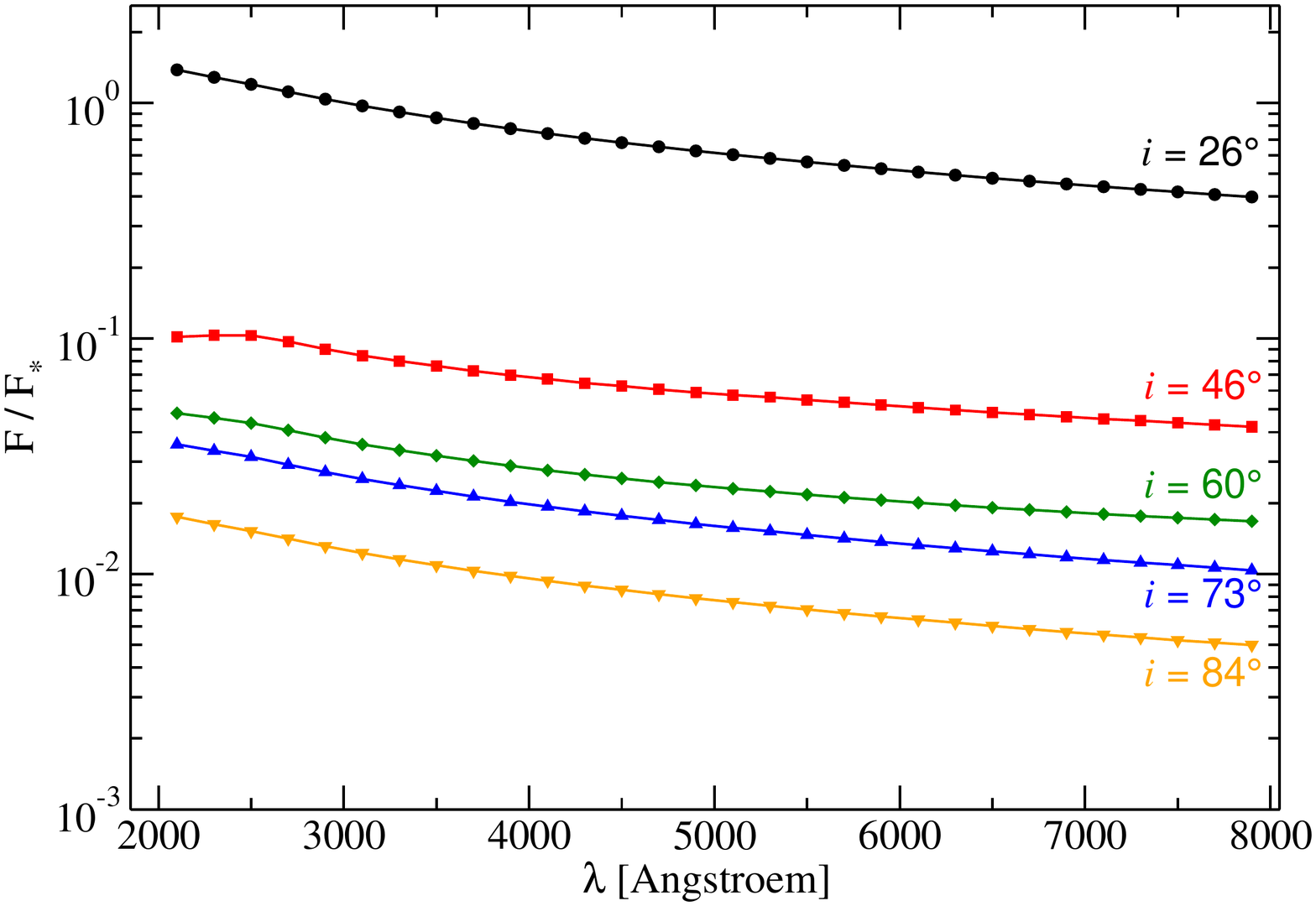}
      \caption{Modeling an electron-filled double cone combined with an
        optically-thick dusty torus, both with a half-opening angle of
        $\theta_{c}~=~30^\circ$ measured with respect to the symmetry
        axis. Results are shown for different viewing angles \textit{i}.
        \textit{Top}: Polarization \textit{P}; \textit{Bottom}:
        the fraction, F/$F_{\rm *}$ of the central flux.}
     \label{Fig6.5}%
   \end{figure}

The spectropolarimetric modeling is presented in Fig.~\ref{Fig6.5}.
The results reported in Sect.~\ref{sec:inditorus} show that at face-on
viewing angles the polarization due to scattering off of a dusty torus
is low and rather independent of wavelength. Here we obtain almost the
same picture because the photons pass through the optically thin wind
mostly by forward or backward scattering. When the line of sight crosses
the torus horizon, the dilution by the unpolarized source flux is suppressed
and the polarization reaches higher values. With further rising inclination,
there are two effects increasing the normalization of the polarization
spectrum: Firstly, there are more dust scattering events necessary to
escape from the torus funnel. The systematic multiple scattering ``sharpens''
the perpendicular polarization. Note that in many other situations
multiple scattering is a depolarizing effect. The polarization ``sharpening''
established here is closely linked to the narrow funnel-geometry and
the large optical depth of the torus, which makes only scattering off
of its surface important. The second effect relates to the outflow in
which stronger polarization is produced for higher $i$. This is due to
the shape of the polarization phase function of Thomson scattering.

In contrast to the modeling of a torus alone (see Sect.~\ref{sec:inditorus}),
here the \textit{F/F$_{\rm *}$} spectrum decreases towards longer
wavelengths at all viewing angles considered; we already discuss in
Sect.~\ref{sec:combequattorus} that additional electron scattering inside
the torus funnel tends to decrease the spectral slope. In this respect,
the polar outflows are even more efficient than the equatorial disc.
The effect is again related to the change in scattering phase function
from the UV to the optical waveband: the electrons inside the torus funnel
scatter primary photons towards the torus inner surfaces. Since these
surfaces have a receding shape towards the exit of the funnel, the
scattered photons impinge it at a rather grazing angle. In the UV, the
scattering phase function largely favors scattering into a cone of
$\sim 30^\circ$ half-opening angle around the forward direction.
This gives many photons a good chance to be scattered only once by the
dust and then to escape from the torus funnel. Optical photons are more
often scattered to a direction that leads back into the funnel and thus
they are slightly more likely to be absorbed. This explains why the
scattered spectrum of the torus is stronger in the UV than in the optical
waveband (even though the albedo in the UV is slightly lower than in
the optical).

As for the case of a scattering torus alone, the normalization of the
\textit{F/F$_{\rm *}$}-spectrum decreases towards a higher viewing
angle because the visible scattering surfaces of the funnel become
rapidly smaller. Also, the scattering efficiency inside the polar
outflows is lower for scattering towards higher $i$ than towards a pole-on
direction. At all type-2 inclinations, a dim feature of dust reprocessing
in the UV is visible and traces the radiation component that is scattered
into the line of sight by the torus.

   \begin{figure}
   \centering
      \includegraphics[trim = 5mm 5mm 0mm 10mm, clip, width=8cm]{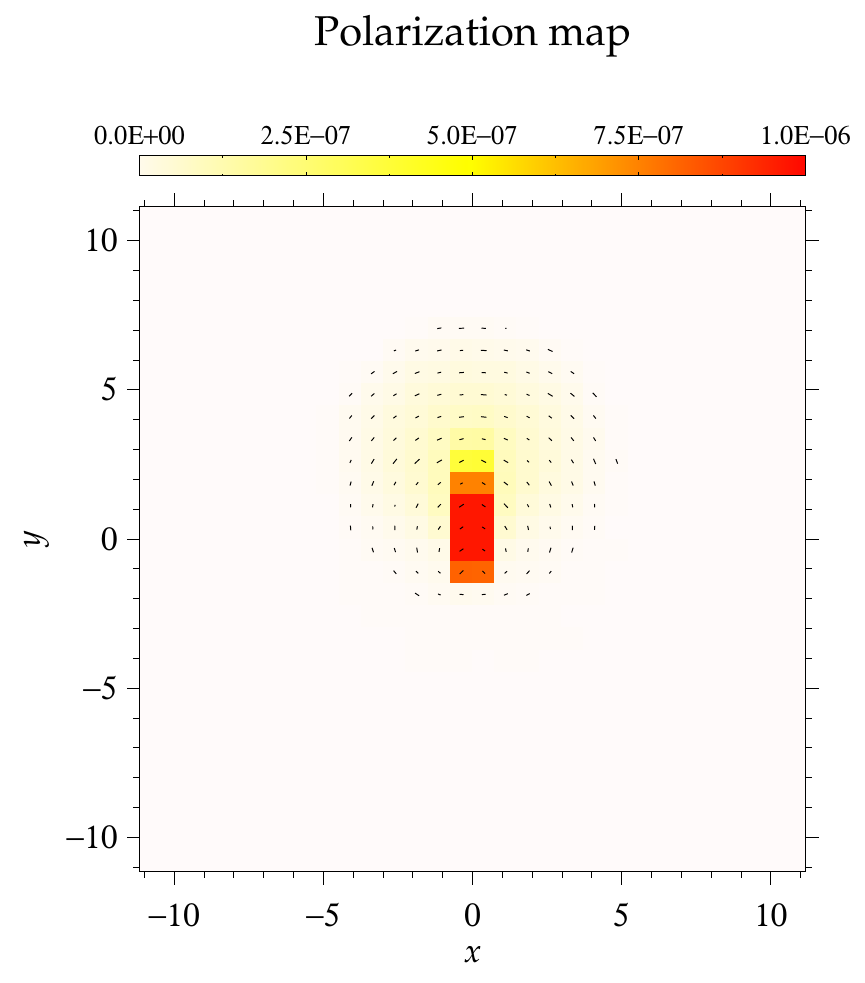}
      \includegraphics[trim = 5mm 5mm 0mm 10mm, clip, width=8cm]{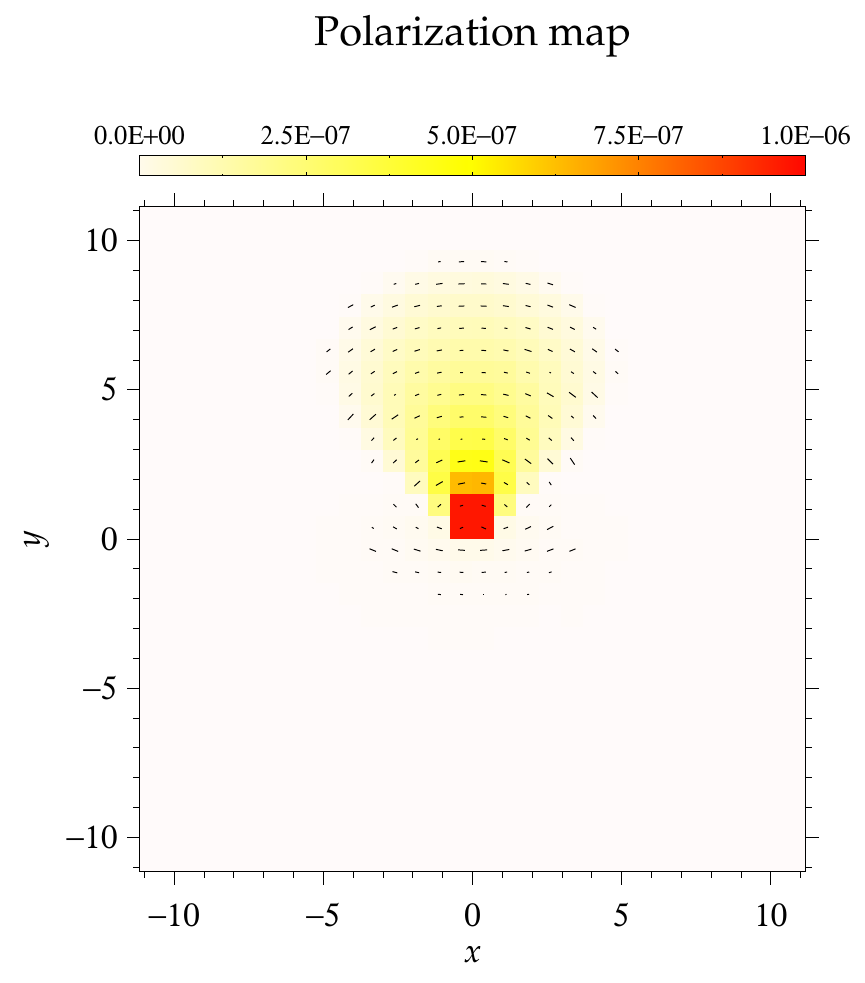}
      \includegraphics[trim = 5mm 5mm 0mm 10mm, clip, width=8cm]{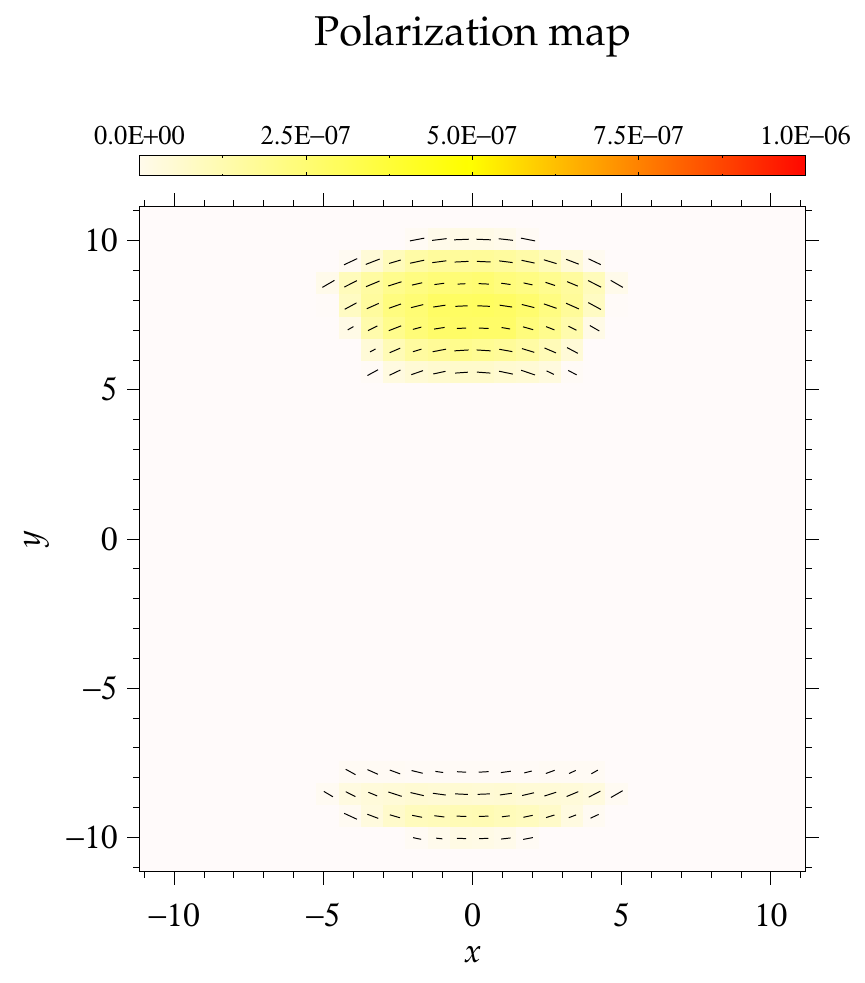}
      \caption{Modeling the polarized flux, $PF/F_{*}$, of electron-filled, polar
	       outflow combined with an optically-thick, dusty torus; $PF/F_{*}$
	       color-coded and integrated over all wavelengths.
	       \textit{Top}: face-on image;
	       \textit{Middle}: image at $i \backsim 45^\circ$;
	       \textit{Bottom}: edge-on image.}
     \label{Fig6.6}%
   \end{figure}

The polarization maps are shown in Fig.~\ref{Fig6.6}. At pole-on view,
the spatial maximum that is related to the far-cone (seen in reflection)
is less extended than the one related to the near cone (seen in transmission);
compared to the pole-on image of an isolated, electron-filled outflow
(see Sect.~\ref{sec:e-cone}), the situation is reversed. This shows how
the reflected flux from the far cone is partially blocked by the optically-
thick torus. Nevertheless, a significant polarized flux comes from the
central region of the model, which is due to the combined scattering inside
the bases of the double cone and off of the inner torus walls, both producing
perpendicular polarization. The two scattering regions thus reinforce each
other in terms of polarization efficiency and the net polarization is, of
course, perpendicular.


At an intermediate viewing angle, the double cone is more significantly
hidden and the far-cone is no longer visible at all. A large amount of
polarized flux still comes from the near cone and compared to the isolated
outflow, the gradient in $PF/F_{*}$ between the base and the outer regions
is more shallow. This is related to the collimating effect of the torus
funnel efficiently channeling photons (back) towards the outflow.
Moreover, we point out the presence of a secondary scattering component
coming from regions of the dusty torus that are not directly exposed to
the source irradiation. When comparing with the polarization map at
intermediate view towards an isolated torus (see Sect.~\ref{sec:inditorus}),
this component becomes particularly visible on surfaces of the torus
that are on the near side with respect to the line of sight.
The photons constituting this component have first undergone back-scattering
inside the outflow and then they were scattered off of the torus surface and
towards the observer. Such a process is possible because the dusty torus
has a significant average albedo of $\sim 0.57$ in the optical/UV band.

Finally, at edge-on view, the more distant areas of both, the upper and
the lower parts of the double cone are visible in reflection again.
The center of the image shows no polarized flux due to the entirely opaque
torus. The two extensions of the outflow scatter photons around the opaque
torus and thereby produce strong polarization at a perpendicular
orientation. Visualizing the polarized flux therefore enables
us again to have the ``periscope view'' at the hidden nucleus.

\section{Modeling the unified scheme of AGN}
\label{sec:AGN}

To approach the unified AGN scheme, we now build up a complex model
composed of three radiatively coupled reprocessing regions: around
the point-like, emitting source we arrange an equatorial electron
scattering disc, polar electron outflows, and an obscuring dusty torus.
The parameters of the model are summarized in Table~\ref{tab:agn_param}.
We investigate the polarization spectra and images for this particular
model and then explore the parameter space somewhat further by varying
the geometry and optical depths of the equatorial and polar scattering
regions.

\begin{table*}[]
 \begin{center}
   \begin{tabular}{|c|c|c|}
   \hline {\bf flared disc}               & {\bf dusty torus}              & {\bf polar outflows}\\
   \hline $R_{\rm min} = 3.10^{-4}$ pc      & $R_{\rm min} = 0.25$ pc          & $R_{\rm min} = 1$ pc\\
          $R_{\rm max} = 5.10^{-4}$ pc     & $R_{\rm max} = 100$ pc           & $R_{\rm max} = 10$ pc\\
          half-opening angle = 10$^\circ$  & half-opening angle = 30$^\circ$ & half-opening angle = 30$^\circ$\\
          equat. optical depth = 1        & equat. optical depth = 750     & vertical optical depth = 0.03\\
   \hline
   \end{tabular}
   \caption{Parameters for a thermal AGN model consisting of three scattering regions.
	    The polar outflows and torus half-opening angles are measured with
	    respect to the vertical symmetry axis of the torus. The half-opening
	    angle of the flared disc is measured from the equatorial plane.}
  \label{tab:agn_param}
 \end{center}
\end{table*}

   \begin{figure}
   \centering
      \includegraphics[width=10.5cm]{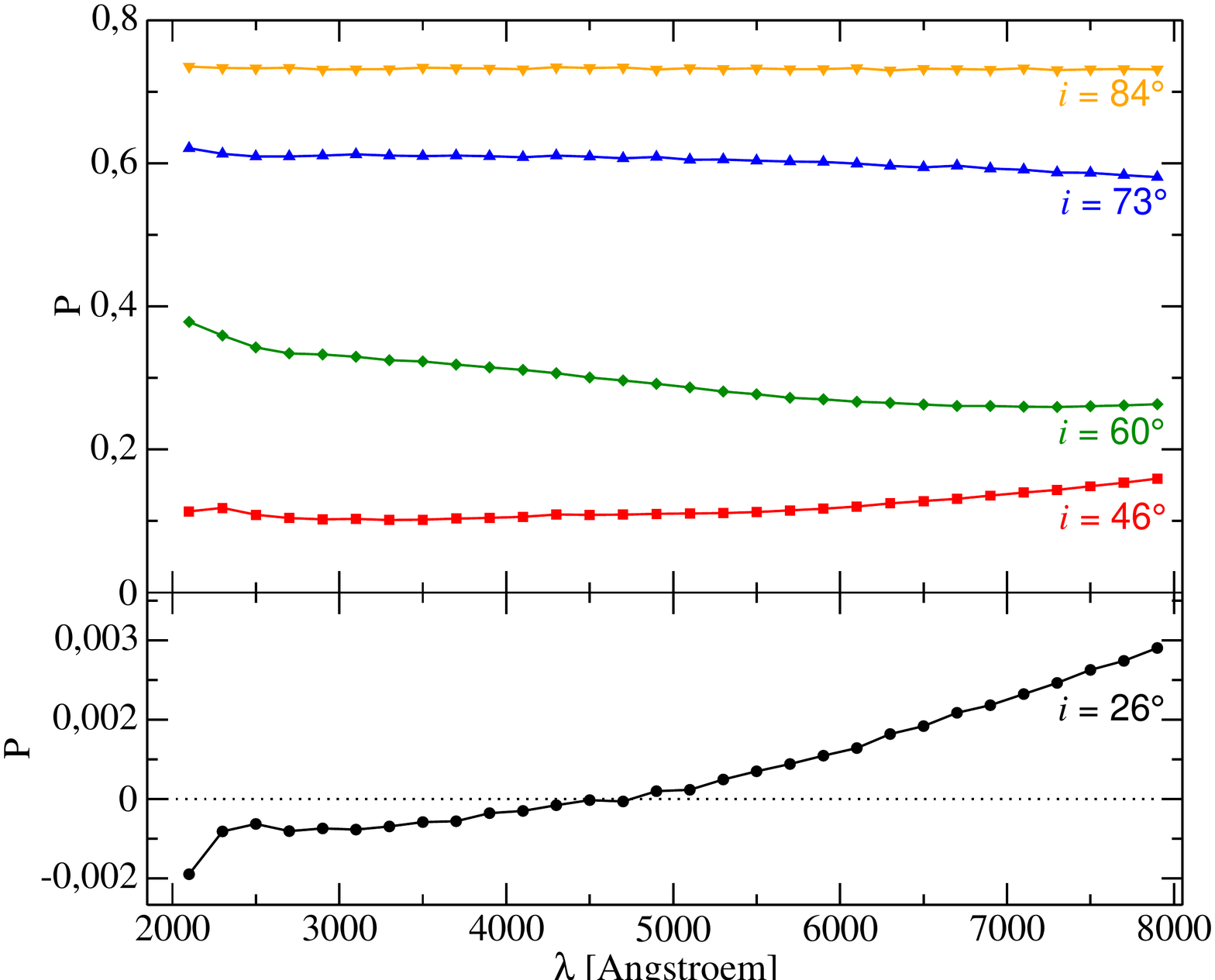}
      \includegraphics[width=10.5cm]{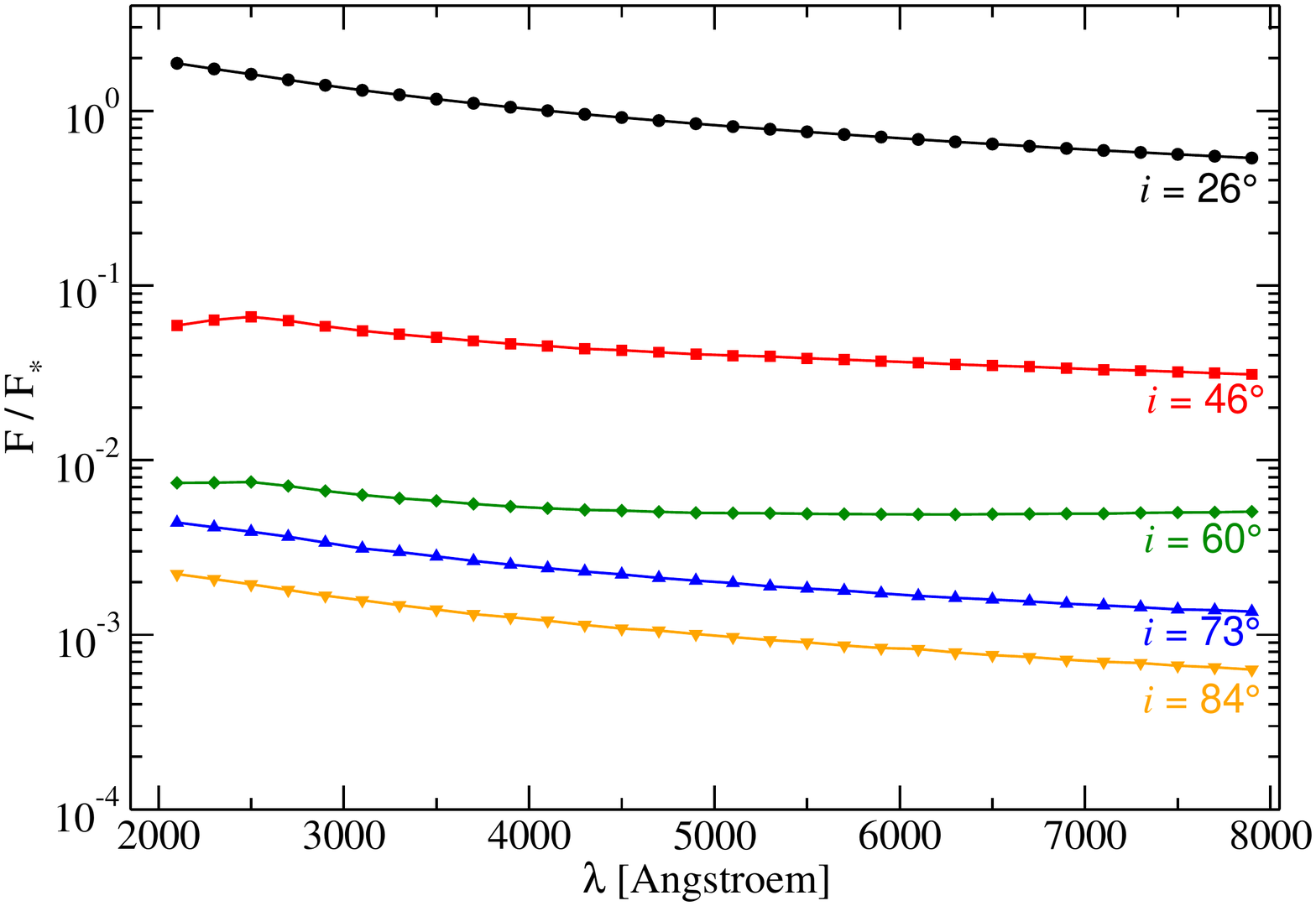}
      \caption{Modeling the unified scheme of a thermal AGN by three reprocessing regions (see text).
        \textit{Top}: Polarization, \textit{P}, as a function of viewing inclination, \textit{i};
        \textit{Bottom}: the fraction, $F/F_{\rm *}$, of the central flux $F_{\rm *}$.}
     \label{Fig7.0}%
   \end{figure}

\subsection{Spectral modeling results}

The spectral results for the model described in Tab.~\ref{tab:agn_param}
are shown in Fig.~\ref{Fig7.0}. At pole-on view, \textit{P} is negative
in the UV (parallel polarization) but of very low magnitude as the system
appears to be almost axis-symmetric. Similarly to the polarization
spectrum obtained for the combination of a dusty torus and an equatorial
scattering disc, a sign inversion is detected between smaller and longer
wavelengths. With respect to the results of Fig.~\ref{Fig6.3} (top), the
transition wavelength is shifted due to the additional presence of the polar
outflows. The exact wavelength at which the polarization angle switches also
depends on the adopted outflow geometry and optical depths. At intermediate
viewing angles, the effects of Mie scattering are visible by a weak feature
in the UV band and a slight slope of the polarization spectrum over optical
wavelengths. Also the $F/F_{\rm *}$ spectrum reveals the influence of
wavelength-dependent dust scattering and absorption in the torus funnel by
a gradually decreasing flux towards longer wavelengths, which is caused by
the additional electron scattering inside the torus funnel (see the discussion
in Sects.~\ref{sec:combequattorus} and \ref{sec:combtoruswind}). A small peak
of the flux around 2175~\AA~ is due to carbonaceous dust and visible at low
and intermediate viewing angles. When further increasing \textit{i},
the radiation becomes dominated by wavelength-independent, nearly perpendicular
electron scattering inside the polar winds and \textit{P} can achieve high
values of more than 70\%.

   \begin{figure}
   \centering
      \includegraphics[trim = 5mm 5mm 0mm 10mm, clip, width=8cm]{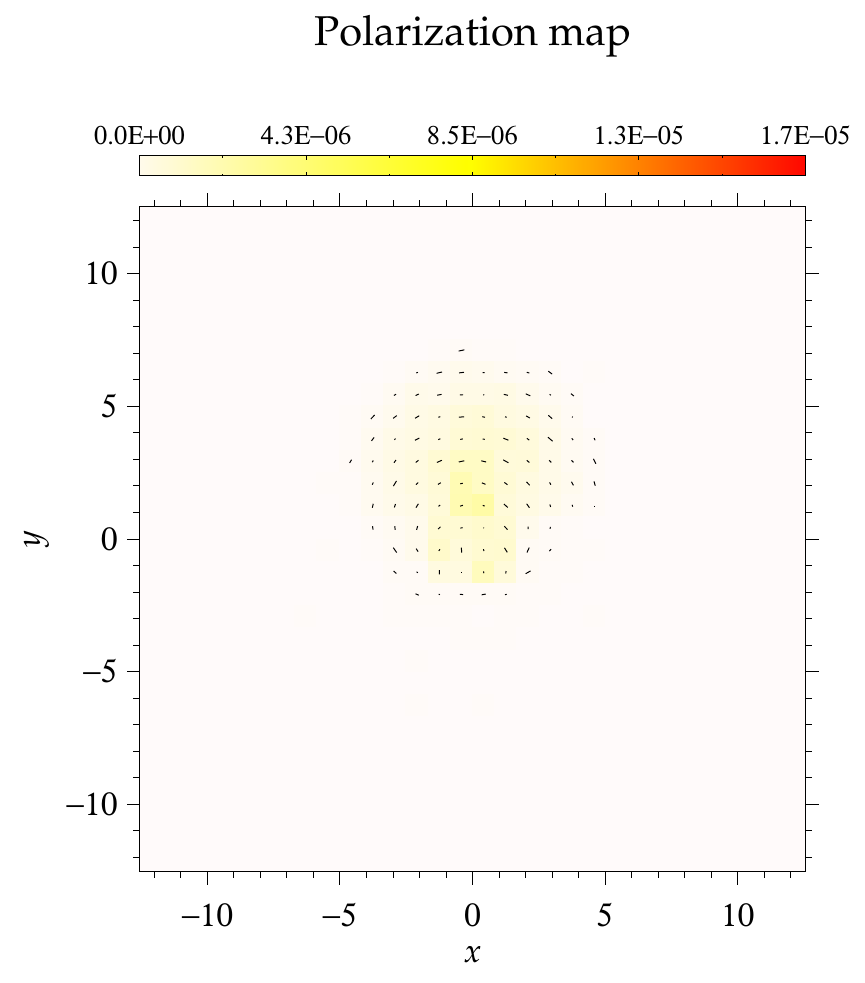}
      \includegraphics[trim = 5mm 5mm 0mm 10mm, clip, width=8cm]{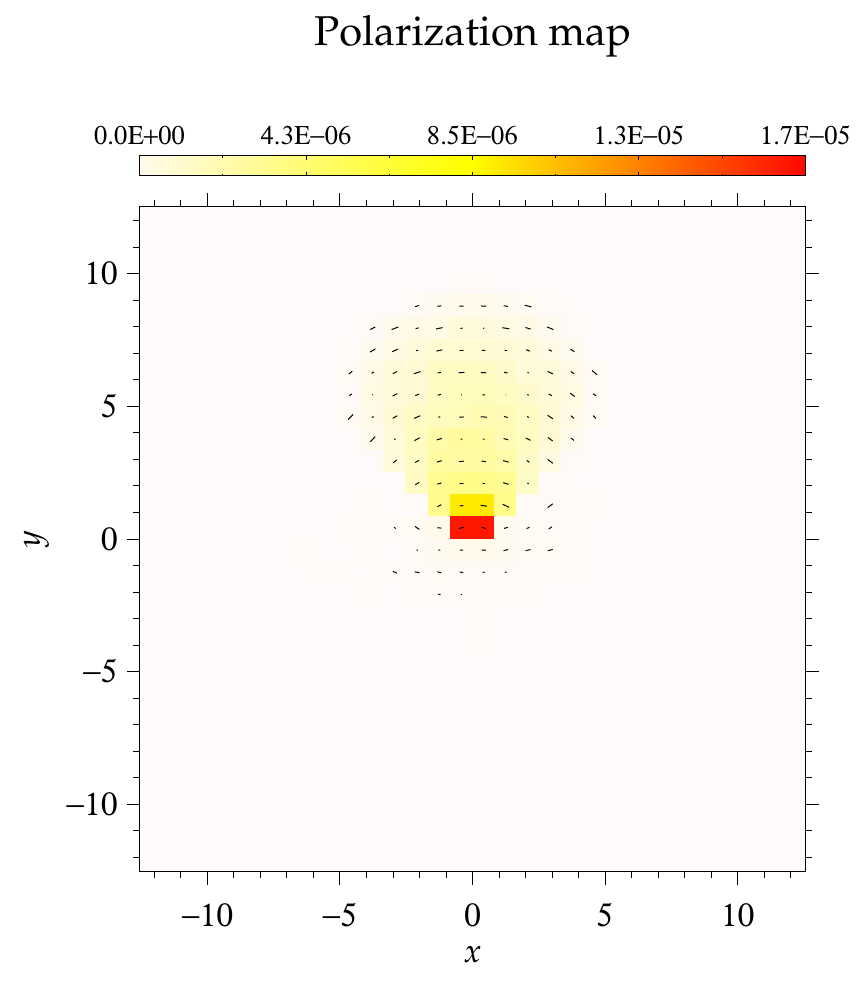}
      \includegraphics[trim = 5mm 5mm 0mm 10mm, clip, width=8cm]{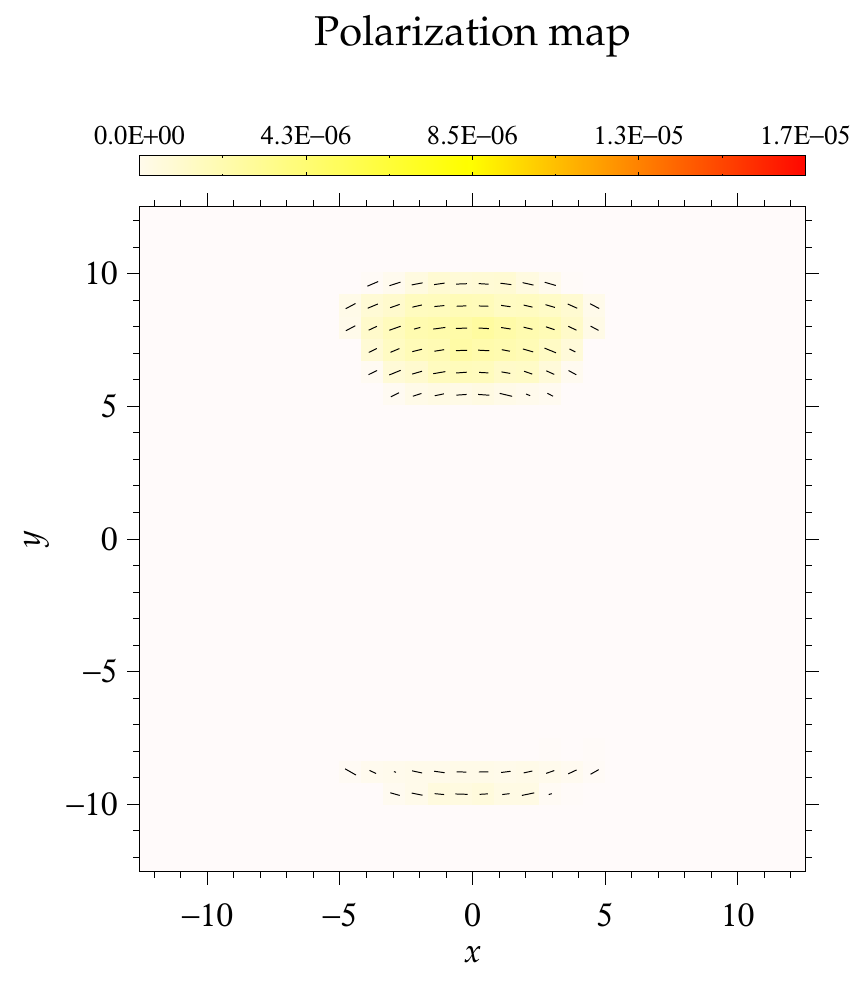}
      \caption{Modeling the polarized flux, $PF/F_{*}$, induced by complex reprocessing in
	       a unified model of AGN. We combine an electron filled, equatorial scattering
	       disc, electron filled polar outflows, and an optically-thick dusty torus;
	       $PF/F_{*}$ is color-coded and integrated over all wavelengths.
	       \textit{Top}: face-on image;
	       \textit{Middle}: image at $i \backsim 45^\circ$;
	       \textit{Bottom}: edge-on image.}
     \label{Fig7.1}%
   \end{figure}

\subsection{Wavelength-integrated polarization images}

The polarization maps of our thermal AGN model are presented in Fig.~\ref{Fig7.1}.
At pole-on view (Fig.~\ref{Fig7.1}, top), the distribution of the polarized
flux is somewhat similar to the one obtained for the combination of an
equatorial disc and a dusty torus only (Fig.~\ref{Fig6.4}, top); the effect
of the polar outflows on the polarization at this viewing angle is rather small;
with the strongest polarized flux coming from the equatorial scattering disc,
the net polarization is parallel -- as it is observed in many type-1 AGN.

At intermediate viewing angles, the equatorial disc and the primary source
are hidden by the optically-thick torus. Nevertheless, the polarization
produced at the base of the near cone (seen in reflection) is influenced
by scattering inside the equatorial disc. The equatorial disc scatters a
certain fraction of the primary radiation towards the outflows thereby
inducing parallel polarization. The perpendicular polarization caused by
the secondary scattering inside the electron-filled double-cone is thus
weakened when compared to the model that does not include the equatorial
scattering disc (see Fig.~\ref{Fig6.6}, middle).

The base of the far cone is not visible because it is hidden behind the
opaque torus. Instead we detect, as in the modeling presented in
Fig.~\ref{Fig6.6} (middle), a low polarized flux scattered off the near inner
surfaces of the torus. The photons of this flux have been back-scattered from
the polar winds onto the torus and then towards the observer. The overall
polarization position angle at intermediate viewing angles is $\gamma = 0^\circ$.

At edge-on view (Fig.~\ref{Fig7.1} below), only the most upper and lower
parts of the double cone appear above and below the body of the torus,
which completely hides the central region of the model. The polarization
effects due to the equatorial disc are now largely exceeded by scattering
in the outflows at almost perpendicular scattering angles. The resulting
net polarization is therefore perpendicular and of high degree.
Note that the adopted viewing angle in Fig.~\ref{Fig7.1} is close but
not exactly equal to $90^\circ$, which explains the asymmetry between
the top and bottom part of the polar winds.

\begin{figure}
   \centering
      \includegraphics[trim = 5mm 5mm 0mm 10mm, clip, width=8cm]{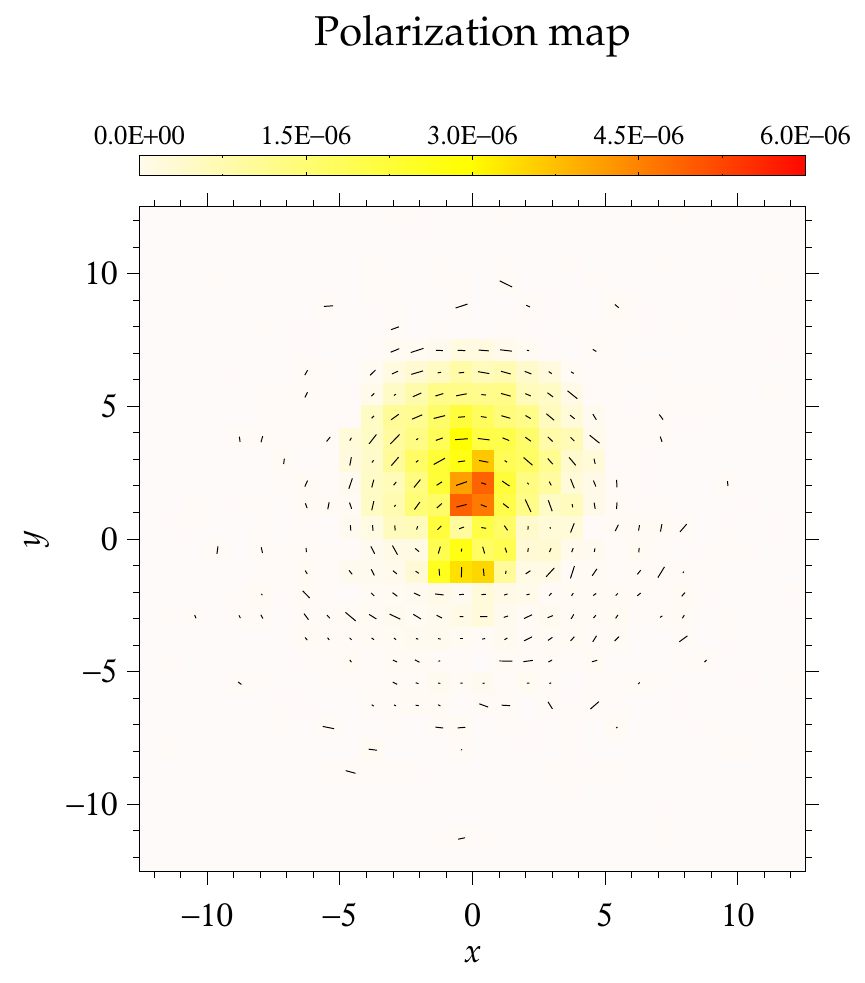}
      \includegraphics[trim = 5mm 5mm 0mm 10mm, clip, width=8cm]{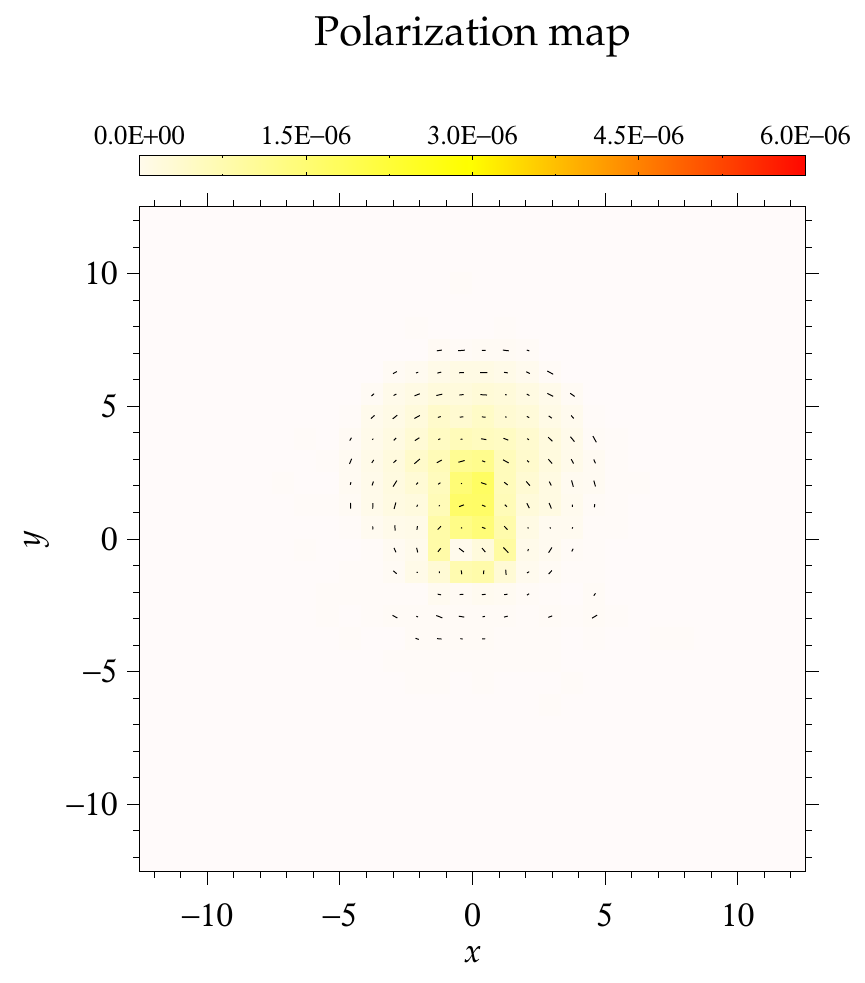}
      \caption{Modeling the polarized flux, $PF/F_{*}$, of an AGN composed of an
        equatorial scattering disc, electron-filled, polar outflows and an optically
        thick dusty torus; $PF/F_{*}$ is color-coded and integrated over all wavelengths.
	\textit{Top}: pole-on image at $\backsim$ 2175 \AA;
	\textit{Bottom}: pole-on image at $\backsim$ 7500 \AA.}
     \label{Fig7.2}
   \end{figure}

\begin{figure}
   \centering
      \includegraphics[trim = 5mm 5mm 0mm 10mm, clip, width=8cm]{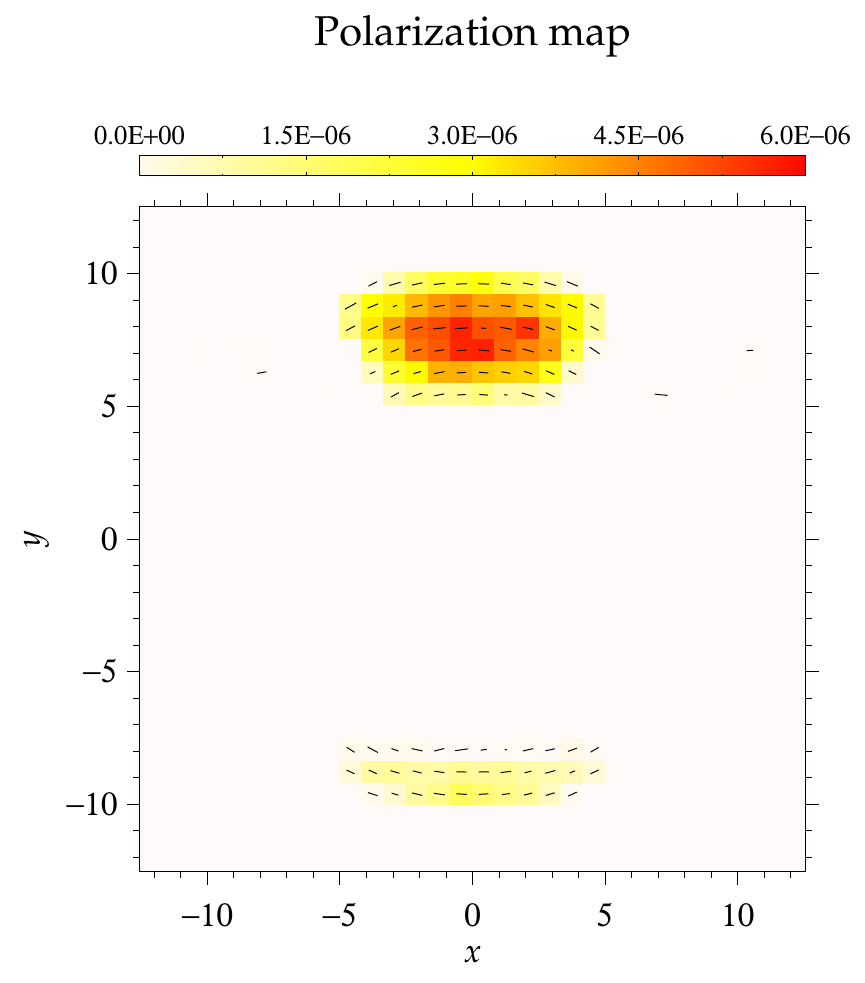}
      \includegraphics[trim = 5mm 5mm 0mm 10mm, clip, width=8cm]{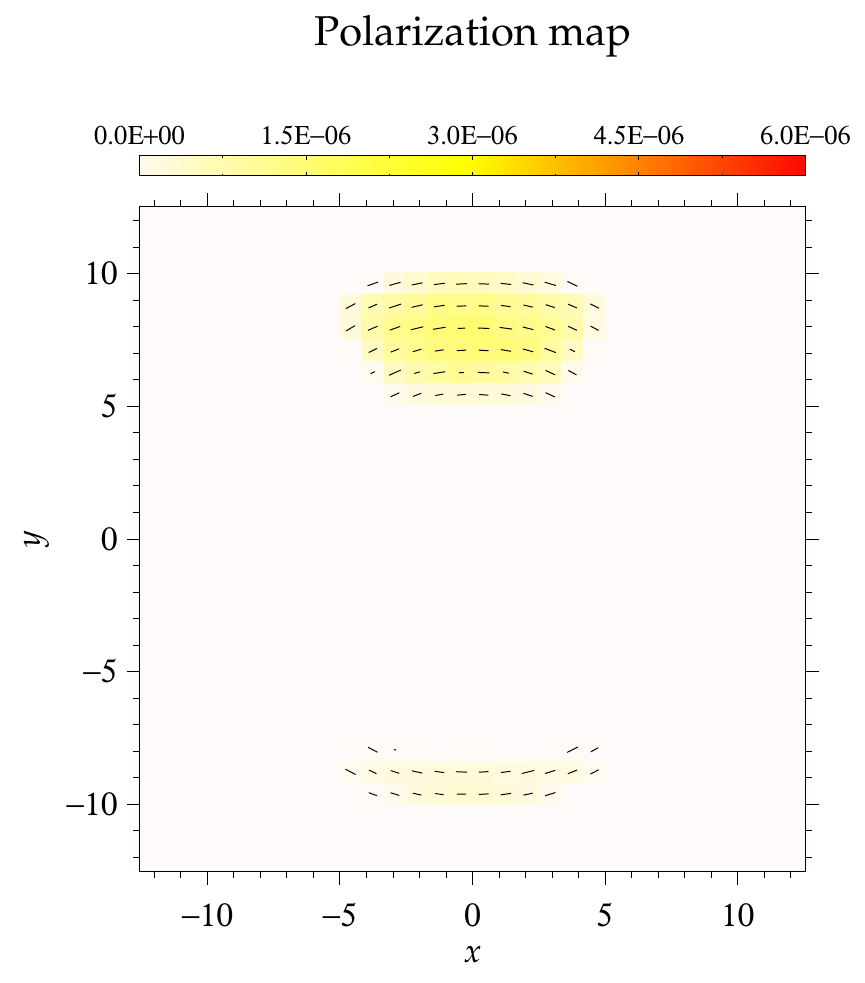}
      \caption{Modeling the polarized flux, $PF/F_{*}$, of an AGN composed of an
        equatorial scattering disc, electron-filled, polar outflows and an optically
        thick dusty torus; $PF/F_{*}$ is color-coded and integrated over all wavelengths.
	\textit{Top}: view at $i \backsim 45^\circ$ and $\lambda \backsim$ 2175 \AA;
	\textit{Bottom}: view at $i \backsim 45^\circ$ and $\lambda \backsim$ 7500 \AA.}
     \label{Fig7.3}
   \end{figure}

In Fig.~\ref{Fig7.2} and \ref{Fig7.3} we respectively show the pole-on and
edge-on polarization map at two different wavelengths, 2175~\AA~(UV, top)
and 7500~\AA~(optical, bottom). For all viewing directions a higher $PF/F_{*}$
is observed at 2175 \AA, which mainly is due to the larger spectral
flux $F/F_*$ in the UV because the polarization is almost wavelength-independent
(see Fig.~\ref{Fig7.1}). In Sects.~\ref{sec:combequattorus} and \ref{sec:combtoruswind},
we have explained the importance of additional electron scattering inside the funnel
for the resulting polarization. The polarization maps further illustrate the mechanism.
At a pole on viewing angles, the polarized flux in the UV comes from a larger
surface area around the torus funnel showing that photons being scattered to
these positions have a higher probability to escape to the observer than in
the optical. Some of these UV photons are then scattered again in the polar
outflows and redirected to the observer, which is why the scattered UV-flux
from the winds is more significant at an edge-on view.

\subsection{The impact of geometry and optical depth}
\label{sec:AGNoptdep}

Starting from our base line model of a thermal AGN, we investigate how the
spectropolarimetric results depend on several crucial model
parameters. We compute a grid of models varying the half-opening angle
of the dusty torus and the polar winds as well as the optical depth of
the electron scattering regions. We consider a common half-opening
angle of the torus and the polar outflow and vary it between
$30^\circ$ and $60^\circ$ thereby implicitly assuming that the torus
always collimates the outflow. The different optical depths assumed in
the winds and the equatorial scattering regions can be interpreted as
different mass transfer rates in both the accretion and the ejection
flow.

A major motivation for our model grid is to explore the behavior of
the polarization dichotomy between type-1 and type-2 thermal
AGN. Bearing in mind the results obtained in Sect.~\ref{sec:combined},
we choose a base line model that optimizes the production of parallel
polarization at type-1 lines of sight. This means in particular that
we limit the spatial extension of the polar winds that produce only
perpendicular polarization; in all modeling cases, the outer parts of
the winds still reach out of the torus funnel but their contribution
to the polarization at a type-2 viewing angle remains moderate.
Also, we explore a broad range of optical depths for the wind reaching
down to low values (0.03 -- 1) and we do not include additional dusty,
polar scattering regions located further out. We study a range of
optical depths (0.1 -- 5) and half-opening angles ($10^\circ$, $20^\circ$,
and $30^\circ$) for the equatorial scattering disc thereby covering
its maximum efficiency to produce parallel polarization (see Paper~I).

   \begin{figure*}[h]
   \centering
      \includegraphics[trim = 0mm 55mm 0mm 0mm, clip, width=22cm]{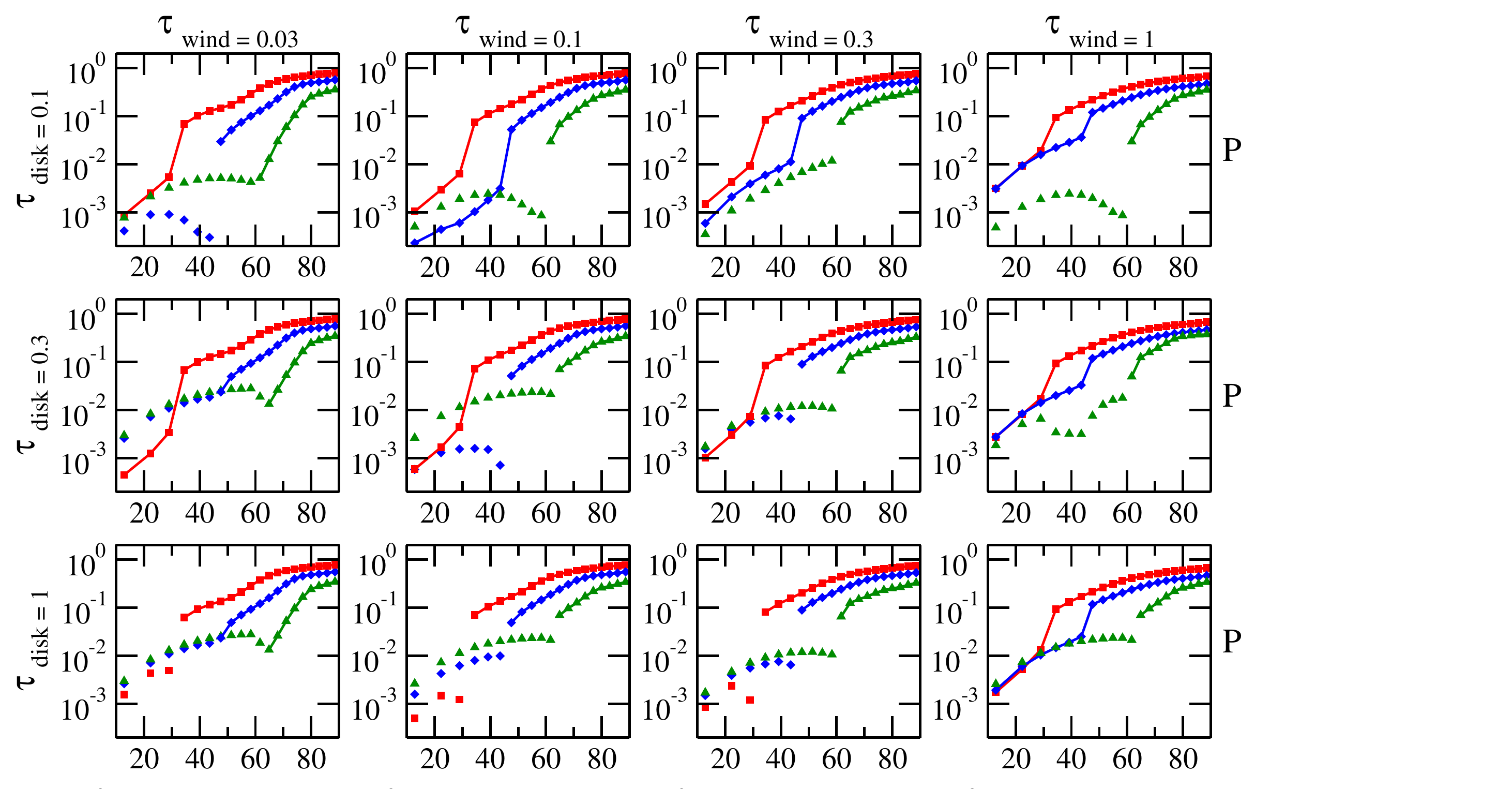}
      \includegraphics[trim = 0mm 0mm 0mm 106mm, clip, width=22cm]{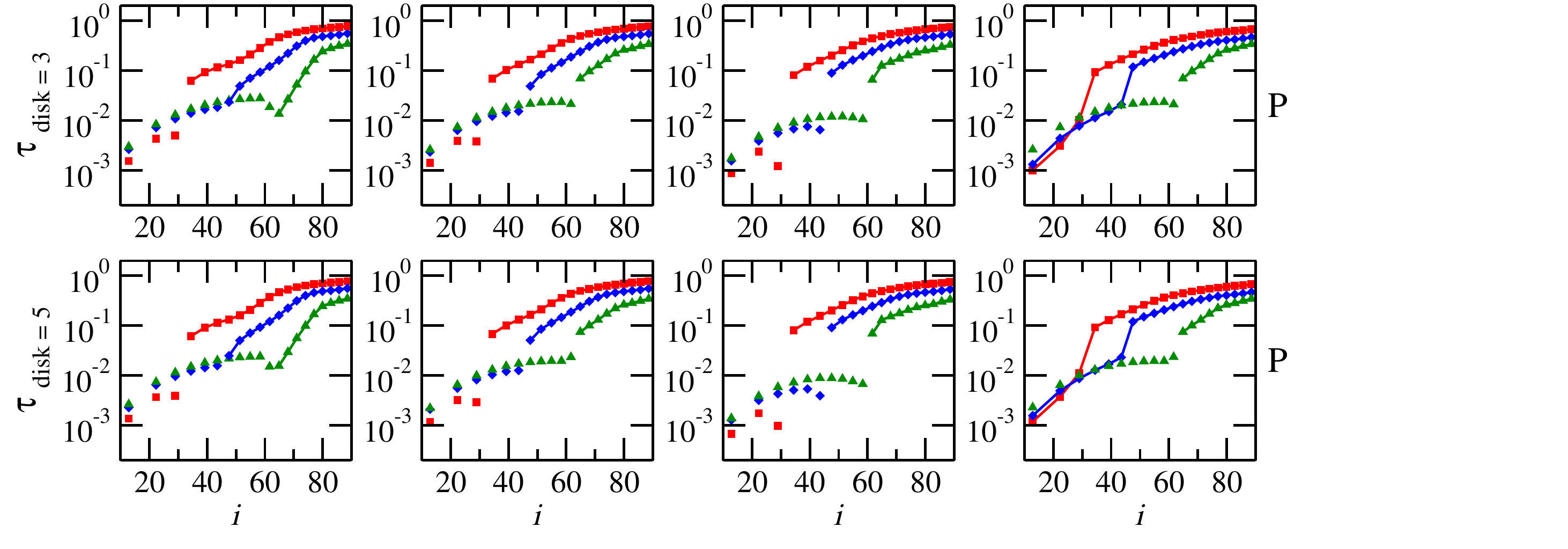}
      \caption{Resulting percentage of polarization, $P$, as a function
           of viewing angle, $i$, for a complex AGN model (see text).
           The half-opening angle of the equatorial scattering disc is
           set to $10^\circ$. \textit{Legend:} The different curves
           denote a common half-opening angles of the torus and the
           polar winds of $30^\circ$ (red squares), $45^\circ$
           (blue diamonds), and $60^\circ$ (green triangles with points
           to the top). The isolated symbols indicate a polarization
           position angle $\gamma = 90^\circ$ (parallel), connected
           symbols stand for $\gamma = 0^\circ$ (perpendicular).
           \textit{From left to right}: increasing the polar outflows
           optical depth from 0.03 to 1; \textit{From top to bottom}:
           increasing the optical depth of the equatorial disc from
           0.1 to 5.}
     \label{Fig7.4}%
   \end{figure*}

   \begin{figure*}[h]
   \centering
      \includegraphics[trim = 0mm 55mm 0mm 0mm, clip, width=22cm]{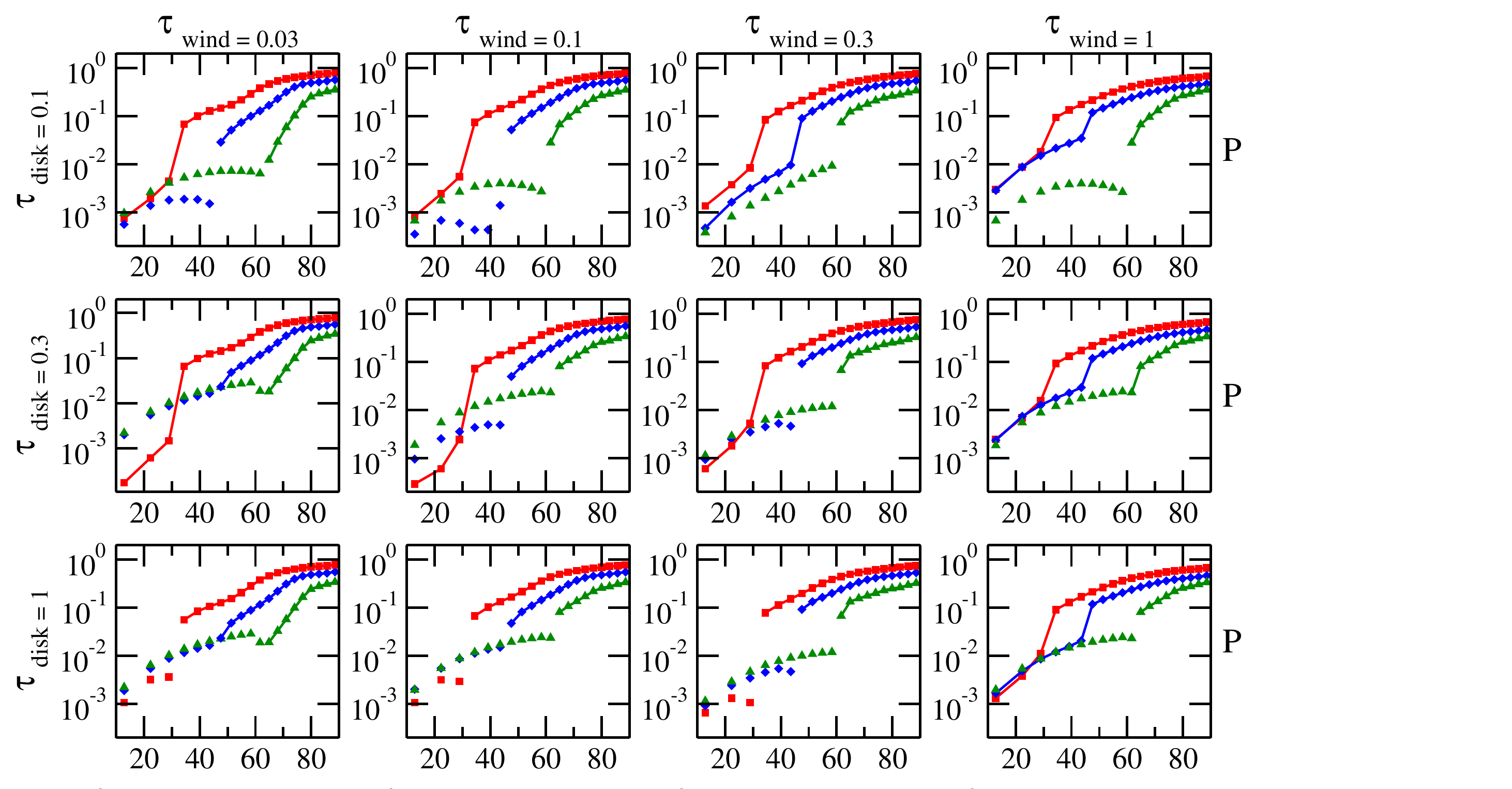}
      \includegraphics[trim = 0mm 0mm 0mm 106mm, clip, width=22cm]{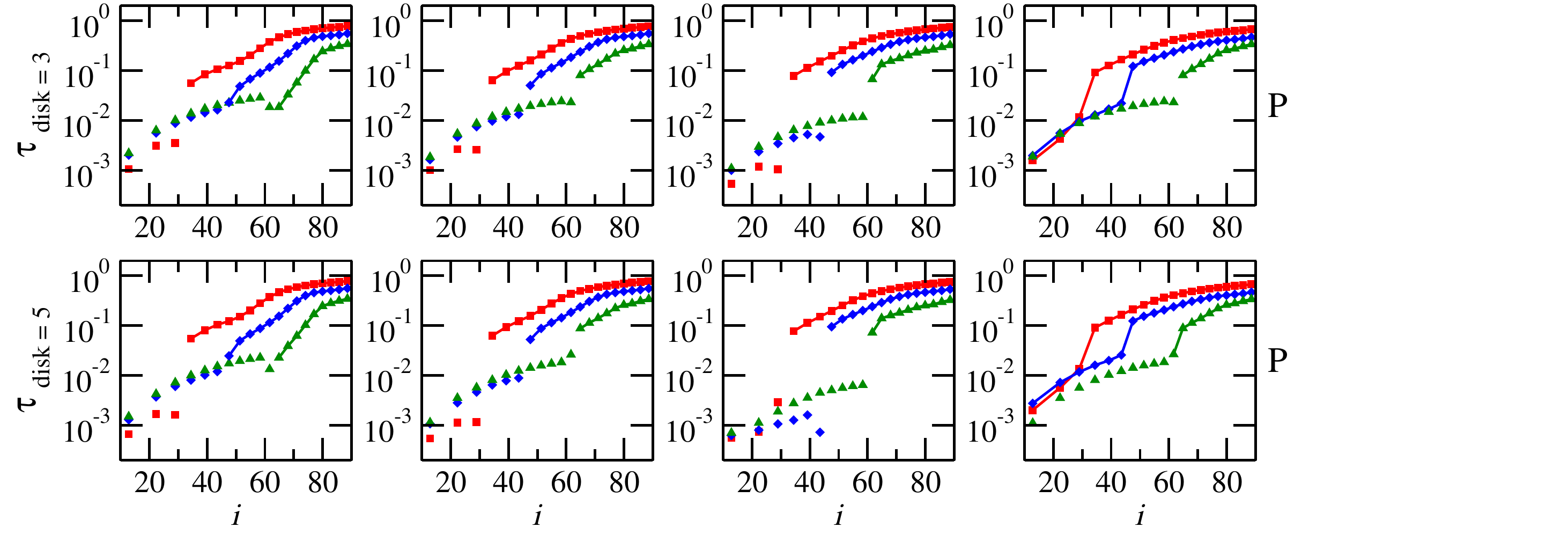}
      \caption{Resulting percentage of polarization, $P$, as a function of
           viewing angle, $i$, for a complex thermal AGN model (see text).
           The half-opening angle of the equatorial scattering disc
           is set to $20^\circ$. The legend is as in Fig.\ref{Fig7.4}}
     \label{Fig7.5}%
   \end{figure*}

   \begin{figure*}[h]
   \centering
      \includegraphics[trim = 0mm 55mm 0mm 0mm, clip, width=22cm]{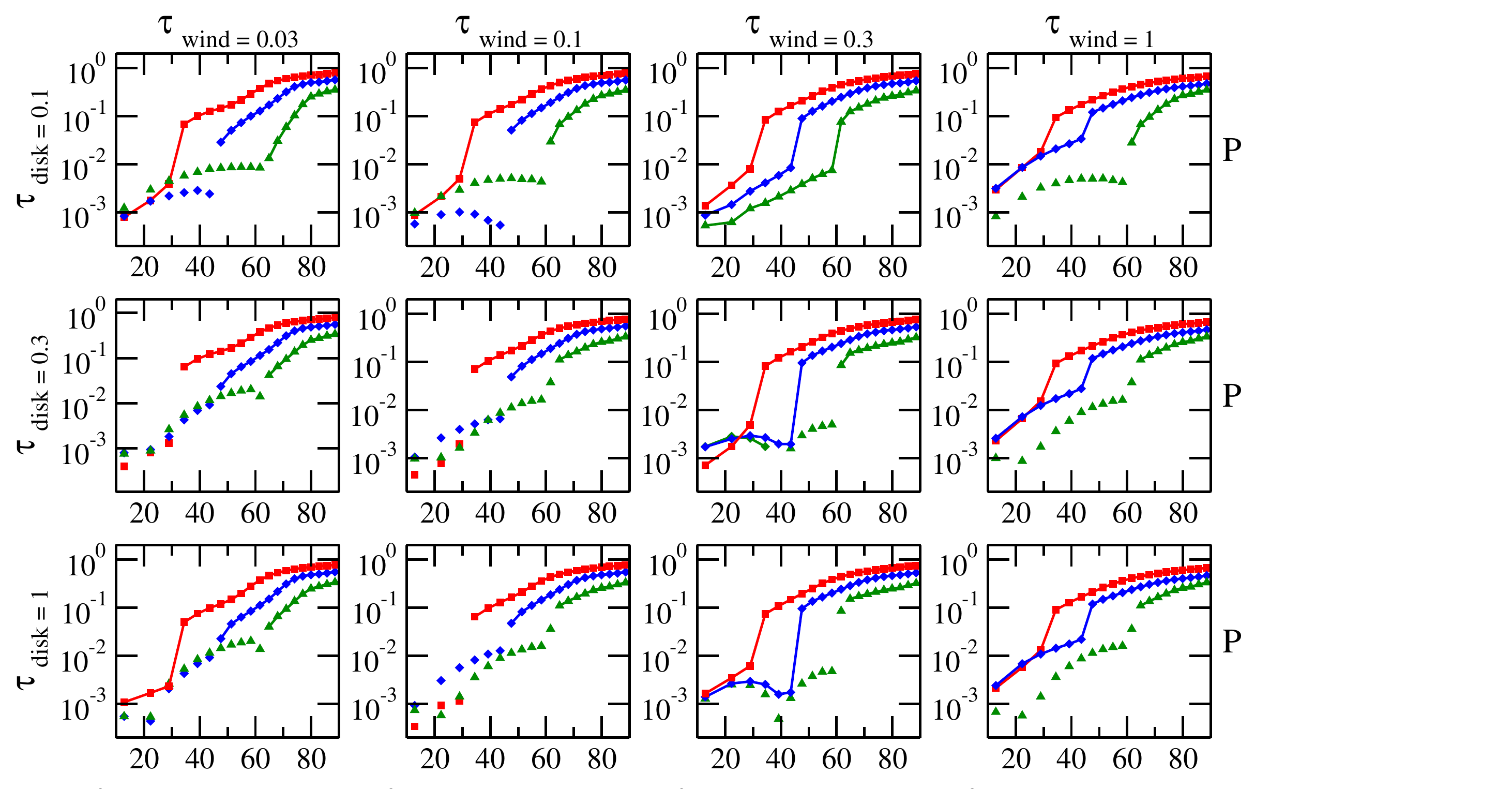}
      \includegraphics[trim = 0mm 0mm 0mm 106mm, clip, width=22cm]{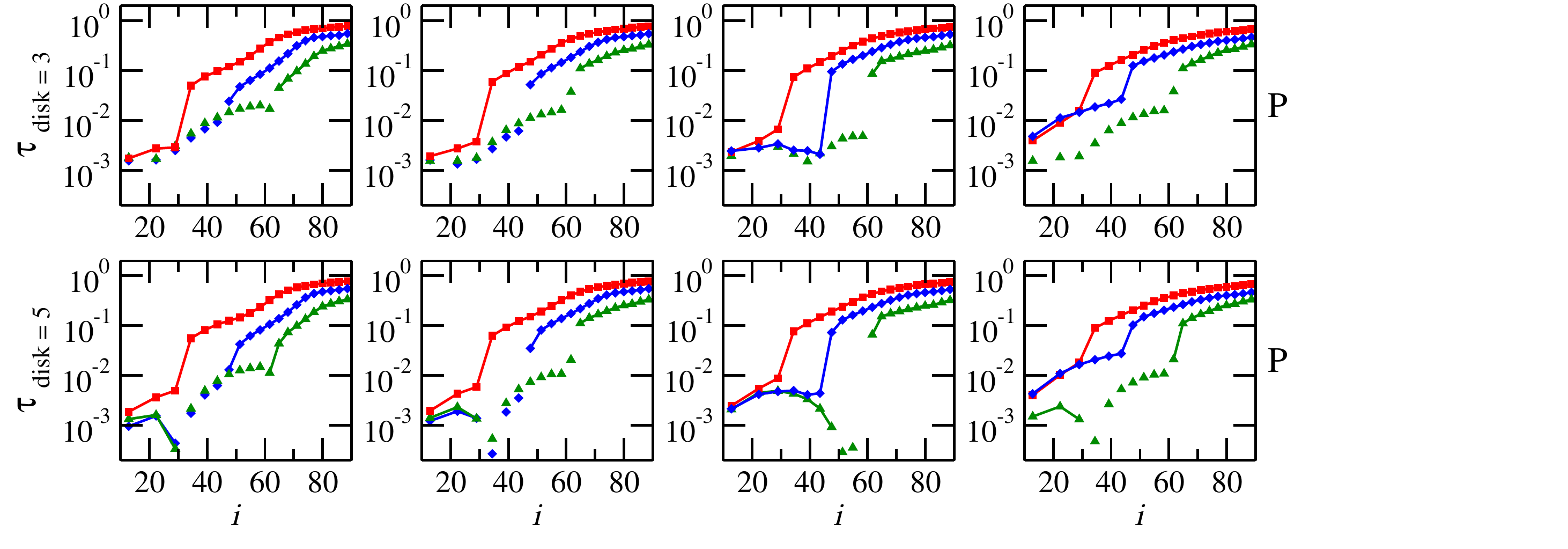}
      \caption{Resulting percentage of polarization, $P$, as a function of
           viewing angle, $i$, for a complex thermal AGN model (see text).
           The half-opening angle of the equatorial scattering disc
           is set to $30^\circ$. The legend is as in Fig.\ref{Fig7.4}}
     \label{Fig7.6}%
   \end{figure*}

We present our results in Figs.\ref{Fig7.4}, \ref{Fig7.5} and \ref{Fig7.6}
for the three half-opening angles of the equatorial scattering region, respectively.
It turns out that the wavelength-dependence of $P$ is rather low therefore the
absolute value of $P$ is averaged in our grid. The geometry of the model
strongly influences the polarization response : a narrow torus and skinny
polar outflows (half-opening angle of $30^\circ$ from the axis) produce
strong perpendicular polarization, as was discussed in \citet{Kartje1995}
and Paper~I. A wide opening angle of the object (half-opening angle of
$60^\circ$ from the axis) acts in the opposite way because it weakens the
perpendicular polarization coming from the polar outflows and, at the same time,
the wide torus produces parallel polarization and thereby reinforces the
polarization signature of the equatorial scattering disc.

Note indeed that for a half-opening angle of $60^\circ$, parallel polarization
is detected at nearly all type-1 lines of sight. Exceptions from this rule
occur for an optically and geometrically thick equatorial disc. If the half-opening
angle of the disc exceeds $30^\circ$, multiple scattering sets in;
then, the mechanism producing parallel polarization becomes less effective and
may lead to a polar-scattered object \citep[see the discussion in][]{Smith2004}
with a relatively low but perpendicular polarization at type-1 viewing angles.
This behavior occurs in particular when the optical depth of the outflow exceeds
0.3 and/or the optical depth of the equatorial disc is larger than 3.

Polar-scattered AGN are very likely to exist also for lower torus half-opening
angles ($\le 45^\circ$ from the axis). They occur when the outflows become
sufficiently optically thick ($\tau \sim 1$) or when the equatorial scattering
is too optically thin ($\tau \sim 0.3$ and smaller). In all other cases,
the usual polarization dichotomy is reproduced: the transition from a parallel
to a perpendicular polarization angle happens when the observer's line of
sight towards the primary source crosses the torus horizon. This behavior
is not much affected by the geometry of the equatorial scattering disc as
long as its half-opening angle stays below $20^\circ$ and the winds are not
too optically thick. It turns out that the most efficient equatorial scattering
geometry to induce parallel polarization in a complex type-1 AGN is obtained
for a disc half-opening angle between $10^\circ$ and $20^\circ$. The exact
position of this optimum depends on Thomson optical depths for equatorial
and polar scattering.

With increasing type-2 viewing angle, the percentage of the perpendicular
polarization rises, which is related to a more favorable scattering angle
for photons coming from the polar outflows. Apart from the scattering geometry,
the behavior of the polarization dichotomy depends on the Thomson optical
depths of both the polar outflow and of the equatorial scattering region.
The observed polarization angle at a given line of sight results from a
competition between the polarization efficiency of, on the one hand side,
the equatorial scattering region always producing parallel polarization
and, on the other hand side, the polar outflows that always imprint
perpendicular polarization. Due to its significant albedo of $\sim 0.57$,
scattering off of the opaque torus has also an important impact,
but here the position angle of the polarization depends on the torus
opening angle.

The polarization percentage at type-2 viewing angles as a function of
the outflow's optical depth does not vary much, in particular for small
or moderate opening angles of the object. The higher the Thomson optical
depth of a medium, the more likely it is that an incident photon is
scattered and polarized. Up to a certain limit the polarization induced
by the medium thus rises, but if it becomes too dense, multiple scattering
starts to depolarize the radiation. Hence, increasing the wind's optical
depth leads to a slightly higher $P$ value until multiple scattering
and depolarization set in and $P$ decreases again. Investigating the
impact of the flared disc's optical depth, we find that optically thin
discs ($\tau \le 0.1$) are inefficient in producing parallel polarization
at type-1 views. To produce parallel polarization, the optical depth of
the disc should be higher than unity. The strongest parallel polarization
is obtained for a scattering disc with $1 < \tau < 3$.

Our results show that the perpendicular polarization of optically-thick winds 
($\tau \sim 1$) predominates over the parallel polarization
coming from the equatorial disc except when the model has a large opening angle.
For a more moderate and especially for a small opening angle, a net parallel
polarization can be produced only when the outflows are sufficiently
optically thin and, at the same time, the optical depth and the half-opening angle
of the equatorial scattering region are in the right range.
The interplay between polar and equatorial scattering may put constraints
on the optical depth of the accretion flow if the optical depth and the
geometry of the outflow in a given type-1 AGN can be estimated independently,
for instance from the shape of UV and X-ray absorption lines. If it
turns out that the effective optical depth in the outflow is above 0.3
and if at the same time the AGN reveals parallel polarization, we can
conclude that a flattened accretion flow between the dusty torus and the
accretion must be optically thick and in the range $1 < \tau < 3$.



\section{Discussion}
\label{sec:discuss}

An important motivation of this work has been to investigate the net polarization
due to radiative coupling between the continuum source and different
reprocessing regions of an AGN. 

We consider electron scattering in an equatorial disc and in polar outflows
as well as dust reprocessing by an obscuring torus. At first approach, it is
reasonable to assume that the torus funnel collimates the polar outflows,
which simplifies this multi-parameter problem. As a result, four parameters
remain: the half-opening angles of the equatorial disc and of the torus/winds
as well as the optical depths of the two electron scattering regions.
Note that the torus is always considered to be optically thick. It turns
out that the net polarization as a function of the viewing angle is most
sensitive to the half-opening angle of the system and to a lesser extend
to the geometry of the scattering disc, at least as long as it does not
exceed a half-opening angle of $20^\circ$ measured from the equatorial
plane. A thicker equatorial disc favors multiple scattering and a partial
disappearance of the parallel polarization. The optical depths of both
electron scattering have a significant impact once they exceed $\sim 0.3$.

When compared to the observed optical/UV polarization, our systematic
modeling can thus put constraints on certain AGN properties, as we
further discuss in the following.

\subsection{Comparison with previous modeling work}

There are a number of radiative transfer codes that include optical/UV polarization
and that are applied in AGN research. Some of them only refer to continuum
polarization \citep[see e.g.][]{Brown1977,Manzini1996,Kartje1995,Wolf1999,Watanabe2003}
while others include details about the polarization of spectral lines \citep[see e.g.][]{Young2000,Wang2007}.
Earlier works often apply semi-analytical or single-scattering methods.
With greater computational power, Monte-Carlo methods that include multiple scattering and
more arbitrary geometries became suitable. In this section, we relate and compare our
results to previous modeling work that is similar to ours.

To our knowledge, \citet{Wolf1999} were the first and so far the only authors to
simulate polarization images in the context of AGN. Their Monte Carlo code is based
on earlier work by \citet{Fischer1994} that also was an important inspiration during
the early development of {\sc stokes}. In their application to AGN, \citet{Wolf1999}
included Mie scattering by two types of dust and electron scattering. Several
reprocessing regions were investigated and gave rise to theoretical spectra and
images of the intensity and polarization as a function of the viewing angle.

The AGN model adopted by \citet{Wolf1999} is composed of: (1) an obscuring dusty
torus, (2) an electron-filled inner region and bi-conical outflow, and (3) an outer,
dusty, bi-conical outflow. The authors showed that multiple scattering has an impact
on the polarization as soon as the optical depths involved are higher than $0.1$.
A comparison to the geometry that we adopted in our work is not straightforward as
there are differences in the details. For instance, the dusty torus employed by \citet{Wolf1999}
includes a cylindrical funnel. However, in our modeling we find a similar dependence
of the polarization percentage on the viewing angle and also a good match in
polarization images. We would like to point out that \citet{Wolf1999} also investigate
the effects of electron scattering inside the torus funnel. Their figure~13 shows
that the resulting polarization spectrum becomes more concave for more centralized
distributions of electrons. This is in agreement with our results and in
Sects.~\ref{sec:combequattorus} and \ref{sec:combtoruswind} we gave an explanation
of the effect based on the Mie scattering phase function and scattering geometry.
Note that \citet{Kartje1995} also considered the effects of additional electron
scattering inside the torus but he did not vary the geometry of the electron distribution.

By also including a discussion on the net polarization angle as a function of viewing
direction, we here extend the modeling done by \citet{Wolf1999}, who did not discuss
the polarization dichotomy. Producing parallel polarization for type-1 objects
must be very hard in their adopted geometry as it lacks an equatorial scattering disc.
Equatorial scattering, however, is included in the Generic Scattering Model by
\citet{Young2000} and was further applied by \citet{Smith2004}
to explain strong, yet systematic variations of the polarization across broad emission
lines. Aside from the interesting dynamical constraints the model provided for the BLR,
it also predicts a sequence in viewing angle for the continuum polarization.
This sequence runs from type-1 AGN with parallel polarization to the so-called
polar scattering dominated type-1 objects with perpendicular polarization
(see Sect.~\ref{sec:PolScattDom}) and ends at the type-2 AGN, again with perpendicular
polarization.

The Generic Scattering Model analytically combines the polarized flux from the equatorial
and polar scattering regions in a semi-analytical manner assuming single scattering.
In \citet{Smith2004}, the authors chose a Thomson optical depth of ~0.2 for both electron
scattering regions, a torus half-opening angle of $45^\circ$ and an equatorial
scattering disc with a half-opening angle of $20^\circ$ or $30^\circ$. Such a configuration
produced the strongest polarized flux with parallel polarization at type-1 viewing angles.
Our Figs.~\ref{Fig7.5}~and~\ref{Fig7.6} indicate that for a comparable case modeled
with {\sc STOKES}, the resulting polarization at type-1 viewing angles is somewhat low
compared to the observed range of generally $\sim 0.3\% - 1.3\%$ \citep[see Table ][]{Smith2002}
or can even adopt a perpendicular polarization angle. One should take into account,
however, that in our model we use a different geometrical shape for the equatorial
scattering disc, a uniform density in all scattering regions, and we also consider the
effects of multiple dust and electron scattering. Therefore, a comparison between both
modeling schemes is not straightforward. We agree with \citet{Smith2004} in that a rather
flat equatorial scattering disc is required (half-opening angle $<= 20^\circ$),
but we prefer it to have a larger optical depth. Also, we find that systems with a torus
half-opening angle significantly larger than $45^\circ$ are better suited to produce the
observed amount of parallel polarization in type-1 objects. We discuss in more detail the
implications of our modeling on the polarization dichotomy in the following section.


\subsection{Polarization at type-1 and type-2 viewing angles}
\label{sec:discuss-1-2}

In Sect.~\ref{sec:AGNoptdep}, we show that for all modeling cases the
polarization detected at a type-2 angle is strong and perpendicular.
Even the most efficient equatorial scattering cannot force the net polarization
to be parallel. If a thermal type-2 object with parallel polarization is observed
it cannot be explained by our current model and a possible re-interpretation
then would be that the small scale radio axis does not coincide with
the symmetry axis of the dusty torus and/or the outflow.

The modeled perpendicular polarization at type-2 viewing angles covers
a range of 30\% to 80\% for edge-on viewing angles. This would exclude
the vast majority of Seyfert-2 galaxies that were analyzed by \citet{Kay1994}
and mostly showed polarization up to only 10\%. Under the reasonable
assumption that the distribution of viewing directions towards Seyfert
galaxies is uniform, our modeling therefore suggests that the half-opening
angle of most Seyfert-2 galaxies should be large, i.e., at least $60^\circ$.
This conclusion would be consistent with the evolution model for Seyfert
galaxies and the type-1/type-2 number counts given by \citet{Wang2007} but
exceed the ones by \citet{Schmitt2001}. On the other hand, it is not clear
that the assumption of a universal half-opening angle for all dusty tori
is realistic.

The situation is different for Seyfert-1 galaxies. The spectropolarimetric
data collected by \citet{Smith2002} and \cite{Smith2004} show low polarization
percentages mostly staying below 1\%. However, it can have a parallel,
perpendicular and sometimes also intermediate position angle. On the modeling side,
the situation is equally ambiguous: narrow (i.e., geometrically thick) tori
and outflows favor the production of perpendicular polarization for type-1 AGN
while systems with wide (i.e., geometrically thin) tori and winds rather
produce type-1 AGN with parallel polarization, even when the outflows are
optically thick. For the case of narrow tori, ionized outflows with an electron
optical depth ranging from 0.03 to 1 still allows us to produce parallel
net polarization at a type-1 viewing angle. Denser winds ($\tau_{\rm V} \ge 1$)
then switch the polarization to perpendicular. In any case the resulting
polarization percentage is low and therefore matches the observed values
rather well.


However, it is noteworthy to point out that some peculiar type-1 AGN
like Mrk 231 \citep{Gallagher2005} and many Warm Infrared Ultraluminous AGN in the 
survey of \citet{Hines1994} show a relatively high, up to 4\% polarization degree. Our grid of models
is actually not able to reproduce such high percentage of polarization despite
the extended space of parameters. It indicates that the clue to this high polarization
level must be different from usual optical depth and half opening angle considerations.
In blazar AGNs, synchrotron emission can produce
degrees of polarization superior to 30\% in the UV/optical range while being strongly variable 
in time \citep{Smith1996}. Considering extra-galactic, non-blazar, radio-loud objects the 
amount of polarization is lower, as in the case of OI 287 exhibiting $P \sim$ 8\% \citep{Angel1980}. 
In addition to the lack of polarization variability in time, \citet{Rudy1988} and \citet{Goodrich1988}
proved using spectropolarimetric observations of OI 287 that the emission lines are polarized in a 
similar way to the continuum, ruling out a synchrotron origin.
In addition to scattering, dichroic extinction might also increase the level linear polarization
in face-on AGN if the dust grains are aligned by magnetic fields. \citet{Kartje1995}
showed that for not too optically-thick ($\tau_V \le 10$) dusty torii with partially aligned grains,
the expected percentage of polarization might rises by 1.5\%. Yet, in denser torii, scattering
and dichroic extinction will partially cancel, causing $P$ to increase more slowly. As our models present
optically-thick torii ($\tau_V = 750$), dichroic extinction will not be sufficient to explain
the large polarization percentages detected in \citet{Gallagher2005} or \citet{Hines1994}.
A third explanation may come from IRAM PdBI observations of the extended outflows in Mrk 231 
\citet{Cicone2012}. They discovered a possible extension of the CO(1-0) transition blue wing a 
few arcseconds to the north which stays unexplained. If we assume that the ionized polar outflows
sustain the same half opening angle and direction as the massive AGN-driven winds
that can be considered as a naturally extended, dust mixed extension \citep{Antonucci1993,
Cracco2011}, the polar outflows in Mrk 231 might be non axisymmetric,
similarly to NGC 1068 \citep{Raban2009,Goosmann2010}. As the system
is seen from the pole, the overall picture would be less symmetric due to the
winds inclination and then create an higher polarization degree.
However, while the explanation of the high $P$ detected in particular type-1 AGN might be 
an addition of the last three mechanisms proposed above, it is important to recognize that we have 
restricted the analysis in this paper to {\em axially-symmetric} situations while real AGNs are 
certainly messier and simple asymmetry could be the explanation. The significant long-term variability 
of the degree of polarization in Mrk~231 while the angle stays constant is qualitatively similar to 
what \citet{Gaskell2012} discuss for NGC~4151. As with NGC~4151, polarimetric reverberation mapping 
could give important information about the location of the main scattering region responsible for the 
relatively high polarization in Mrk~231.

The equatorial scattering region is important in determining the net
polarization position angle for a type-1 view. If the half-opening angle
of the system is intermediate or small ($< 45^\circ$ from the axis),
equatorial scattering with sufficient Thomson optical depth can still
lead to parallel polarization at a type-1 viewing angle. The degeneracy
between the impact of the torus/winds on the one hand side and the equatorial
scattering region on the other must be resolved by taking into account
additional observables. We mentioned before that the analysis of UV and
X-ray absorption lines in a type-1 AGN may constrain the optical depth
in the outflow, but possibly also the spectral slope of the polarization
can give a hint to the amount of equatorial scattering. When comparing the
polarization spectra for a type-1 AGN in Fig.~\ref{Fig6.3} and Fig.~\ref{Fig7.0}
to the one in Fig.~\ref{Fig6.5}, it turns out that electron scattering
inside the polar cones produces a flat polarization spectrum, whereas
the equatorial scattering induces a concave shape in $P$. Comparing the
broad-band continuum shape of the spectral flux to the polarization
percentage can thus help to detect electron scattering that occurs deep
inside the torus funnel.

In our modeling of AGN polarization we assumed very favorable conditions
for the production of parallel polarization: the spatial extension of the
ionization cones is relatively small and we do not consider additional
dust scattering in farther away, polar regions. Still, under these
favorable conditions it turned out that the torus half opening should
be rather large and the optical depth of the equatorial scattering
region has to be of the order of 1 -- 3 to maximize the percentage
of parallel polarization in type-1 views. Lower ($\tau < 0.3$) optical
depths will lead to the disappearance of parallel signatures and
higher values ($\tau > 3$) will decrease $P$ due to depolarization effects.

\subsection{Constraining particular AGN classes}

\subsubsection{Bare and naked AGNs}

The orientation unification model \citet{Antonucci1993} explains
the observed differences between type-1 and type-2 Seyfert galaxies as
originating from orientation effects, but it is now recognized
that orientation is not the whole explanation \citep{Antonucci2011}.
For example some Seyfert-1 AGN show weak or even no intrinsic warm absorption 
along the observer line-of-sight \citep{Weaver1995, Patrick2011}. These ``bare'' 
AGN allow for direct observation of the central engine

\citet{Hawkins2004} reported the presence of ``naked'' Seyfert-2 AGN, resulting
from a survey of about 800 quasars and emission line galaxies. This new subclass
is characterized by the weakness or even absence of the BLR and is coupled
to strong optical variability, suggesting that the inner regions are not anymore
hidden by an obscuring media. Those unabsorbed Seyfert 2 galaxies, which are part of a more general class called
``non hidden BLR'' type-2 AGNs = NHBLR type-2s \citep{Gliozzi2007}, were mostly studied in the X-ray
domain \citep{Panessa2009, Gliozzi2010}, i.e., without the help of spectropolarimetric
tools.

For both of these subclasses of AGNs, spectropolarimetric
measurements are not available and the absence of the cited reprocessing regions is
based on X-ray analysis. In this paper, we present a modeling case
(see Sect.~\ref{sec:combtoruswind} for the ``naked'' model and Sect.~\ref{sec:combequattorus}
for the ``bare'' case) that could be compared to optical and UV observations.
Such a broadband spectropolarimetric comparison could bring more information about
the real nature of ``naked'' and ``bare'' AGNs as well as
constraining the real morphology of these particular galaxies.

As is discussed at length in \citet{Antonucci2011}, it is often not clear whether
the failure to detect a hidden BLR in polarized flux means that the AGN in a
non-thermal AGN lacking a Big Blue Bump and BLR or whether the failure is due
the absence of a suitable scattering region to give us the needed ``periscopic''
view of the central regions. Polarization modeling can help us assess these two
possibilities.

\subsubsection{Polar scattering dominated AGN}
\label{sec:PolScattDom}

A particular group of Seyfert-1 galaxies are known as
\textit{polar scattering dominated AGN} \citep{Smith2002}. In these type-1
objects the resulting polarization angle is perpendicular and relatively weak.
In our reprocessing model, polar scattering dominated AGN generally have more
collimated outflows and tori with steep inner surfaces. Possibly, these
objects also reveal a low polarized flux induced by (inefficient) equatorial
scattering. However, if the polarization induced by equatorial scattering is
strong, i.e., the equatorial scattering region is geometrically thin and has
an optical depth of $1 < \tau < 3$, then the outflows must have a sufficiently
column density to still produce a net polarization that is perpendicular.

Spectropolarimetric observations show that the optical polarization of
polar-scattered AGN typically rises towards the blue. According to \citet{Smith2004},
this wavelength-dependence is due to dust extinction occurring along a line
of sight that passes very close to the torus horizon, while still denoting
a type-1 viewing angle. Material form the uppermost layers of the torus then
produces a visible extinction of $1 < A_{\rm V} < 4$ and causes the gradient
in $P$. Our modeling shows such a trend only for a torus half-opening angle
of $45^\circ$. It leads to the very opposite spectral slope of $P$ for more narrow
tori. Thus, also the observed visible extinction in polar-scattering dominated
AGN indicates that the dusty torus should be rather wide than narrow. Note that
for a torus half-opening angle of $60^\circ$ no polar scattering dominated
AGN exist in our model.

We will return to this issue in future work when we analyze in more detail the
wavelength-dependence of our models also for clumpy media (see below), dust
scattering in the polar regions, and different prescriptions for the dust.
We have shown already that the polarization spectrum is quite sensitive to the
dust composition and grain size distribution \citep{Goosmann2007c,Goosmann2007b}.

\subsection{More general geometries for the reprocessing regions}

We have investigated the effects on our model results, when deviating from
the adopted scales and geometry of certain scattering regions. This includes,
in particular, the size of the obscuring torus and the geometrical shape of
the equatorial scattering ring. It turns out that a more compact torus with
the same geometry and half-opening angle does not significantly alter the
results. The same holds true when replacing the toroidal scattering ring in
the equatorial disc by a flared disc.

Our modeling assumes a uniform constant density across a given scattering
region. This a considerable simplification. Observations of ionized outflows
in AGN show that the winds should be at least partially fragmented; Using the
spectropolarimetric capacities that were available on $HST$, \citet{Capetti1995}
obtained spectropolarimetric and polarization imaging data of NGC~1068 in the
UV. The authors detected a clumpy structure in polarization percentage and
position angle. In most AGN, the ionized outflows cannot be spatially resolved
and the polarized flux from a clumpy structure is integrated and only gives
a net polarization for the entire outflow. Note that by following this principle,
\citet{Ogle2003} measured an average optical depth for the ionized outflows
of NGC~1068 from spatially resolved X-ray observations with the HETGS on-board
$Chandra$. The ratio of scattered X-ray continuum and emission line flux from
the central region of the AGN gave an estimate of the Thomson optical depth
of $0.27 \pm 0.08$. This confirmed the hypothesis that X-ray, NLR and the
optical/UV/X-ray reprocessing regions are likely to coincide.

There are more and more indications from observations and modeling in the
IR that also the obscuring torus should have a clumpy structure
\citep{Nenkova2008a,Nenkova2008b,Nenkova2010,Schartmann2008,Hoenig2010,Heymann2012}.
For the future, a clumpy torus will be important to add to our modeling.
We currently conduct preliminary studies for the expected optical/UV polarization
of a centrally irradiated torus with a fragmented structure. So far,
we find that, on the one hand side, the normalization of the polarization
spectra is significantly lower, while its spectral shape does not change much
compared to the case of a uniform density torus. On the other hand side,
a clumpy torus allows more flux to escape from the central funnel, even at
type-2 viewing angles that are strongly obscured by a uniform-density torus.
Since the polarization efficiency is measured by the polarized flux emerging
from the torus we thus expect that the torus remains an important scattering
region even when it has a fragmented structure. Nonetheless, the lower
polarization produced by a clumpy torus should lead to a weaker net polarization
detected at type-2 viewing angles (see Sect.~\ref{sec:AGN}). This would help
to bring our modeling results to better agreement with the observed
spectropolarimetric data for Seyfert-2 galaxies \citep{Kay1994}.

More general considerations about a clumpy but continuous accretion
flow are currently discussed in the literature
\citep[see][and references therein for an overview]{Elitzur2007} and therefore
should be considered systematically in modeling work such as the one
presented here. The clumpy nature of the dusty torus might also resolve the
known issue that a geometrically thick torus lacks stability against self-gravity
and therefore cannot be in hydrostatic equilibrium. The torus could rather
be interpreted as a dynamical region of flowing, optically-thick clouds.
Note, however, that our modeling results and the observed parallel polarization
in type-1 AGN as well as the moderate perpendicular polarization in type-2
objects anyway argue against a geometrically thick torus (see Sect.~\ref{sec:discuss-1-2}).

Finally, we should emphasize that in this work we always assume the axes and
half-opening angles of the dusty torus and the ionized outflows to be identical.
A multi-wavelength analysis done by \citet{Raban2009} suggests, however,
that the bi-conical outflow in NGC~1068 should be inclined by $\sim 18^\circ$
with respect to the axis of the obscuring torus. The collimation effect on
the ionized outflows might thus not be an obvious (although greatly simplifying)
assumption in our modeling. We are going to investigate the consequences of
such a misalignment for the observed polarization spectra and images in a
forthcoming paper (Marin et al., in prep.). This work is part of a multi-wavelength
modeling attempt that also includes predictions for future X-ray polarimetry
observations of NGC~1068 \citep{Goosmann2010}. Some preparatory work on the
optical/UV polarization signatures of the misaligned outflows in NGC 1068 at
different azimuthal and polar viewing angles can be found in \citet{Marin2012}.
It shows that the misalignment can lead to slightly higher parallel polarization
in type-1 AGN and to systematically lower perpendicular $P$ at a type-2 view. The
latter effect would approach the model to the observational results by \citet{Kay1994}.


\section{Summary and perspectives}
\label{sec:conclu}

We have upgraded the radiative transfer code {\sc STOKES} by implementing a
more efficient random number generator and by adding the capability to model
polarization imaging. The new version of the code is again made publicly available.
We tested the new routines by reanalyzing the polarization signature of individual
reprocessing regions, which then served as a basis for more sophisticated model
setups.

We have combined three reprocessing components in order to analyze the complex
radiative coupling occurring in Seyfert galaxies: (1) an ionized, equatorial
scattering disc, (2) an optically- and geometrically-thick, dusty torus, and (3)
an ionized, bi-conical outflow along the polar directions. We then computed the
resulting polarization as a function of wavelength, projected position on the sky,
and viewing angle. Our modeling shows that including multiple scattering effects
inside a given scattering region and between several scattering regions is important.

As limiting cases, the model grid we present also covers polarization due to scattering in
so-called ``bare'' and ``naked'' AGN.

In the general
case of a unified model, it is possible to reproduce the observed dichotomy of the
polarization angle for thermal AGNs within a limited range of geometries and optical depths:

\begin{enumerate}

\item A flat, equatorial scattering region with $1 < \tau < 3$ is required to
generate the observed parallel polarization in type-1 AGN. This indicates
that the accretion flow at the outer edge of the accretion disc
in Seyfert-1 galaxies is optically thick.

\item A wide half-opening angle ($\sim 60^\circ$) for the torus helps to produce
parallel polarization, whereas narrow tori and/or a higher optical depth of the
polar outflow produce polar-scattering-dominated AGNs 
(The same AGNs show perpendicular polarization when seen at a type-1 viewing angle).

\item At type-2 viewing angles, all the cases we modelled produced perpendicular
polarization that often is significantly stronger than the observed polarization
percentage. Again a wide half-opening angle of the torus and the wind helps to lower
this discrepancy with respect to the observations. It is likely, though, that a
more important moderation of the perpendicular polarization is due to a clumpy
structure.

\item Detailed observations and modeling of the polarization spectrum may give
hints on the presence of an equatorial electron scattering inside the torus funnel.
We found that the equatorial scattering region leads to a more concave polarization
spectrum than dust scattering off the inner torus walls alone. This type of modeling
can help to explore further the nature of AGNs that intrinsically seem to lack a BLR.

\end{enumerate}

In this paper we have explored, in a preliminary and so-far purely theoretical manner,
the effects of multiple reprocessing in AGNs. In ongoing and future work
we are also considering irregular reprocessing media and non-axisymmetric setups. 
We will then proceed to compare our modeling to spectropolarimetric and imaging observations
of individual objects.

\acknowledgements The authors would like to thank the anonymous referee for useful and constructive comments.
This research has been supported by the French GdR PCHE, the mutual exchange
program maintained by the French CNRS and the Academy of Sciences of the Czech Republic.
FM is grateful for financial and technical support from the Czech project COST-CZ LD12010. 
RG is grateful to the University of Valpar\'iso for their hospitality.
MG is grateful for support from the GEMINI-CONICYT Fund of Chile through project N{\degr}32070017 and 
FONDECYT N{\degr} 1120957. Finally, MD acknowledges support from the project RVO:67985815 and COST-CZ LD12010.

\bibliographystyle{aa}
\bibliography{biblio}

\end{document}